\documentclass[twocolumn]{aastex63}
\usepackage{graphicx}
\usepackage{epsfig}
\usepackage{times}
\usepackage{natbib}
\usepackage{url}
\usepackage{color}
\usepackage{amssymb,amsmath}
\usepackage{hyperref}
 \usepackage{booktabs}
 \usepackage{float}
\usepackage{subfigure}
\usepackage[normalem]{ulem}
\newcommand{\Mm}{{\mathrm{\, Mm}}}
\newcommand{\km}{{\mathrm{\, km}}}
\newcommand{\kms}{{\mathrm{\, km \; s^{-1}}}}

\newcommand{\densunits}{{\mathrm{\, kg \; m^{-3}}}}

\newcommand{\secs}{{\mathrm{\; seconds}}}
\newcommand{\mins}{{\mathrm{\; minutes}}}
\newcommand{\G}{{\mathrm{\, G}}}
\newcommand{\mysize}{0.3333}
\newcommand{\mysizec}{0.235}
\DeclareMathOperator\erf{erf}
\newcommand{\vlong}{{v_\mathrm{long}}}

\newcommand{\csound}{{c_\mathrm{sound}}}

\newcommand{\vtransz}{{v_\mathrm{trans, \, z}}}
\newcommand{\vtransy}{{v_\mathrm{trans, \, y}}}

\newcommand{\zdip}{{z_\mathrm{dip}}}

\newcommand{\vcm}{{v^{cm}}}
\newcommand{\vcmpar}{{v_\parallel^{cm}}}
\newcommand{\vcmperp}{{v_\perp^{cm}}}
\newcommand{\vcmperpy}{{v_{\perp\, y}^{cm}}}
\newcommand{\vcmperpz}{{v_{\perp\, z}^{cm}}}

\newcommand{\vshock} {{V_\mathrm{sh}}}

\newcommand{\separ} {W_d}

\shorttitle{LAO triggered by a jet}
\shortauthors{Luna and Moreno-Insertis}

\begin{document}

\title{Large-amplitude prominence oscillations following the impact by a coronal jet}

\author[0000-0002-3841-313X]{Manuel Luna}
\affiliation{Departament F{\'i}sica, Universitat de les Illes Balears, E-07122 Palma de Mallorca, Spain}
\affiliation{Institute of Applied Computing \& Community Code (IAC$^3$), UIB, Spain}
\affiliation{Instituto de Astrof\'{\i}sica de Canarias, E-38200 La Laguna, Tenerife, Spain}
\affiliation{Departamento de Astrof\'{\i}sica, Universidad de La Laguna, E-38206 La Laguna, Tenerife, Spain}
\correspondingauthor{Manuel Luna}
\email{manuel.luna@uib.es}
\author{Fernando Moreno-Insertis}
\affiliation{Instituto de Astrof\'{\i}sica de Canarias, E-38200 La Laguna, Tenerife, Spain}
\affiliation{Departamento de Astrof\'{\i}sica, Universidad de La Laguna, E-38206 La Laguna, Tenerife, Spain}

\begin{abstract}
Observational evidence shows that coronal jets can hit prominences and set them in motion. The impact leads to large-amplitude oscillations (LAOs) of the prominence. In this paper we attempt to understand this process via 2.5D MHD numerical experiments.
In our model, the jets are generated in a sheared magnetic arcade above a parasitic bipolar region located in one of the footpoints of the filament channel (FC) supporting the prominence. The shear is imposed with velocities not far above observed photospheric values; it leads to a multiple reconnection process, as in previous jet models. Both a fast Alfv\'enic perturbation and a
slower supersonic front preceding a plasma jet are issued from the reconnection site; in the later phase, a more violent (eruptive) jet is
produced. The perturbation and jets run along the FC; they are partially reflected at the prominence and partially transmitted through it.
There results a pattern of counter-streaming flows along the FC and oscillations of the prominence.  The oscillations are LAOs (with amplitude above $10\kms$) in parts of the prominence both in the longitudinal and transverse directions.
In some field lines, the impact is so strong that the prominence mass is brought out of the dip and down to the chromosphere along the FC. Two cases
are studied with different heights of the arcade above the parasitic bipolar region, leading to different heights for the region of the prominence
perturbed by the jets.  The obtained oscillation amplitudes and periods are in general agreement with the observations.
\end{abstract}


\section{introduction}\label{sec:introduction}

Jets and prominences are two different phenomena frequently appearing in the
solar atmosphere. Both types of objects have been studied separately in the
past decades; their mutual interaction is a comparatively recent
topic. There is already a number of observations reporting the apparent
impact of a jet onto a prominence. Yet, there is so far no in-depth
theoretical study of such encounters. The solar prominences (filaments when
observed on the disk) are embedded in a much larger magnetic structure called
the filament channel \citep[FC;][]{martin_conditions_1998,
 gaizauskas_filament_1998}. FCs are regions of sheared magnetic field
\citep{hagyard_quantitative_1984, moore_flare_1987,
 venkatakrishnan_evaluation_1989} above photospheric polarity inversion
lines \citep[PIL;][]{babcock_suns_1955}. Magnetic loop arcades overlie the
FCs, reaching high above them and with roots on both sides of the channel
\citep{martin_conditions_1998,tandberg1995}. The term FC was originally used
to name the region around the PIL where the chromospheric fibrils are aligned
with the PIL. However, nowadays there is consensus to use this term to refer to the entire magnetic field rooted in that region, which extends into the corona and contains the low-lying lines with dips able
to support the heavy prominence mass against gravity \citep[see reviews
 by][]{Mackay10, gibson_coronal_2015}.

Although known for a much shorter time than prominences, coronal jets are
also common phenomena in the Sun. They are understood to follow from a
rearrangement of the magnetic field structure in the atmosphere caused by,
e.g., magnetic flux emergence from beneath the surface or other processes which drive the footpoints of the field lines through photospheric flows, stress the coronal field and may lead to the (often violent) release
of magnetic energy as a jet or mini-flare. Irrespective of the precise
configuration or evolutionary pattern leading to the jets, reconnection is
assumed to be at the core of the process, causing the abundant conversion of
magnetic energy into kinetic and internal energy of the plasma in the jets.
Most of the previous studies focused on jets occurring in active regions (AR)
or coronal holes (CH). However, flux-cancellation and shearing motions
 continually occur near the PILs below FCs, thus jet-like activity is also
 expected in the neighborhood of filaments. In fact, jets are often
detected near solar filaments, possibly inside the filament channel, and they
may have appreciable effects on the matter constituting the prominence and
the supporting magnetic field \citep[e.g.,][]{chae_formation_2001,chae_formation_2003,wang_transient_2013}.
Further observations of the appearance of jets in the neighborhood of
 prominences are reported by \citet{zirin_production_1976},
 \citet{wang_jetlike_1999}, \citet{liu_production_2005},
 \citet{wang_formation_2018}, and \citet{panesar_network_2020}, among
 others.
  
In this paper we are interested in studying the impact of coronal jets on preexisting prominences, and, in particular, the oscillations launched by the jets in them. Solar prominences are very dynamical
structures that show a large variety of motions including
counterstreaming flows, different instabilities, and oscillations
\citep[][]{Mackay10}. Prominence oscillations have been measured with
 velocities from the threshold of observation up to $100 \kms$
\citep[see the recent review by][]{Arregui2018aa}. A particular set of
oscillations are large-amplitude oscillations (LAOs) with velocities above 10$\kms$ in which a large portion of the filament oscillates.
The triggering of LAOs has been attributed to energetic events such as
Moreton or EIT waves
\citep{Moreton60b,eto_relation_2002,Okamoto04,gilbert_filament-moreton_2008,Asai12,liu_observation_2013},
EUV waves
\citep{shen_chain_2014,xue_transverse_2014,takahashi_prominence_2015}, shock
waves \citep{shen_chain_2014}, nearby subflares and flares
\citep{jing_periodic_2003,jing_periodic_2006,vrsnak_large_2007,li_sdoaia_2012},
and the eruption of the filament
\citep{isobe_large_2006,Isobe07,pouget_analyse_2007,chen_sohosumer_2008,foullon_ultra-long-period_2009,bocchialini_oscillatory_2011} .
Recently, \citet{luna2018} presented a catalog of almost 200 prominence oscillation events, with nearly half of them being LAOs. The authors found that LAO events frequently occur in the Sun with, on average, one LAO event every two days on the visible side of the Sun. %

A detailed observational description of a jet that triggers LAOs in a 
prominence was presented by \citet{luna_observations_2014} and \citet{zhang_simultaneous_2017}. In both cases, the jet source is magnetically connected to the prominence and the jet flows propagate along the field lines.
\citet{luna_observations_2014} showed a series of jets near the PIL that hit the prominence producing substantial periodic motions in a large fraction of the filament. The jetting activity occurred in the form of collimated plasma emanating from outside the prominence and traveling towards it with velocities up to $95 \kms$. 
The primary jet produced substantial displacements of the prominence plasma with velocities ranging from $10$ to almost $50\kms$. On the other hand, in the observations by \citet{zhang_simultaneous_2017} a jet was seen to be generated in a C2.4 flare in AR 12373 and to propagate toward a prominence with a projected speed of $224\kms$. The prominence was located some 225 Mm
away from the active region.
Using HMI LOS magnetograms the authors found that the jet was associated with the
cancellation of a bipolar region at the jet base; after moving along
 the magnetic field lines, most of the jet hit the prominence. The impact
produced simultaneous large-amplitude transverse and longitudinal oscillations in the
prominence.

Motivated by these observations, we have carried out a theoretical study to investigate the physics governing the interaction between a jet generated outside a preexisting prominence and the prominence
itself. To that end we use 2.5D numerical simulations with the MANCHA3D code \citep{felipe_magneto-acoustic_2010},
modeling the launching and propagation of jets along an FC that eventually impact and set in oscillation the hosted prominence. As a candidate for the filament channel that is to host the prominence we use a simple superposition of magnetic dipoles. The resulting configuration has a central dipped field where we load the prominence mass in a preliminary phase of the experiment.

To trigger the jet we get inspiration from the extensive literature dealing
with models for the launching of coronal jets 
(e.g., 
\citealt{Yokoyama_Shibata_1996}, 
\citealt{Moreno-Insertis08},
\citealt{Nishizuka_giant_2008},
\citealt{Archontis_Hood_2008,Archontis_Hood_2012, Archontis_2013},
\citealt{Pariat09, Pariat10, pariat_model_2015, pariat_model_2016},
\citealt{Moreno-Insertis_2013},
\citealt{Wyper_2016a}, \citealt{Wyper_2016b, Wyper2018aa}, \citealt{karpen_reconnection-driven_2017}).
Direct predecessors of the model used in the present paper to launch the jets
are the experiments of \citet{Moreno-Insertis_2013} and
\citet{Wyper2018aa}. In those experiments, a magnetic arcade straddling a
photospheric polarity inversion line experiences reconnection at its top,
i.e., in the interface with the overlying coronal field, leading to the
emission of a quiescent (or non-eruptive) jet. In \citet{Wyper2018aa}, the
arcade is already given as part of the initial condition and it is sheared along the
PIL via boundary driving at the photosphere; in \citet{Moreno-Insertis_2013} and
\citet{Archontis_2013}, instead, the arcade is the result of the emergence of
a bipolar region from below the photosphere, and the shear along the PIL occurs 
as a natural outcome of the process of flux emergence. The shear increases the magnetic energy of the arcade, which grows in height while its legs of
opposite magnetic sign get closer to each other; as a result, in a second
phase, a vertical current sheet is formed between the two sides of the arcade right
above the PIL, and reconnection ensues; this then leads to the formation of a twisted flux rope contained between the sheet and the overlying coronal field. The flux rope
is expelled upwards, yielding a violent eruptive jet that propagates along the field lines of the overlying coronal field.
In fact, the basic problem of the consequences of magnetic arcade shearing, like the explosive, flare-like reconnection occurring in a vertical current sheet within the arcade,
has been studied for many years in the context of the theory of filament eruptions, CMEs, and jets. Further papers dealing with this problem are, e.g.,
\citet{mikic_etal_1988}, 
 \citet{Van_Ballegooijen_Martens_1989}, 
\citet{sturrock1989},
\citet{Antiochos_1990},
\citet{moore_roumeliotis_1992},
\citet{Antiochos_etal_1999_CMEs},
\citet{moore_etal_ssr_2010},
\citet{Manchester_2001}, \citet{Manchester_etal_2004},
\citet{Archontis_Torok_2008},
\citet{Aulanier_etal_2012},
\citet{Karpen_etal_2012, karpen_reconnection-driven_2017}.

In the current paper we are primarily interested in the action of the jets on a prominence once they travel a considerable distance away. So, based on the results of the 3D models mentioned above, we adopt a configuration that efficiently leads to
this kind of process: to the FC field we add a third small
dipole to mimic a parasitic structure located far from the major dip hosting
the prominence. The parasitic dipole produces a dome-shaped fan surface with
a null point at its top. The magnetic arcade underneath the null point is
then sheared in the initial phase of the experiment which, after going
through different stages, finally leads to both a collimated and an eruptive
jet. Sections~\ref{sec:model} and~\ref{sec:jet-synthesis} describe the
general setup of the experiment and various details of the formation of the
jet. The main chapters of the paper are Section~\ref{sec:the-jet}, in which
the interaction of the jet with the prominence is studied, and
Section~\ref{sec:after-jet} in which the subsequent evolution of the
prominence after the jet, including the prominence oscillations, is analyzed.

\section{Initial setup}\label{sec:model}

\begin{figure}[!ht]
	\centering\includegraphics[width=0.45\textwidth]{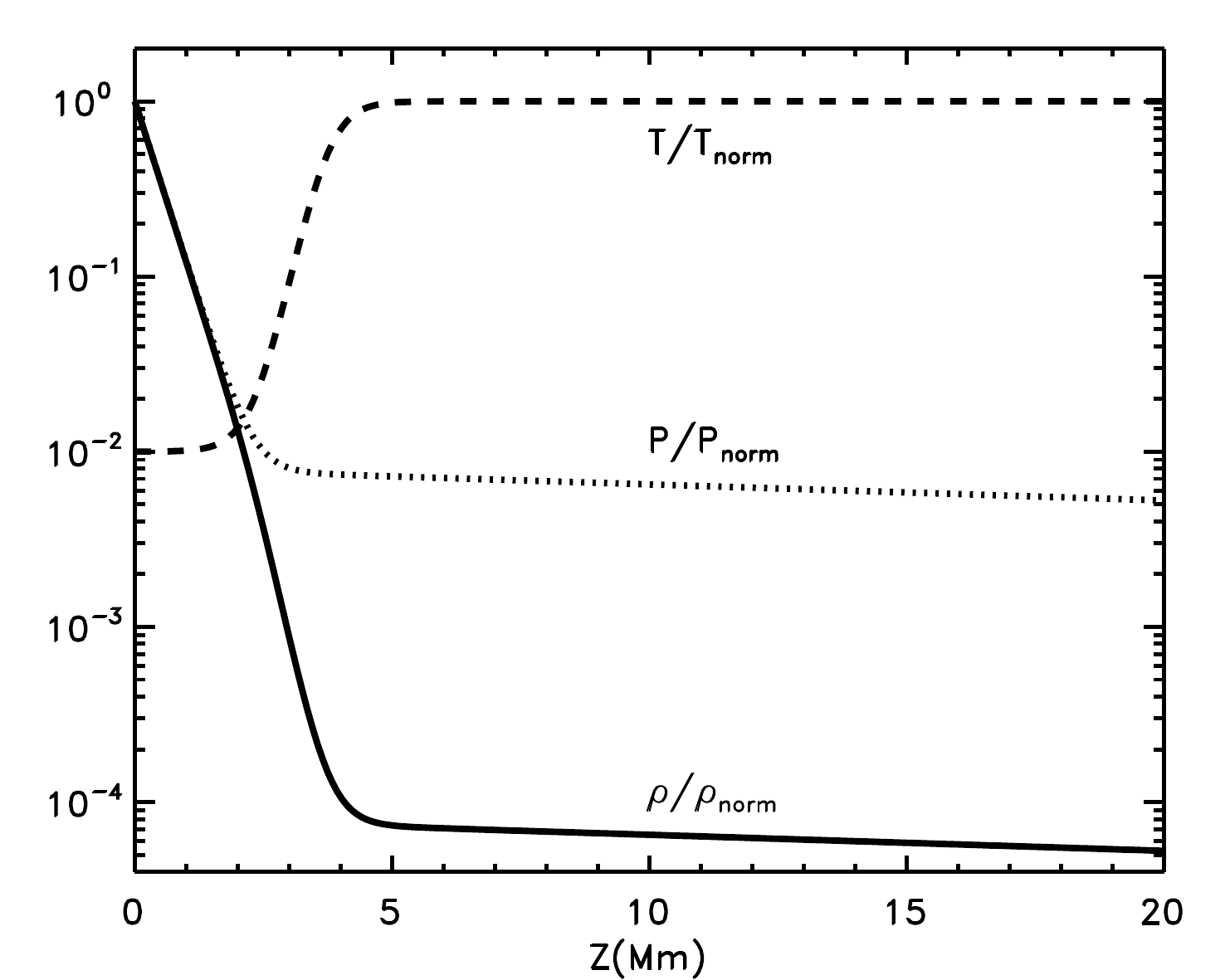}
  	\caption{Distribution of the initial density (solid), pressure (dotted) and temperature (dashed) along the vertical direction, $z$. The curves have been normalized to their respective maximum value, namely: 
  		$\rho_\mathrm{norm}=2.96\times10^{-8}\densunits$, $P_\mathrm{norm}=3.85 \, \mathrm{Pa}$ and $T_\mathrm{norm}=10^{6}$ K. \label{fig:initialstratification}}
\end{figure}

We assume an initial atmosphere consisting of stratified plasma in hydrostatic equilibrium embedded in
a potential magnetic field. The temperature profile is given by
\begin{equation}\label{eq:temperature-profile}
T(z)= T_{0} + \frac{1}{2} \left(T_\mathrm{c}-T_{0} \right) \left[1+ \tanh
\left( \frac{z-z_{c}}{W_z}\right) \right] \, , 
\end{equation}
and is intended to contain a chromosphere, a transition region, and an isothermal corona. We have chosen the following values:
$T_\mathrm{c}=10^{6}$ K, $T_{0}=10^{4}$ K, $W_z=0.625\Mm$,
$z_{c}=4.25\Mm$. Figure \ref{fig:initialstratification} shows the stratification profiles for the density, pressure and
temperature, normalized to their respective maximum value. The computational domain consists of a box of $3000\times500$ grid points and physical dimensions of $144 \Mm \times 24\Mm$, resulting in a spatial resolution of $48\km$. The numerical simulations have been performed with the MANCHA3D code \citep[see][]{felipe_magneto-acoustic_2010}. The code solves the MHD equations in conservative form for mass, momentum, magnetic field, and total energy. Thermal conduction and radiative losses are not considered in this work. The experiments presented in this paper are 2.5D: the $y$ coordinate is ignorable so that the variables depend on $(x,z)$ exclusively; yet, the velocity and magnetic field can have a non-zero y-component.

\begin{figure*}[!ht]
\centering\includegraphics[width=1\textwidth]{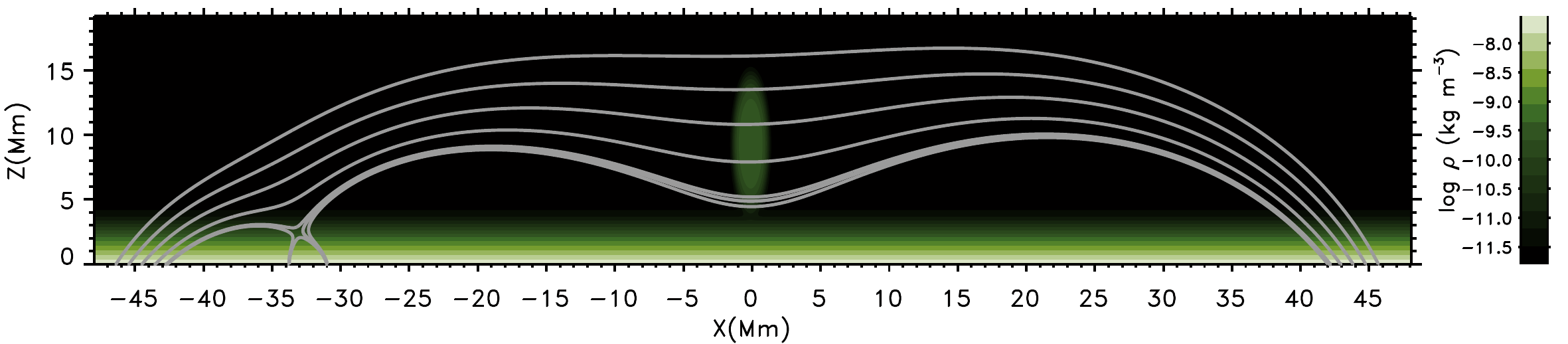}
\caption{Plot of the initial configuration of the filament channel and prominence for the particular configuration called case~1.  The parasitic polarity is apparent around $x \approx -33\Mm$. The color indicates the density of the stratified atmosphere and the lines are the field lines of the force-free initial configuration. The temperature of the atmosphere is defined by Eq. \ref{eq:temperature-profile} and the density and gas pressure are obtained by stratification conditions and ideal gas law.\label{fig:initialconfiguration}}
\end{figure*}

The initial magnetic field (Fig.~\ref{fig:initialconfiguration}) is purely 2D, contained in the $(x,z)$ plane, and given through the superposition of three horizontal magnetic dipoles; the dipoles have moment $m_j (j=1,2,3)$ and are located at $(x_{j},z_{j})=(30,-31.7),
(-30,-31.7), (-32,-3.2) \Mm$, respectively. The field distribution, given in complex
variable form, is, therefore, \citep[e.g.,][]{priest2014}:

\begin{equation}\label{eq:initialfield}
B_{z}+i B_{x} =\sum_{j=1,2,3} \frac{i \, m_{j}}{\left[(x-x_{j}) +
	i(z-z_{j})\right]^{2}} \,. 
\end{equation}

\noindent The strength of the dipole moments $m_j$ can be given in terms of a
reference value $m_{0}=-B_{0} h_\mathrm{b}^{2}$, with 
$B_{0}=30$ Gauss and $h_\mathrm{b}=-31.7 \Mm$. 
Dipoles number 1 and 2 have moment $m_1 = 0.9\,m_{0}$ and $m_2=0.88\, m_{0}$,
respectively.
The configuration resulting from their superposition is a dipped magnetic field system similar to the one used by \citet{luna2016}, which sets the large-scale properties of the FC distribution (see Fig.~\ref{fig:initialconfiguration}). Dipole number 3 introduces a parasitic dipolar structure with a small magnetic moment of opposite sign to the other two; it is discussed later in this section. Parasitic polarities have been routinely observed at the edges of filaments sometimes leading to the apparition of a so-called filament barb \citep[see][]{martin_conditions_1998,aulanier_3-d_1998} or causing dynamical episodes in the filament \citep[e.g.,][]{deng_filament_2002, schmieder_proper_2014}.

To have a prominence already formed when the experiment starts, in a preliminary phase we add mass
to the dipped region of the large-scale magnetic field 
instead of introducing mechanisms of prominence formation
\citep[see review by][]{karpen_plasma_2014}; we use the
simple device of including an artificial source mass term in the continuity
equation for a short period of time, following the method used by
\citet{terradas_magnetohydrodynamic_2013} and \citet{Liakh2020aa}.
The source term has the shape of an elongated Gaussian in space (see
Fig.~\ref{fig:initialconfiguration}) as follows
\begin{equation}\label{eq:source-term}
S_\rho (x,z)=\frac{\kappa \rho_0(x,z)}{t_\mathrm{load}} \, \exp \left[
-\frac{x^2}{\sigma_x^2} - \frac{(z-z_0)^2}{\sigma_z^2} \right] \, , 
\end{equation}
\noindent where $z_0=9.36\Mm$ is the height of the center of the
prominence, $\sigma_x=1\Mm$ and $\sigma_z=4\Mm$. The dimensionless parameter
$\kappa=200$ is the density contrast with respect to the ambient corona
without prominence, $\rho_0$. This source term is applied only during the
initial $t_\mathrm{load} = 160$ s. During this time the weight of the injected mass
increases the curvature of the dipped field lines thus attempting to
regain force equilibrium. In addition, during the mass loading the gas
pressure in the prominence region (given in Eq.~(\ref{eq:source-term}))
remains approximately constant, with the consequence that the temperature decreases, reaching a minimum value
of $5000$ K at the end of the loading process. The maximum density contrast
and temperature are typical values in solar prominences
\citep{labrosse_physics_2010}.

To perturb the FC and prominence configuration, we add the third,
 parasitic dipole mentioned above, which had been labeled $m_3$.
 We will be showing results for two values of the dipole moment, namely 
$m_3=-0.0175\, m_{0}$ and $m_3=-0.015\, m_{0}$ that we henceforth call case~1
and case~2 respectively. The parasitic dipole produces a null point (NP; see
Fig.~\ref{fig:initialconfiguration}), or, in fact, given the symmetry, a line of nulls along the y-axis located at $(x,z) = (-69.46, 2.54)$ Mm
for case~1 and $(x,z) = (-69.84, 2.16)$ Mm for case~2. Those specific values have been selected so that the jets we will be describing in later
sections reach the dipped part of the large-scale configuration where the prominence
will reside. Seen in the $xz$ plane, the lower spine of the null point reaches the bottom plane at
$x=-34\Mm$. On its right, we find a collection of closed loops, an arcade, below the null point, and overarching the PIL located at $x_\mathrm{PIL}=-32\Mm$; 
following the models mentioned in the introduction, to launch a jet,
photospheric driving is applied to the footpoints of the magnetic arcade,
as described in the following.

The initial field structure given by Equation (\ref{eq:initialfield}) is
potential. Free energy is then injected into it by applying footpoint driving through the $y$-component of the velocity, as
follows:
\begin{eqnarray}
\nonumber
&&v_{y}(x,z=0,t)=v_{y 0}(t) \times \\ \label{eq:shear-flox-vy}
&& ~~~~~~~~~\left\{ \exp\left[-\frac{(x-x_{L})^{2}}{\separ^{2}}\right] - \exp\left[-\frac{(x-x_{R})^{2}}{\separ^{2}}\right]\right\} \, , 
\end{eqnarray}
where $\separ=480\km$, $x_{L}= x_\mathrm{PIL}-240\km$ and
$x_{R}=x_\mathrm{PIL}+240\km$.
The value $\separ$ is selected small enough to shear only the region under the parasitic bipole but large enough to power the jet in a finite time.
\newcommand{\Wsh}{W_{sh}}
The function $v_{y 0}(t)$ is set to provide a smooth activation and
deactivation of the shear, as follows:

\begin{equation}\nonumber
v_{y 0}(t) = \, \frac{v_0}{2} \, \left[
\erf\left(4\,\frac{t-t_1}{\Delta} - 2\right)
\\ \label{eq:driving-temporal-dependence}  
\,- \erf\left(4\,\frac{t-t_2}{\Delta} - 2\right)\right]\;.
\end{equation}
This function starts at $t_{1}=200$ seconds and
increases monotonically for $\Delta = 100$ seconds. It is constant and equal
to $v_{0}$ from $t_{1}+\Delta$ to $t_{2}=1200$ seconds and decreases to zero during the next $\Delta$ seconds. The shear velocity, therefore, keeps its maximum value $v_{0}$ for $t_{2}-t_{1}-\Delta=900$ seconds. In this paper we have chosen $v_{0}= 5 \kms$, so the associated motion is subsonic and sub-Alv\'enic and the field evolution is quasi-static. 
This driving is a convenient device to inject free energy into the arcade; it can correspond to the spontaneous shearing across the PIL in a flux emergence situation, or the accumulation of photospheric motions in other scenarios (see Sec.~\ref{sec:introduction}). In actual filaments on the Sun the FC is observed to be sheared. Our experiment in the present paper has no shear in the FC at time zero and, in that sense, it is only a first step toward understanding the interaction of the jets with filaments.
Concerning the shear itself, the value adopted for $v_0$ is larger than expected from photospheric observations; this is chosen so as to speed up the evolution and counter the effect of the artificial numerical diffusivity in the reconnection site. In fact, the extra push is also adequate since our model is started with a potential field configuration whereas in a realistic solar situation the structure is already stressed.

\newpage

\section{Generation of the jets}\label{sec:jet-synthesis}
We first explore the mechanism implemented in the two experiments to
 launch the plasma jets that later on collide with the prominence. The basic
 idea is that the photospheric driving injects energy into the
magnetic arcade overlying the parasitic dipole with the consequence that
 current sheets are formed in successive stages and reconnection starts, the
 process ending up in the ejection of plasma into the large-scale magnetic
 system. To monitor the successive steps leading to
the jets, we start by considering the energy variation in the system,
computed via the following integrals:
\small{
\begin{eqnarray}\label{eq:magnetic-energy}
E_\mathrm{mag}&=&\iint \frac{{B}^{2}(x,z,t)}{2 \mu_{0}} \,dS
- \iint \frac{B^{2}(x,z,t=0)}{2 \mu_{0}} \,dS \,
, \\ \label{eq:internal-energy}
E_\mathrm{int}&=&\iint \frac{p(x,z,t)}{\gamma-1} \,dS
- \iint \frac{p(x,z,t=0)}{\gamma-1} dS \, , \\ \label{eq:kinetic-energy}
E_\mathrm{kin}&=&\iint \frac{1}{2} \, \rho \, {v}^{2}(x,z,t)\, dS \,
, \\ \label{eq:injected-energy}
E_\mathrm{inj}&=&\int_{0}^{t}\int \frac{1}{\mu_{0}} \left[\vec{E} (x,0,t')\times \vec{B}(x,0,t')\right]\cdot \hat{z} \; dx dt' .
\end{eqnarray}}
The magnetic, internal, and kinetic energy contents in
Equations~(\ref{eq:magnetic-energy})-(\ref{eq:kinetic-energy}) 
are calculated as an integral over the $xz$-plane in the domain shown in
Figure~\ref{fig:shearing-region-current-case1};
the result, therefore, has units of energy per unit length along the
$y$-direction. Further, they are 
calculated as variations with respect to the static equilibrium at
time $t=0$. $E_\mathrm{inj}$~is the cumulative energy injected by the
boundary driving via the Poynting flux; the range in $x$ chosen for
that integral is the same as for
(\ref{eq:magnetic-energy})-(\ref{eq:kinetic-energy}).  

\begin{figure}[!ht]
\centering\mbox{\subfigure{\includegraphics[width=0.45\textwidth]{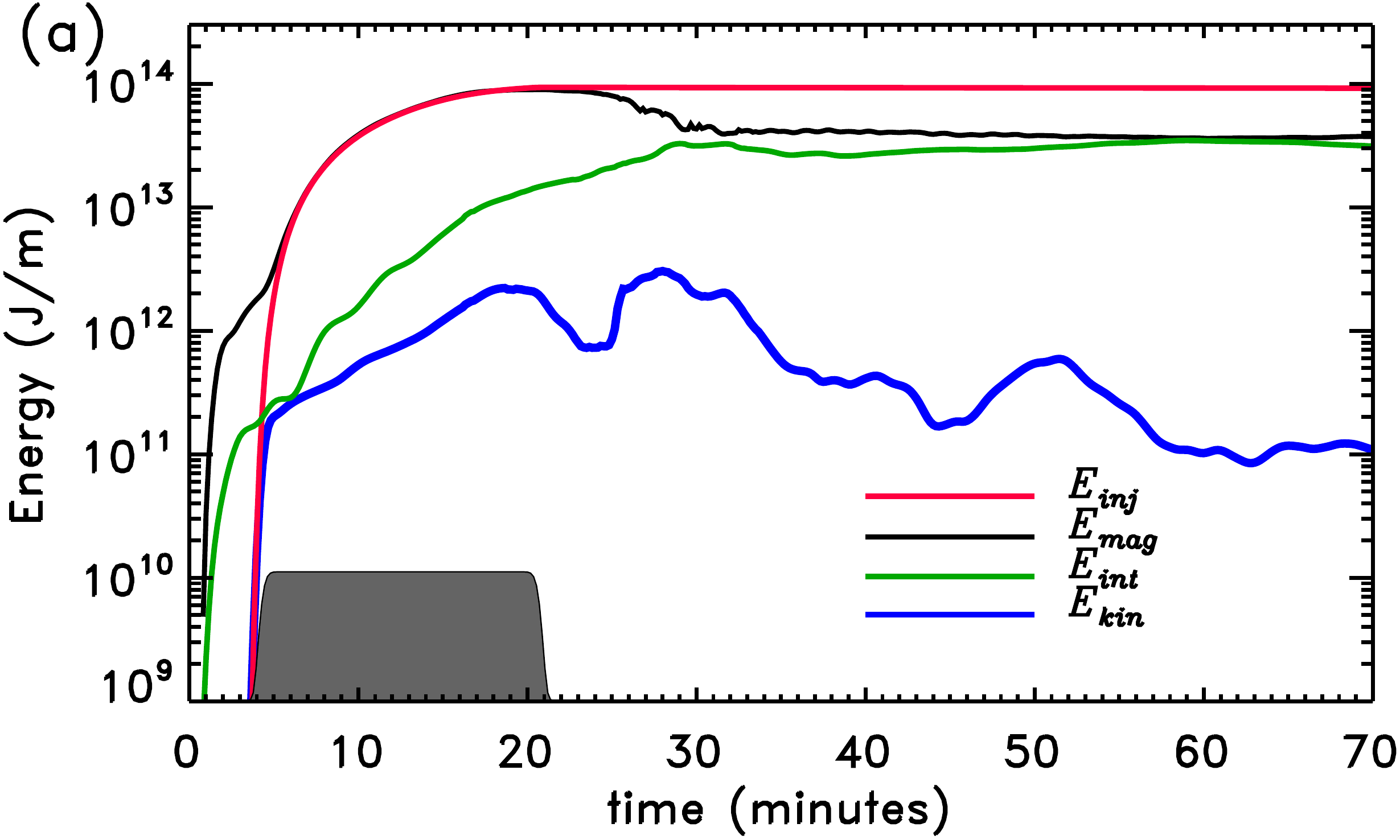}}}\quad
\centering\mbox{\subfigure{\includegraphics[width=0.45\textwidth]{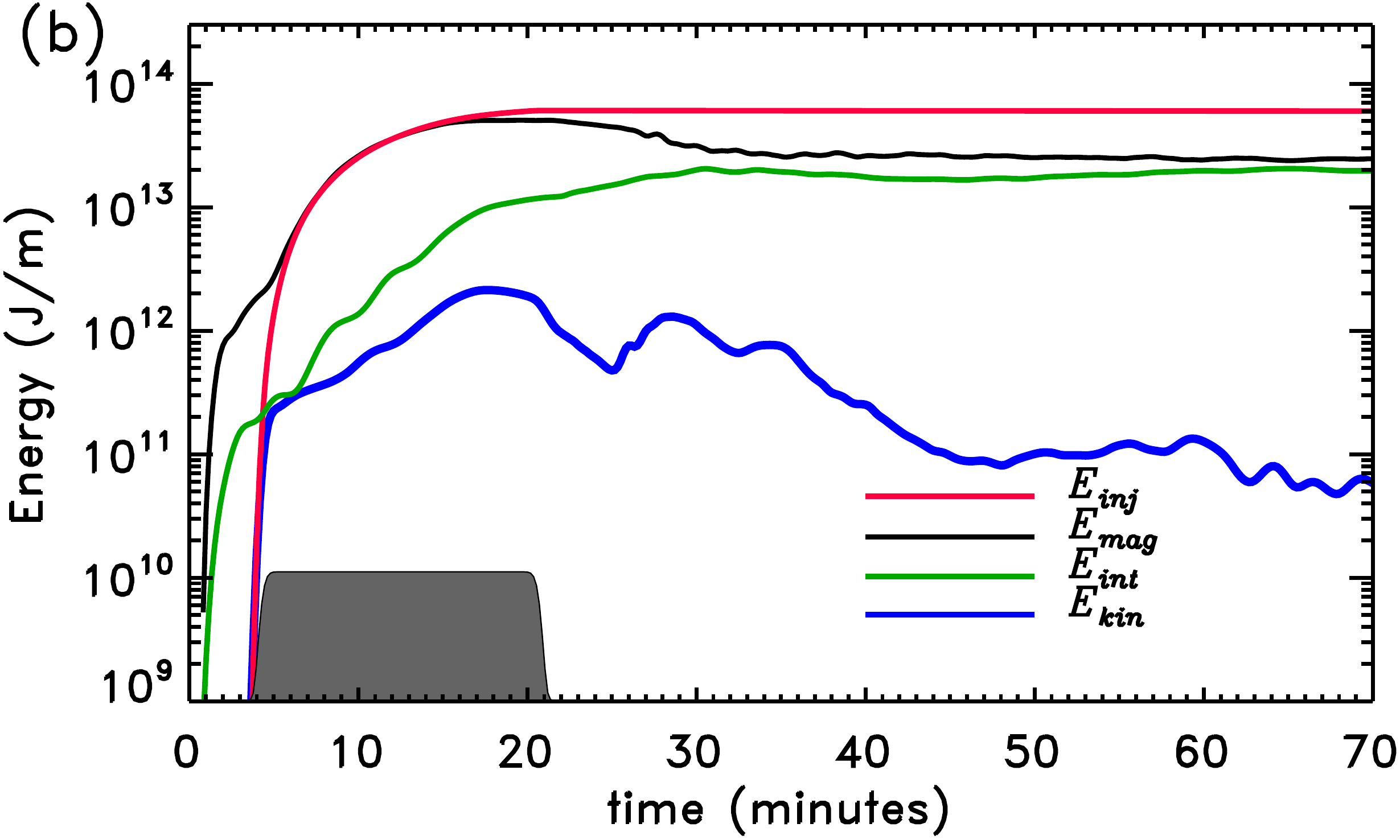}}}
\caption{Plot of the time evolution of the energy integrals
 (\ref{eq:magnetic-energy}) -- (\ref{eq:injected-energy})
in the domain shown in Figure~\ref{fig:shearing-region-current-case1} for (a) case~1 and (b) case~2.
The curves are for the injected energy (red,
 equation~\ref{eq:injected-energy}), 
the magnetic energy (black, Equation~\ref{eq:magnetic-energy}), the internal
energy (green, Equation~\ref{eq:internal-energy})
and the kinetic energy (blue, Equation~\ref{eq:kinetic-energy})
To
mark the time interval when the driving is active, a dashed area showing
$v_{y 0}(t)$ from Equation (\ref{eq:driving-temporal-dependence}) 
normalized to arbitrary units has been
added.
\label{fig:temporal-evolution-energies}}
\end{figure}

Figure \ref{fig:temporal-evolution-energies}(a) shows the energy integrals
(\ref{eq:magnetic-energy})-(\ref{eq:injected-energy}) as a function of time
for case~1 in a format similar to that used by
\citet{Wyper2018aa}. In the first 4 minutes, the time evolution is governed by
the loading of the prominence mass (Sec. \ref{sec:model}), which, as
can be seen, entails a change in magnetic and internal energy.  
The injected energy curve $E_\mathrm{inj}$ (red curve)
 appears when the driving starts, at $t=4$ min (the presence of the driving
 is indicated through a shaded area at the bottom of the figure that shows
 the profile of $v_{y 0}(t)$ in arbitrary units). 
During the first minute of the driving (i.e., until $t=5$ minutes), the
 shear reaches the high-density chromosphere; a small fraction of the
 injected energy goes to increasing the kinetic energy there, but the vast
 majority is turned into magnetic energy.
From the moment when the boundary driving reaches a constant value ($t
\approx 5$ min) until it finishes ($t\approx 20$ min), the flow in
the bottom layers remains steady, and most of the injected energy
 is converted into magnetic energy, with the consequences described in the
 following.

\begin{figure*}[!ht]

\hspace{-0.5cm}\includegraphics[width=1.1\textwidth]{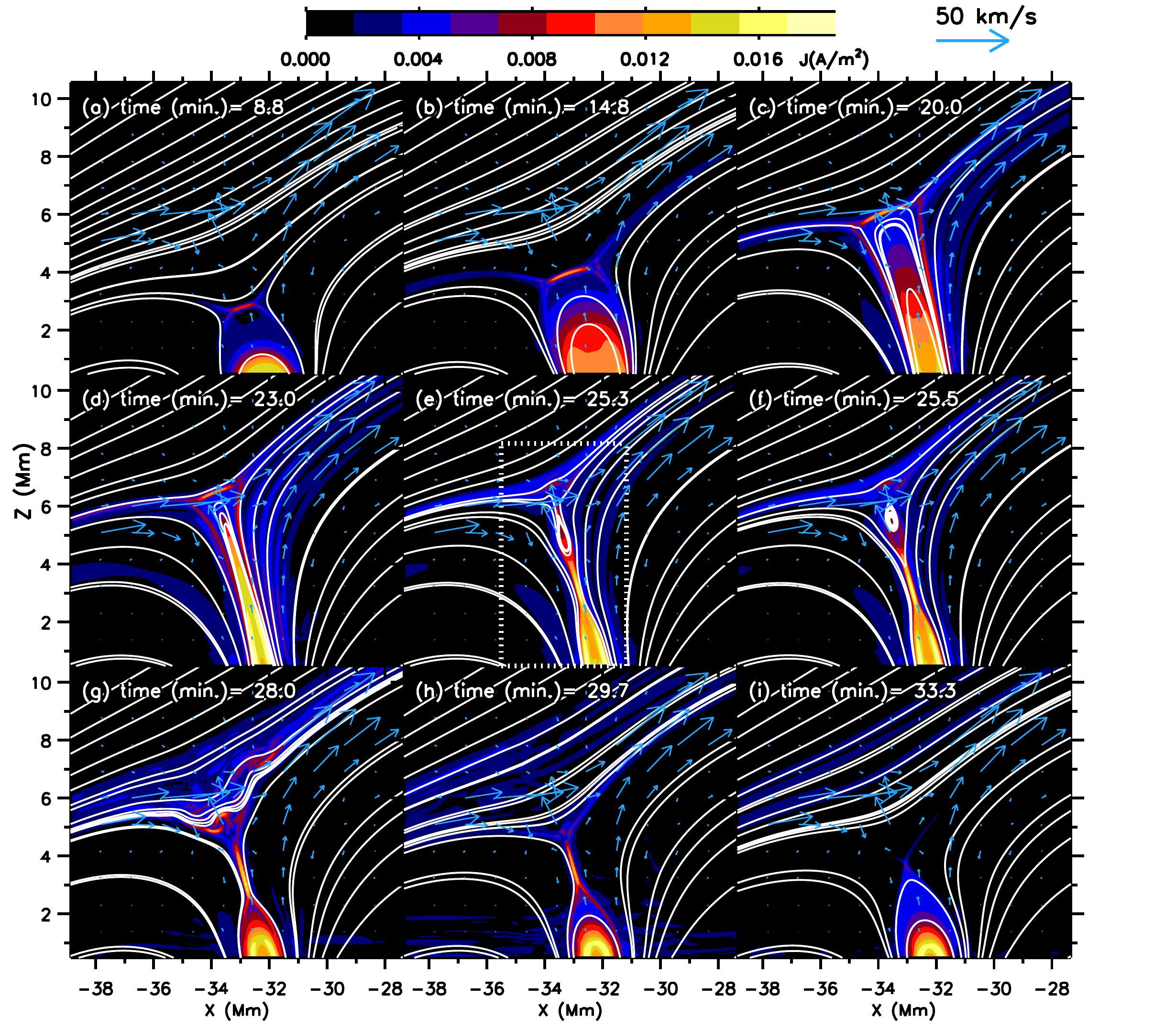}
\caption{Detail of the time evolution of the electric current distribution in the shear region for
 case~1. The white curves are the magnetic field lines projected onto the $(x,z)$ plane. The blue arrows correspond to the projection of the velocity field: the arrow lengths are scaled linearly with speed with the reference size given by the arrow in the top-right corner of the figure, which represents a velocity of $50\kms$. In panel (e), the dotted box encloses the area shown in Fig. \ref{fig:ropeformation}.
\label{fig:shearing-region-current-case1}}
\end{figure*}

\begin{figure*}
\centering\includegraphics[width=0.33\textwidth]{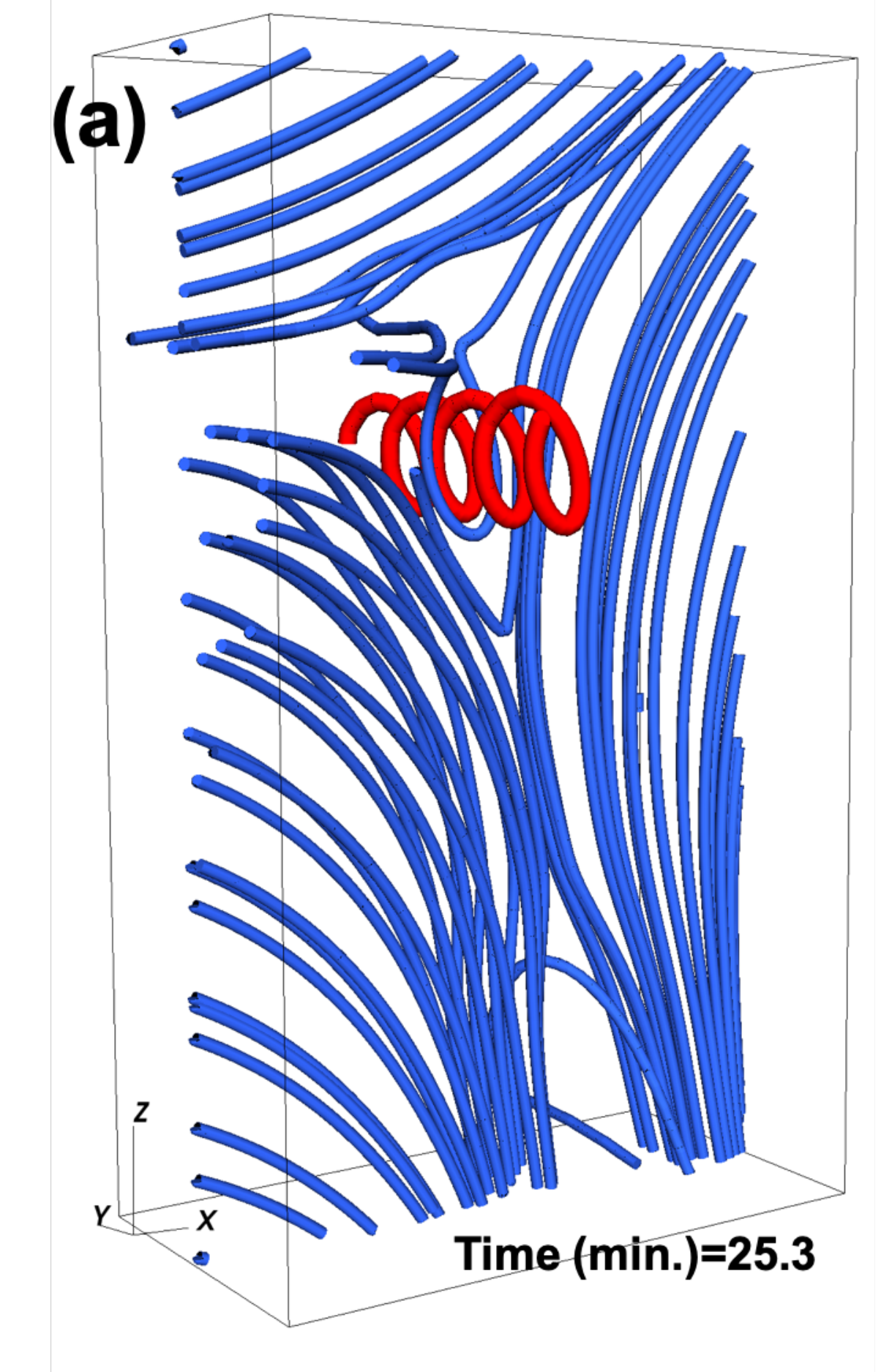}\centering\includegraphics[width=0.33\textwidth]{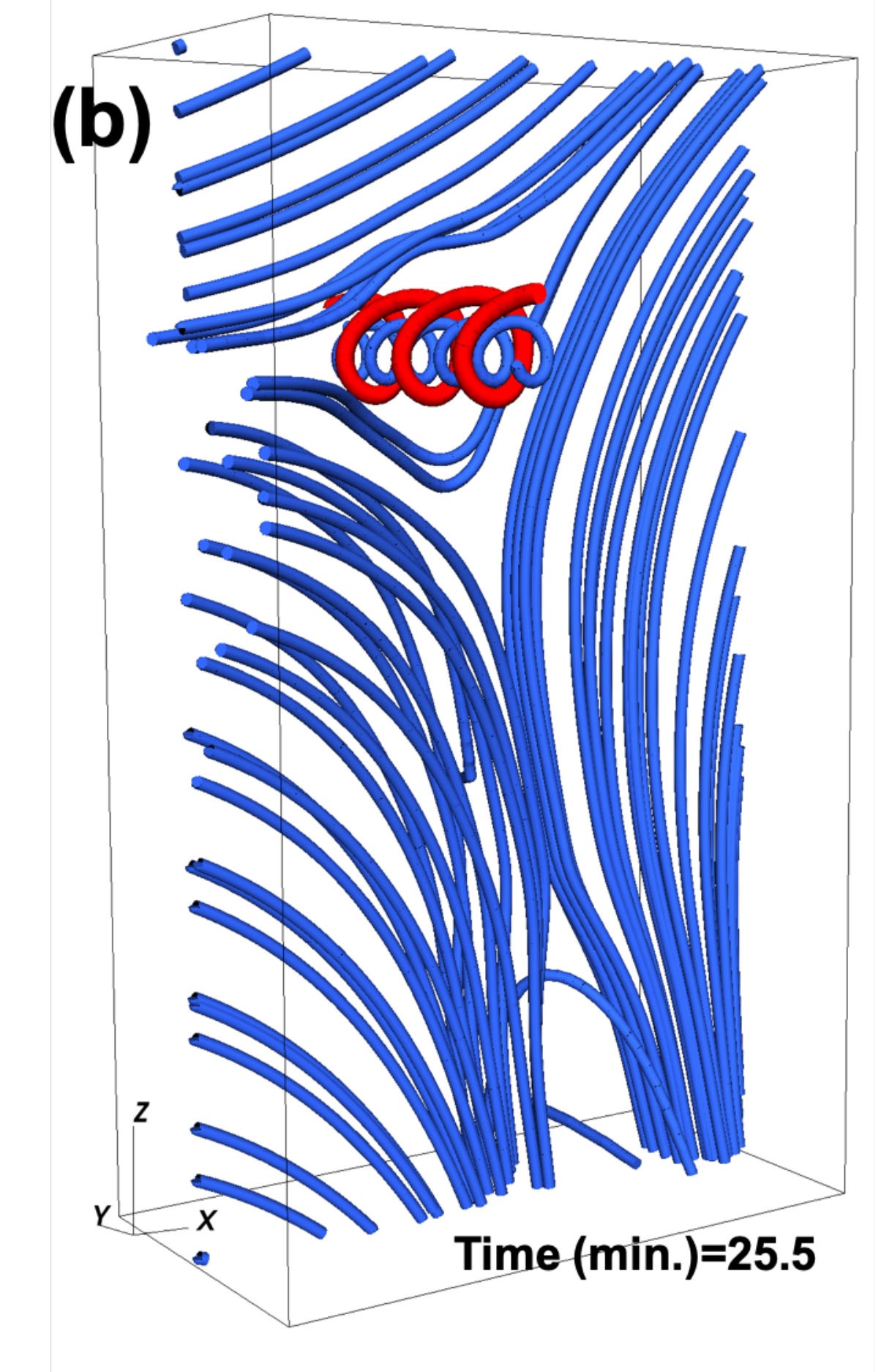}\centering\includegraphics[width=0.33\textwidth]{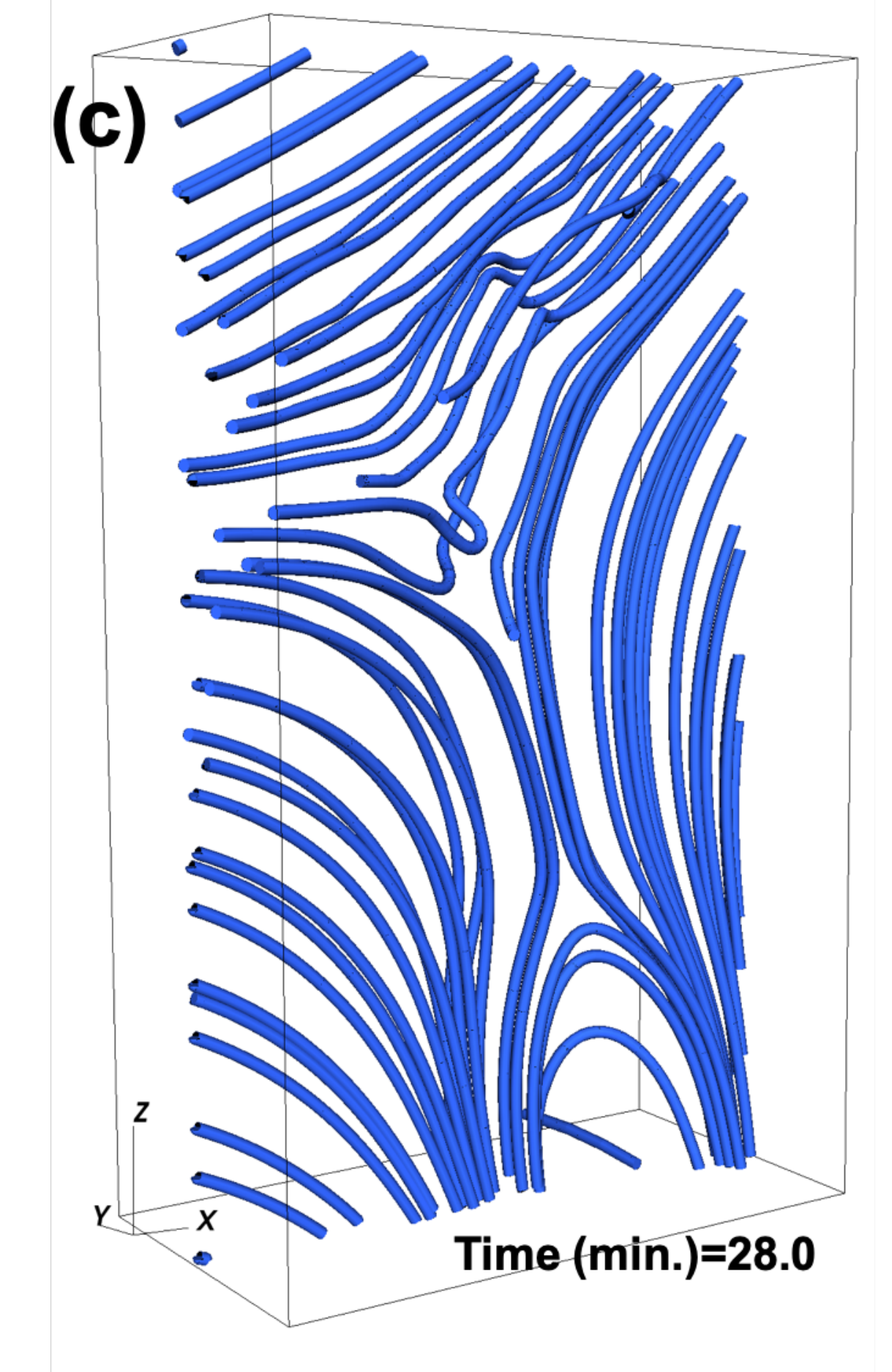}
\caption{Three-dimensional illustration of the magnetic field in the region marked with a white dotted box in Fig. \ref{fig:shearing-region-current-case1}(e) containing the sheared arcade and the current sheets at three different times. The red line is a field line of the plasmoid. The panels in this figure correspond to Figs. ~\ref{fig:shearing-region-current-case1}(e)-(g), respectively. \label{fig:ropeformation}}
\end{figure*}

Figure \ref{fig:shearing-region-current-case1} shows the electric currents in
and around the shearing region. The initial potential arcade above the
parasitic dipole is being turned by the driving into a sheared non-potential arcade with a nonzero electric current. The arcade loops consequently expand and push upward the oppositely directed field above the arcade.
This produces an elongated current sheet at the interface with roughly horizontal
orientation [Fig.~\ref{fig:shearing-region-current-case1}(a) to \ref{fig:shearing-region-current-case1}(d)] and
leads to reconnection, characterized by inflows roughly in the vertical direction and
plasma and magnetic field ejection roughly horizontally, which are
particularly clear in the panels for $t=14.8$ min \ref{fig:shearing-region-current-case1}(b) and $t=23$ min
\ref{fig:shearing-region-current-case1}(d). Following the nomenclature introduced by \citet{Antiochos_etal_1999}, we
shall call this phenomenon the {\it breakout} reconnection and the current
sheet {\it the breakout current sheet} (BCS); it lasts for the whole duration
of the driving and a little beyond it, until about $t=24$ min. The outflows to the right of the reconnection site constitute a comparatively quiescent ejection of plasma along the FC field lines that lead to the first jet launched in the direction of the prominence. During this time, the BCS first grows in length and then shrinks, becoming less important in later times. Below the current sheet, the sheared arcade becomes extremely elongated in the vertical direction.
Concerning the energy, going back to Figure~\ref{fig:temporal-evolution-energies}(a), we see that the vast 
majority of the energy input by the driving goes into magnetic energy; yet, the plasma is heated and accelerated in the current sheet and this leads to the increase in internal and kinetic energy apparent in the figure during this phase.

The second phase begins around $t=23$ minutes; the arcade is so
sheared and vertically elongated that it collapses from the sides
and the oppositely directed field lines start to reconnect in a more or less vertical current sheet (Figs. \ref{fig:shearing-region-current-case1}(d)-(g)). This phase ends at around $t=30$ minutes. In panel~\ref{fig:shearing-region-current-case1}(e),
in the region indicated with a dotted square, we see that a large plasmoid, a horizontal flux rope, is formed as a consequence of the reconnection in the vertical current sheet. To ascertain that it indeed is a flux rope, we have added a 3D view of the magnetic
field in that region (Fig.~\ref{fig:ropeformation}); this
is achieved by showing a vertical slab of $4$ Mm thickness in the
$y$-direction, taking advantage of the independence of the variables with
$y$. The three panels in this figure correspond to Figures \ref{fig:shearing-region-current-case1}(e)-(g) but just show the area inside the dotted rectangle of Figure~\ref{fig:shearing-region-current-case1}(e). The field lines of the flux rope are clearly visible as red and blue helices in Figures~\ref{fig:ropeformation}(a) and \ref{fig:ropeformation}(b). The rope is quickly rising toward the
BCS [Fig. \ref{fig:ropeformation}(b)] where it undergoes reconnection with the overlying magnetic field. This process leads to a large perturbation of the FC even though the twisted-rope structure is lost in the reconnection process [Fig. \ref{fig:ropeformation}(c)]. The perturbation then propagates along the FC as an eruptive jet which, like the first one, ends up colliding with the prominence.
The parameters (location of the null point, amount of driving) have been
chosen in a range that permits the two jets to impact the prominence; case~1,
the case described so far, is such that the eruptive jet impacts the top end
of the prominence, whereas in case~2, described below, it
impacts the prominence center. Figures
\ref{fig:shearing-region-current-case1}(h) and
\ref{fig:shearing-region-current-case1}(i) show that the reconnection
continues after the ejection of the plasmoid and the system finally relaxes to a new approximate equilibrium.
Concerning the energies, the second phase just described is apparent as
a secondary maximum in the kinetic energy curve (blue) between about $t=25$
and $t=30$ minutes. Simultaneously with this maximum, the magnetic energy
decreases, and the internal energy increases by opposite but comparable
amounts: most of the released magnetic energy is used to heat the
plasma. All of the foregoing is a consequence of the reconnection process. 

\begin{figure}[!ht]
\begin{center}
\includegraphics[width=\mysizec\textwidth]{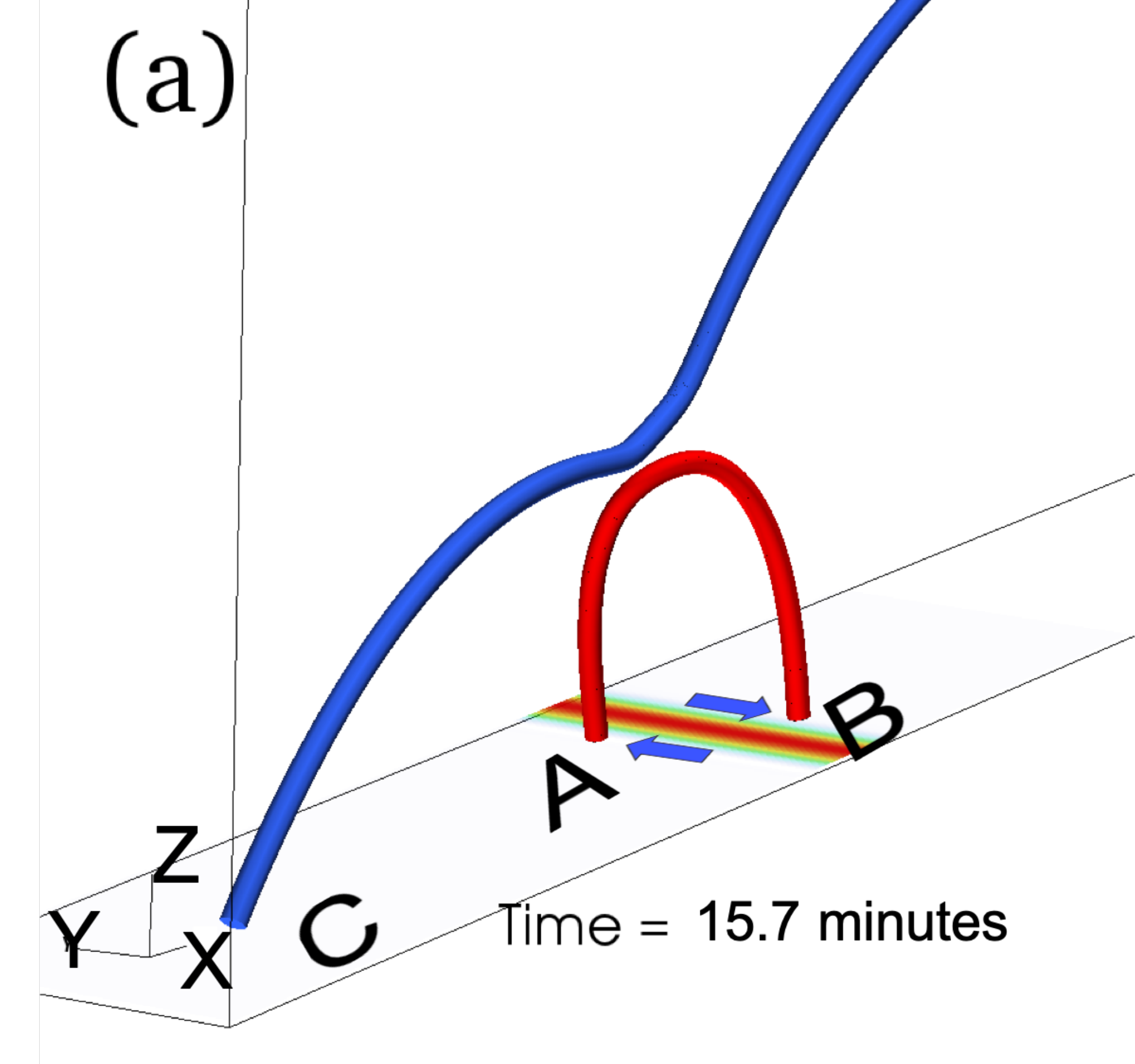}\includegraphics[width=\mysizec\textwidth]{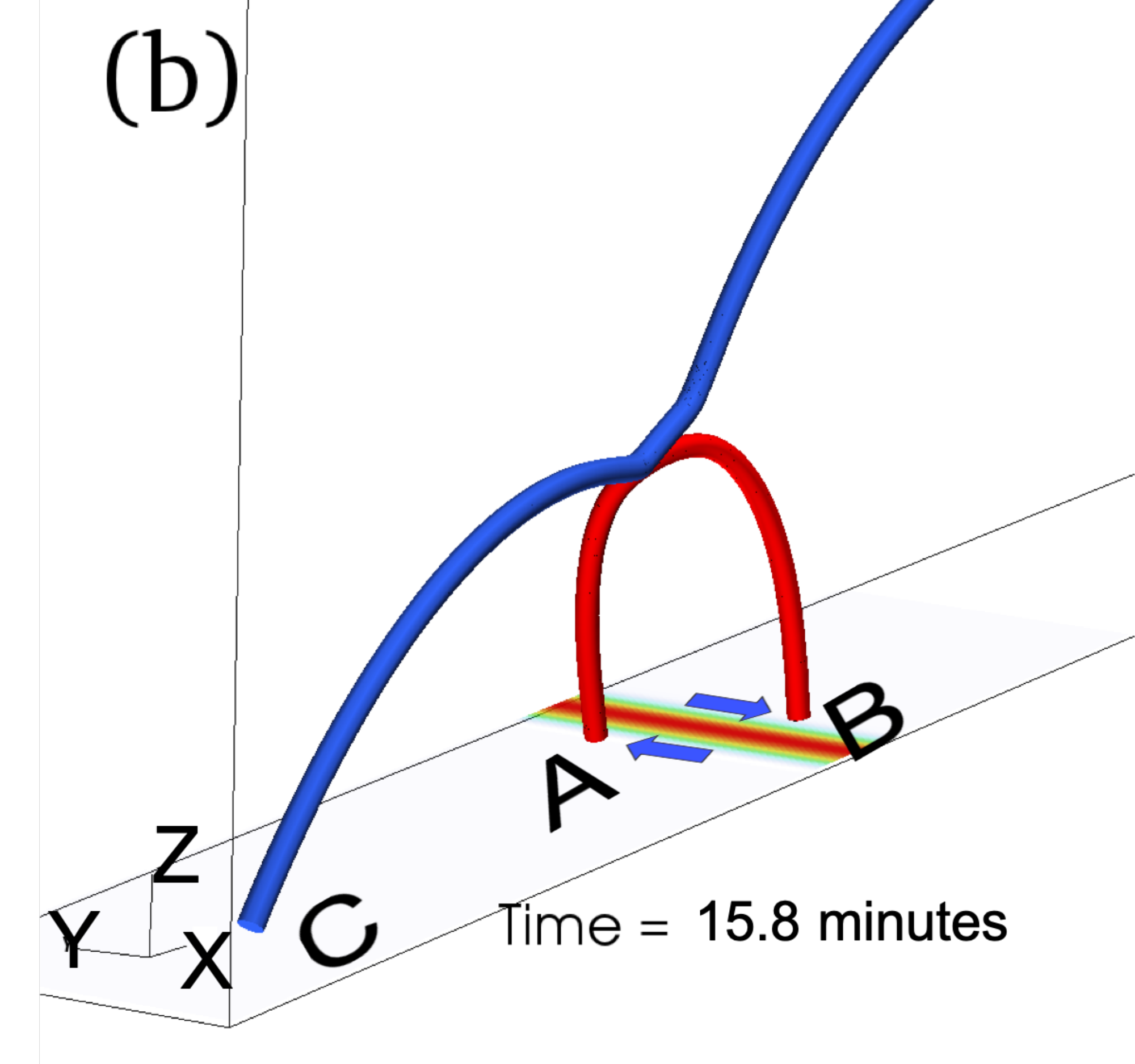}\\
\includegraphics[width=\mysizec\textwidth]{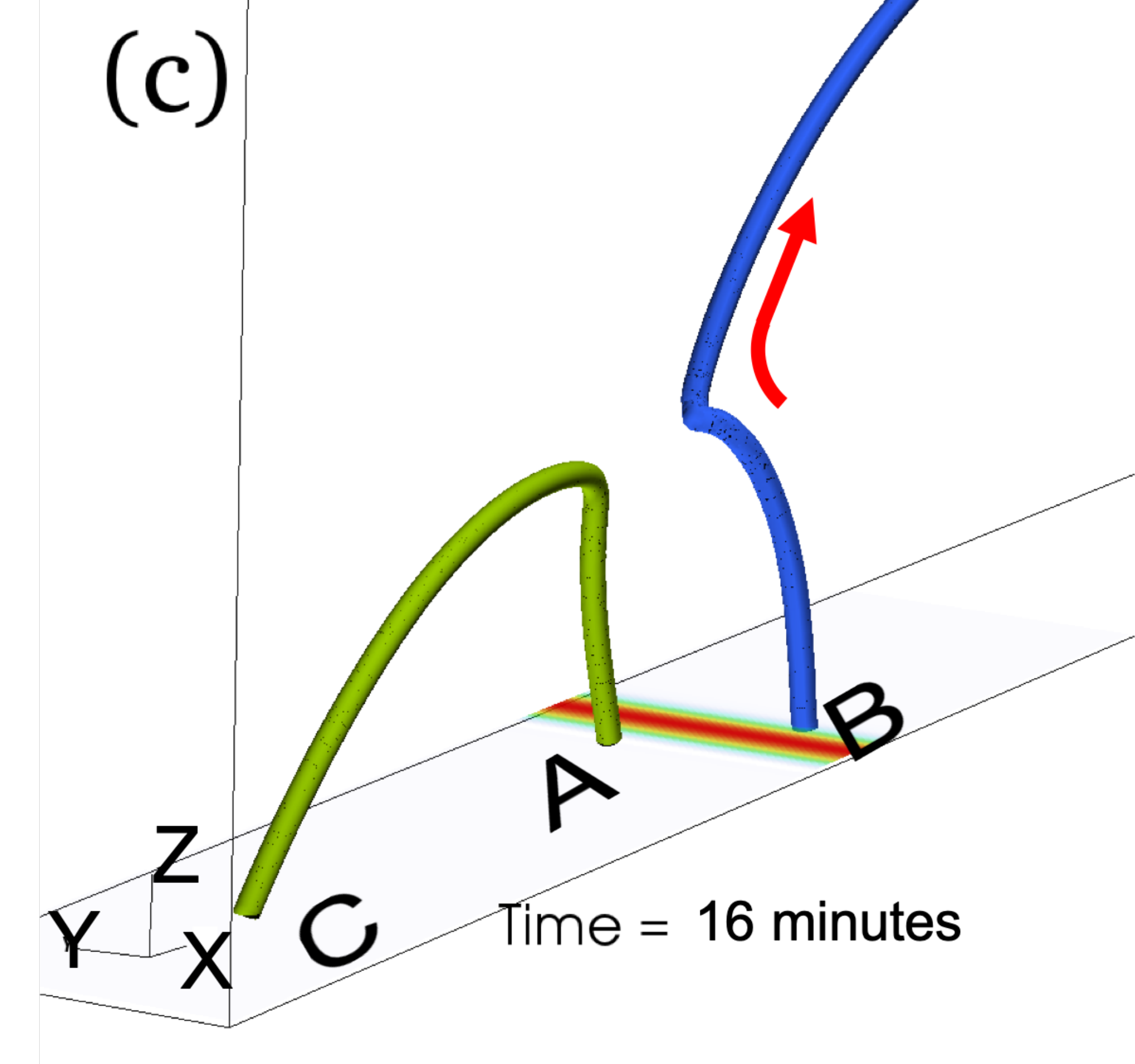}\includegraphics[width=\mysizec\textwidth]{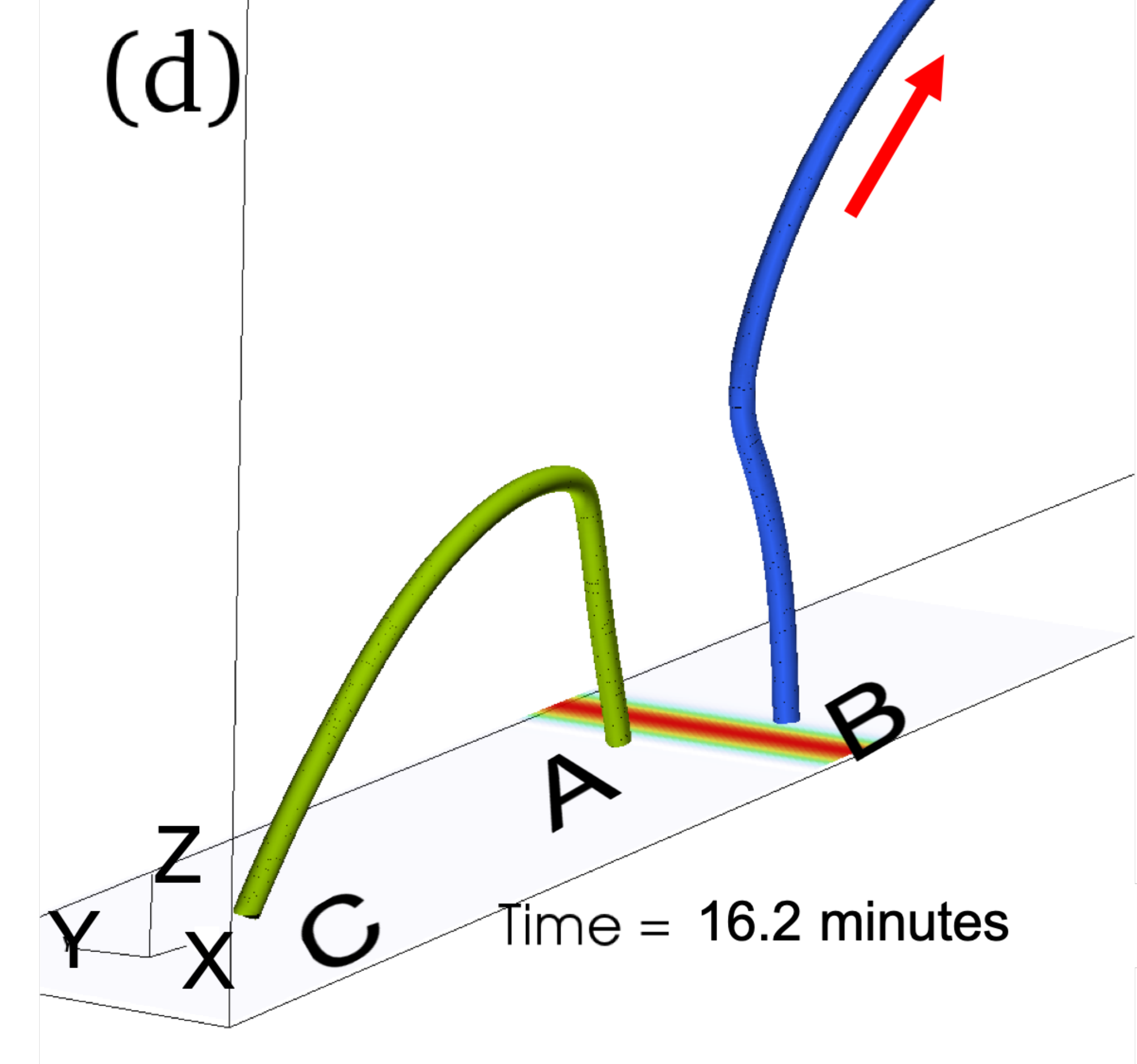}\\
\includegraphics[width=\mysizec\textwidth]{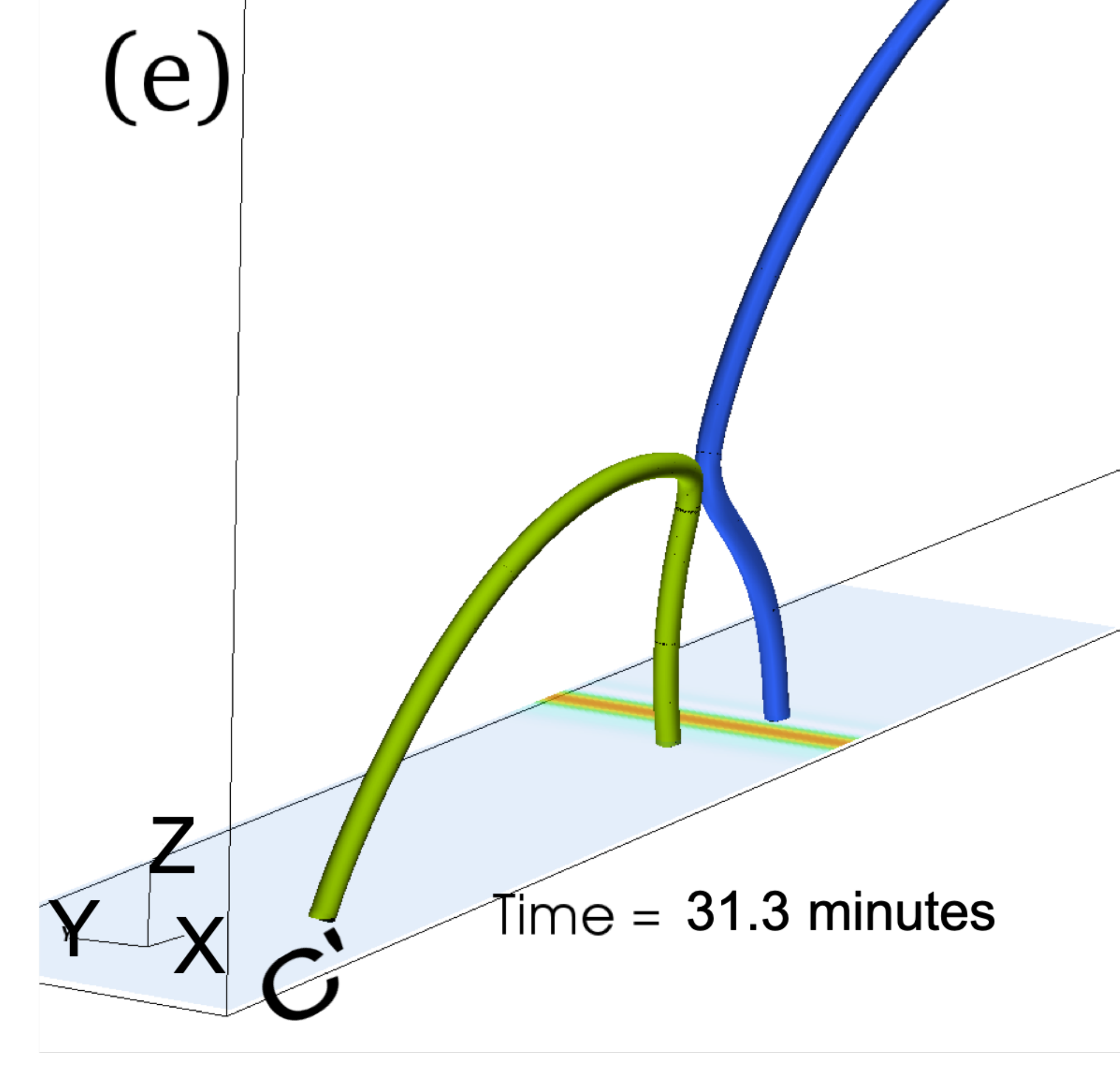}\includegraphics[width=\mysizec\textwidth]{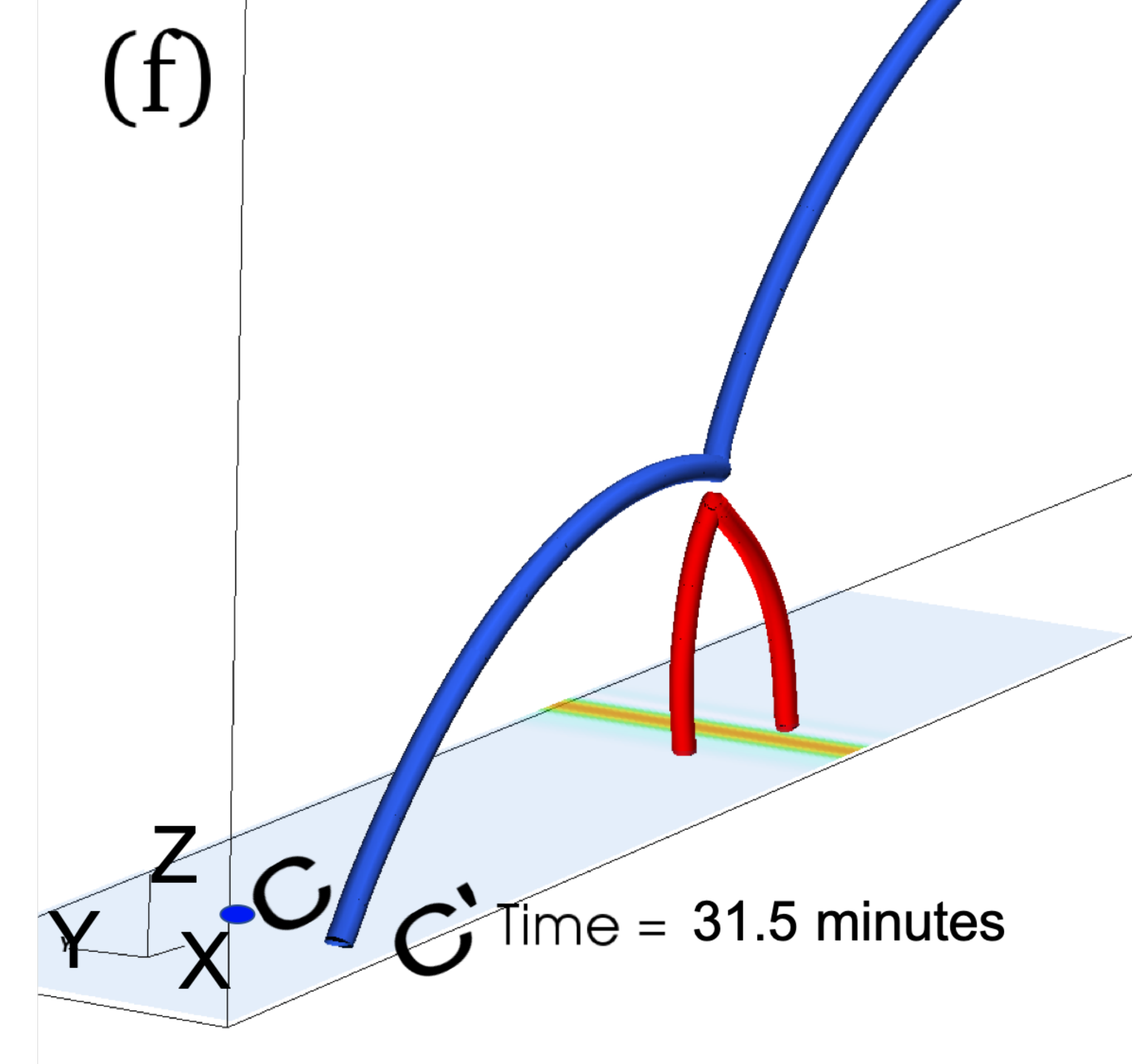}
\end{center}
\caption{Illustration of the double reconnection process suffered by a
  field line of the filament channel.  Blue: magnetic
 field line of the filament channel that supports the prominence. This line is rooted in the point labeled with C at $(x,z)=(-45, 0) \Mm$ and the other end is anchored on the other side of the numerical domain (not
 shown). Red and green: the other field lines involved in the double
  reconnection process. The bottom plane shows the $B_{y}$-component
 injected by the boundary driving. The red arrows in (c) and (d) show how the magnetic curvature propagates upwards along the blue line. See the text for additional
 details. \label{fig:byinjection}}
\end{figure}

To facilitate the visualization of the magnetic
connectivity changes during the jet generation phase described in the
foregoing, we have added another 3D figure (Fig.~\ref{fig:byinjection})
 containing a number of relevant field lines, still for case~1 and thickness
 $4$ Mm in the $y$ direction, as in Figure~\ref{fig:ropeformation}, but now
 for a larger domain in the $xz$ plane (for case~2 the evolution is
 qualitatively identical). The shown $xz$ domain contains the reconnection site and the arcade being sheared. (To appreciate
 the size of the represented domain, compare these panels with the global view
 in Fig. \ref{fig:initialconfiguration}). As per our initial condition,
 at time $t=0$ (not shown), all field lines are contained in the
 $xz$-plane. The blue field line in the six panels is traced from the
 photosphere at a location in the mid vertical plane of the figure and at
 $x=45$ Mm, i.e., far away to the right of the shown domain: this field line
 is part of the filament channel and goes through the prominence where the
 mass is loaded. By tracing it from the photosphere at a site with no
 horizontal motion we manage to show the evolution of the same field line in
 time. In the initial panel \ref{fig:byinjection}(a), for $t=15.7$ min, the other photospheric
 end of the line is located at $x=-45$ Mm, and is labeled point C. That field line is totally outside of the magnetic arcade being sheared: in panel \ref{fig:byinjection}(a), therefore, it continues being fully contained in the vertical midplane along the $x$-axis and has no $B_y$-component. In the same panel, the red field line is a loop of the parasitic dipole; its footpoints are labeled A and B and it is visibly slanted along the $y$ direction,
 as a direct effect of the photospheric driving. Figures(b)-(d) are
 taken at small intervals of less than $20$ seconds after the first one. At
 $t=15.8$ minutes (Fig. \ref{fig:byinjection}(b)) both field lines are about to reconnect at the
 current sheet at the top of the arcade; in panel \ref{fig:byinjection}(c) the reconnection has
 already taken place with a consequent drastic change in surface
 connectivity; the resulting field lines are moving apart sideways,
 pushed by the reconnection outflows.
In Figure \ref{fig:byinjection}(c), the blue field line shows a large degree of torsion and curvature that propagates away from the reconnection site in the form of an Alfv\'enic disturbance. The most important effect for us is the propagation upwards along the line [red arrow in Figs. \ref{fig:byinjection}(c) and
\ref{fig:byinjection}(d)]: the reconnection is injecting $B_y$ component into the filament channel, which, as said above, had none at the
beginning. Inversely, the perturbation is decreasing the $B_y$ component of
the lower part of the line, between the reconnection site and footpoint
B, as apparent when comparing Figures \ref{fig:byinjection}(c) and \ref{fig:byinjection}(d). 

In the later phase [Figs. \ref{fig:byinjection}(e) and \ref{fig:byinjection}(f)], the blue field line suffers a second
episode of reconnection at the almost vertical current sheet described
earlier in the section; here, the reconnection is with another field line
from the opposite side of the sheet, drawn in green in Figure \ref{fig:byinjection}(e) with
footpoint labeled C'. Through the reconnection (Fig. \ref{fig:byinjection}(e)), the blue field line gets connected to point C' and acquires a strong curvature right above the reconnection site, with consequent quick motion upwards.  
Below the reconnection site, the resulting line (drawn in red), which
could be said to be a {\it postflare loop} in analogy with the situation in
flares, is correspondingly moving downwards away from the reconnection
site. We see that the final configuration of the blue field line is slanted
in the $y$ direction.

The time evolution of the different energy integrals in case~2 is
qualitatively similar to case~1 as we can see in Figure
\ref{fig:temporal-evolution-energies}(b). As for case~1, one finds two peaks
of the kinetic energy, here at $t=17\mins$ and $t=28\mins$, associated, respectively, with the initial quiescent (breakout) reconnection and with the more violent second phase of reconnection, which takes place in the vertical current
sheet. 
The NP in case 2 is located at a lower height than in case~1; the photospheric field strength is therefore weaker and this leads to a lower total net flux contained in the arcade and to a smaller total injected energy (red curve) than in case~1. We have calculated the flux contained in the arcade through the two phases of the reconnection: the temporal rate of change is very similar (in relative terms) in both; the quiescent reconnection phase lets the arcade flux shrink by about 90\% in both cases; the reconnection at the vertical current sheet lets the arcade recover up to 60-70\% of the initial flux only, not the full initial magnetic flux.

As said in the introduction, the generation of the jets in the present experiment is parallel to the findings of \citet{Moreno-Insertis_2013}, \citet{Archontis_2013} and \citet{Wyper2018aa}. For comparison, one may check, e.g., the illustration of the two current sheets in Figures~11 and 12 (middle panel in either case) of \citet{Moreno-Insertis_2013}; the flux rope apparent in those figures is hurled and pressed against the upper current sheet and this leads to what those authors describe as the first violent eruption in their experiment (see the animation to their Figure 3, minutes 60 to 69). All of the previous simulations, however, did not have remote connectivity to a prominence or any similar major structure separate from the reconnection site, whose evolution is the objective of our present experiment.

\section{The propagation of the jets and their interaction with the prominence}\label{sec:the-jet}

In this section, we describe how the quiescent and eruptive jets launched at the reconnection sites described in Section~\ref{sec:jet-synthesis} propagate inside the filament channel and how they interact with the prominence. In the 3D jet models of \citet{Pariat09, pariat_model_2016} the beginning of the reconnection process is seen to lead to a nonlinear torsional Alfv\'en wave that communicates to the open field lines the change in orientation of the magnetic field vector that is taking place in the reconnection site. At the rear of the wave, the plasma is compressed, heated, and accelerated. When the plasma $\beta$ is small, the compression front advances with lower speed than the Alfv\'enic perturbation.

In our experiments, due to the small plasma beta in the FC, we see a clear-cut separation between an Alfv\'enic front and a compressive wave that follows with lower velocity. 
Considering the first reconnection process, i.e., the quiescent reconnection across the breakout current sheet, successive pairs of field lines (one from the filament channel and one from the sheared arcade) reconnect and lead to a hybrid field line consisting of a section with a significant $B_y$-component (the part coming from the arcade) and a section contained in the $xz$ plane, coming
from the FC. The {\it corner} between the two sections leads to the launching of an Alfv\'enic perturbation along the field line; the perturbation is incompressible and moves approximately with Alfv\'en speed toward the prominence (as well as downward toward the photosphere). On the other hand, the plasma on the reconnected field line also has a pressure excess compared to the original values in the FC. This launches the quiescent jet that moves toward the prominence, preceded by a shock with supersonic but sub-Alfv\'enic speeds, hence lagging behind the Alfv\'enic perturbation. The following subsections describe those two fronts (Alfv\'enic, supersonic).

\begin{figure}[!ht]
\centering\includegraphics[width=0.48\textwidth]{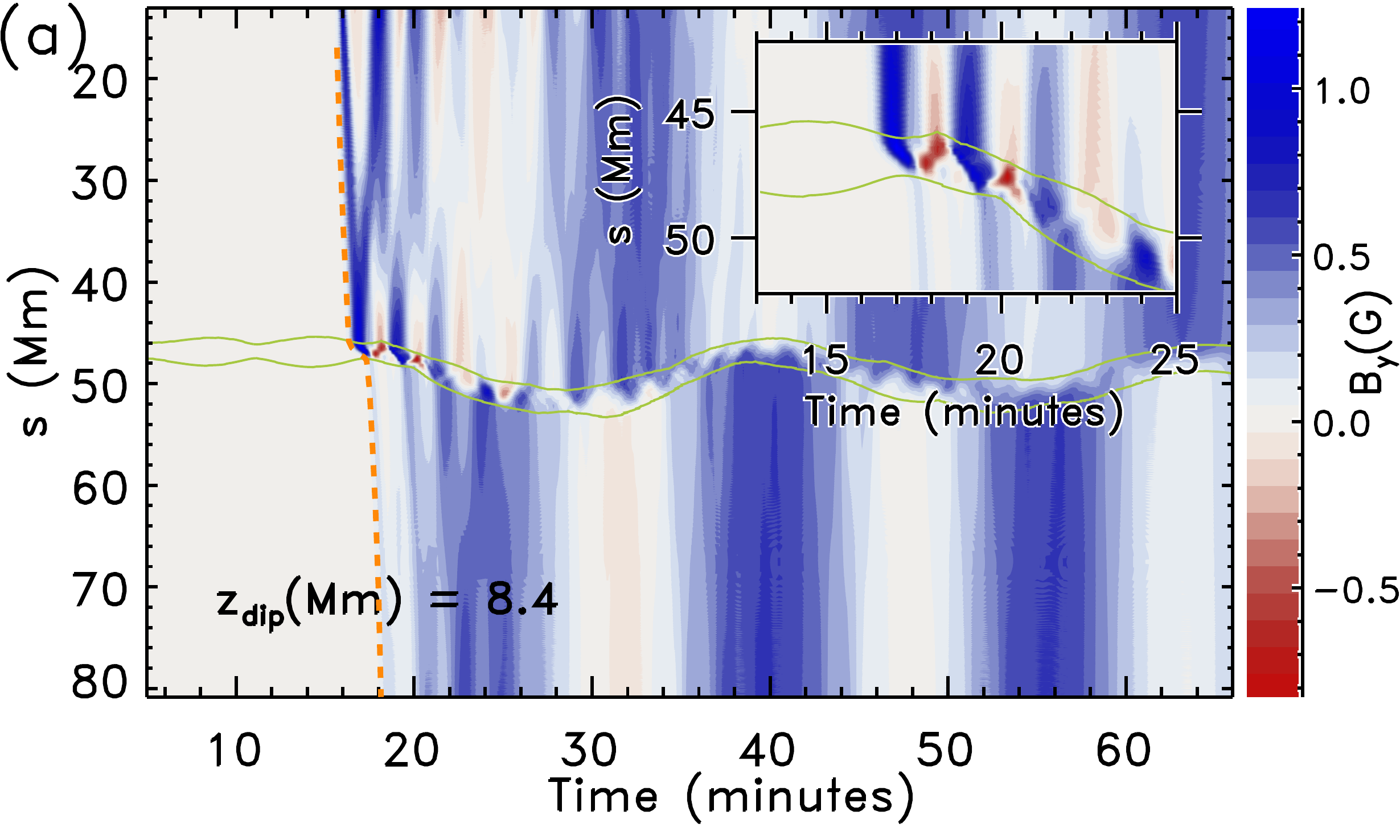}\\
\includegraphics[width=0.48\textwidth]{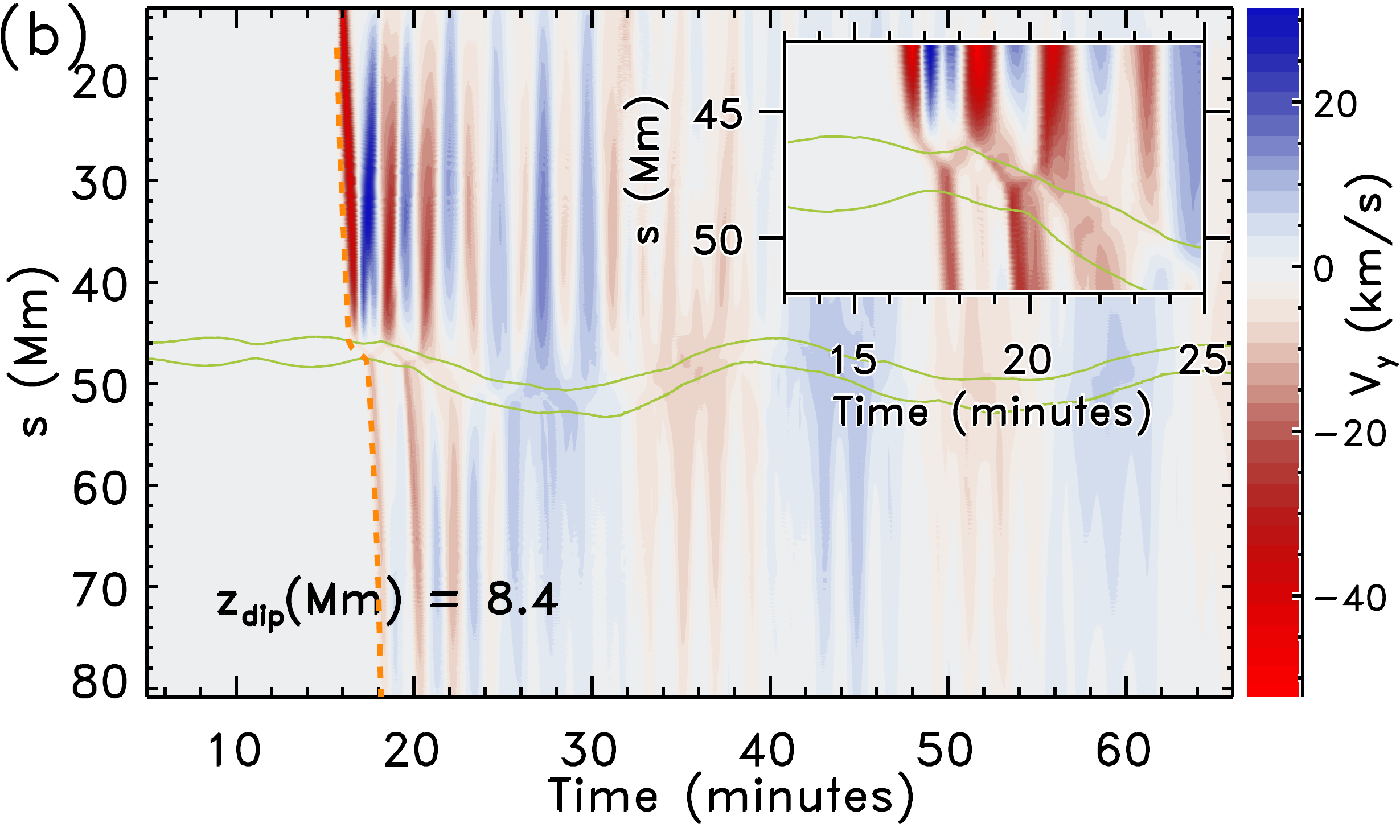}
\caption{Time-distance diagram of $B_y$ (a) and $v_y$ (b) along a selected field line for case~1. The field line has the bottom part of the dip at $\zdip=8.4\Mm$ and $x=0$. The thin green contours delimit the cool prominence plasma. The orange dashed line shows the position with time of the first Alfv\'enic front. The slope of this orange curve shows the propagation velocity of the front.\label{fig:alfvenic-propagation}} 
\end{figure}

\subsection{The Alfv\'enic front}\label{sec:alfvenic front}

\begin{figure}[!ht]
\centering\includegraphics[width=0.45\textwidth]{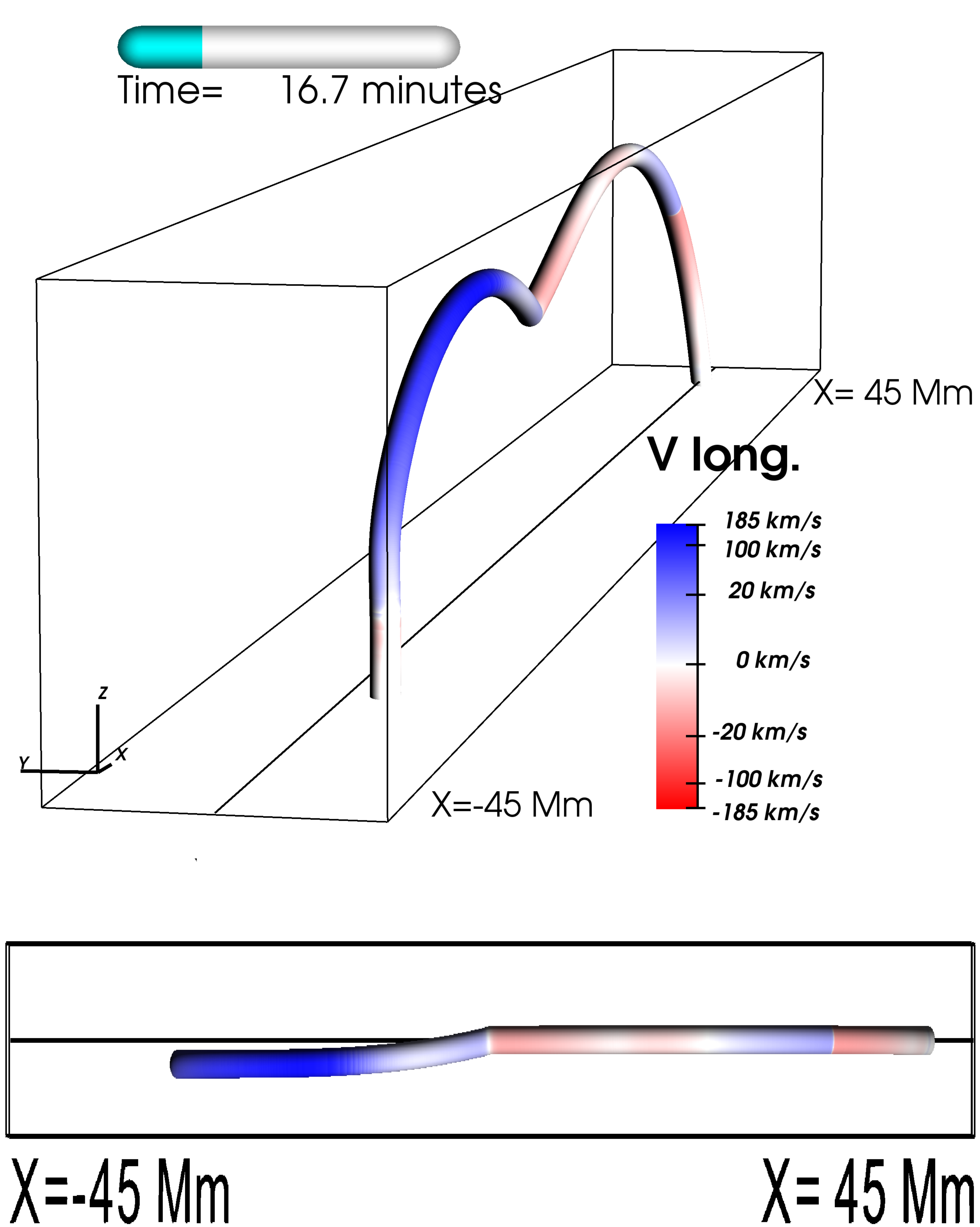}
\caption{Three-dimensional plot of a typical FC field line. In the upper panel one sees the dipped structure with the two footpoints rooted in the chromosphere. In the lower panel a top view is shown. The colors show $\vlong$ along the field line using a logarithmic scale for $|\vlong|$ on either side of the zero. An animation of this figure is available in the online version of the journal. The animation shows the temporal evolution of the field line from the beginning of the simulation to $t=81.5$ minutes; shown in the present figure is the frame at $t=16.7$ minutes. The animation shows that the reconnection produces a sudden horizontal displacement along the $y$-direction of the footpoint initially located at $x=-45\Mm$. A nonlinear Alfv\'enic perturbation propagates back and forth along the field line producing complex motions that finally settle into a standing mode along the horizontal $y$-direction. In the animation we also see the jet plasma flows emanating from the reconnection site propagating along the field line with velocities up to 180$\kms$. The  plasma in these flows ends up moving back and forth along the field line with decreasing velocity (below $30\kms$ at the end of the simulation). \label{fig:3dviews-singleline}}
\end{figure}

To illustrate the propagation of the first front, in Figure~\ref{fig:alfvenic-propagation} a time-distance diagram is presented for $B_{y}$ (top panel) and $v_{y}$ (bottom panel) along a typical field line in the FC. For simplicity we choose for this the blue field line shown in Figure \ref{fig:byinjection}. For later use in this section, the same field line is also shown in Figure \ref{fig:3dviews-singleline}, and the accompanying animation, from two different perspectives.
In Figure~\ref{fig:alfvenic-propagation}, the vertical axis is the distance, i.e., the arc-length parameter, $s$, measured along the field line.
The range of distances covered in the plot goes from $s=17$ Mm to $s=80$ Mm, which corresponds to a range in $x$ roughly between $(-33, 33)$ Mm, with a slight variation in time as the field line moves. The prominence is located between $s=(45,50)$, approximately, and its position, marked with a thin green line, shifts back and forth in time; initially, it is located at the bottom of the dip at $s\approx 47$ Mm.

The Alfv\'enic front resulting from the quiescent reconnection of the chosen field line is seen entering the diagram from the top at $t=16$ min. The front advances with a velocity of $1000$~km~s$^{-1}$ (corresponding to the slope of the segment of the orange dashed line between the top and $s=46$), slightly above the local Alfv\'en velocity, and reaches the prominence at $t\approx 16.5$ min. During that time, the perturbation seems to be evolving toward a switch-on shock, but it is too weak (Alfv\'en Mach number only slightly above 1) as to become a sharp transition, a real shock, before it reaches the prominence. When reaching the prominence, given the large inertia of the latter, the front is partially reflected, so, in the diagram, the advance of the front in the arc-length region above the prominence ($s \gtrsim 46$ Mm) has a V-shape. Looking more in detail we realize that $B_y$ increases from zero coinciding with the passage of the front, but that it returns to zero immediately thereafter, only to be temporarily modified once more when the reverse front sweeps the region again. In other words, the front is trying to straighten the section of the field line between reconnection point and prominence, {\it first according to the position of the new footpoint resulting from the reconnection}, and then according to the attachment point of the field line at the prominence. This is apparent in the bottom panel of the figure, where the forward perturbation has a negative $v_y$ speed, while the opposite applies for the reverse one. This is more clearly seen in Figure \ref{fig:3dviews-singleline}, which is taken at the time of the first arrival of the front at the prominence: in the panel on the bottom, which contains a top-view of the domain, the field line is seen to have been straightened after the passage of the front (the whole evolution of the field line can be seen in the accompanying animation). The reverse front is reflected back again in the chromosphere of the reconnected field line and travels back and forth leading to a pattern of $V$'s thereafter in the figure. There are no clear density, pressure, $B_{x}$, or $B_{z}$ perturbations associated with this front.

When arriving at the prominence, the forward front is partially transmitted into the region of dense plasma. There (see the inset on the top right of the panels in Figure~\ref{fig:alfvenic-propagation}), the Alfv\'en velocity is much smaller; the front steepens to a real switch-on shock of Alfv\'enic Mach number up to $5$ and with $B_y$ attaining the largest values in the whole simulation. Even then, the shock front in the prominence advances with a lower speed than the precursor front outside: this is apparent in the figure, especially in the inset, where the shock velocity is measured to be $22$~km~s$^{-1}$. Also apparent is the reflection of the front at the right-hand side boundary of the prominence, and the (weak) partial transmission to the external domain beyond that. The front is therefore trapped inside the prominence, but it is partially leaked in each internal reflection. Concerning the transmission into the domain to the right of the prominence, there the front propagates again basically with Alfv\'en speed as a non-shocked Alfv\'enic perturbation. When reaching the chromosphere on that side, the front bounces back and leads to a multiple-{\it V-pattern} in Figure~\ref{fig:alfvenic-propagation} on that side of the prominence as well. Seen in time, one discerns a checkered pattern in the top panel of the figure, which corresponds to the establishment of the fundamental Alfv\'en mode oscillating from end to end of the given field line, as explained in Section~\ref{sec:after-jet}. Another remarkable feature apparent in Figure~\ref{fig:alfvenic-propagation} is the fact that the prominence is oscillating sideways along the field line (as seen through the sinusoidal shape of the thin green curves). This oscillation is excited through the impact of the jets, as described in the following subsections and in Section~\ref{sec:after-jet}. 

\begin{figure*}[!ht]
\centering\includegraphics[width=1\textwidth]{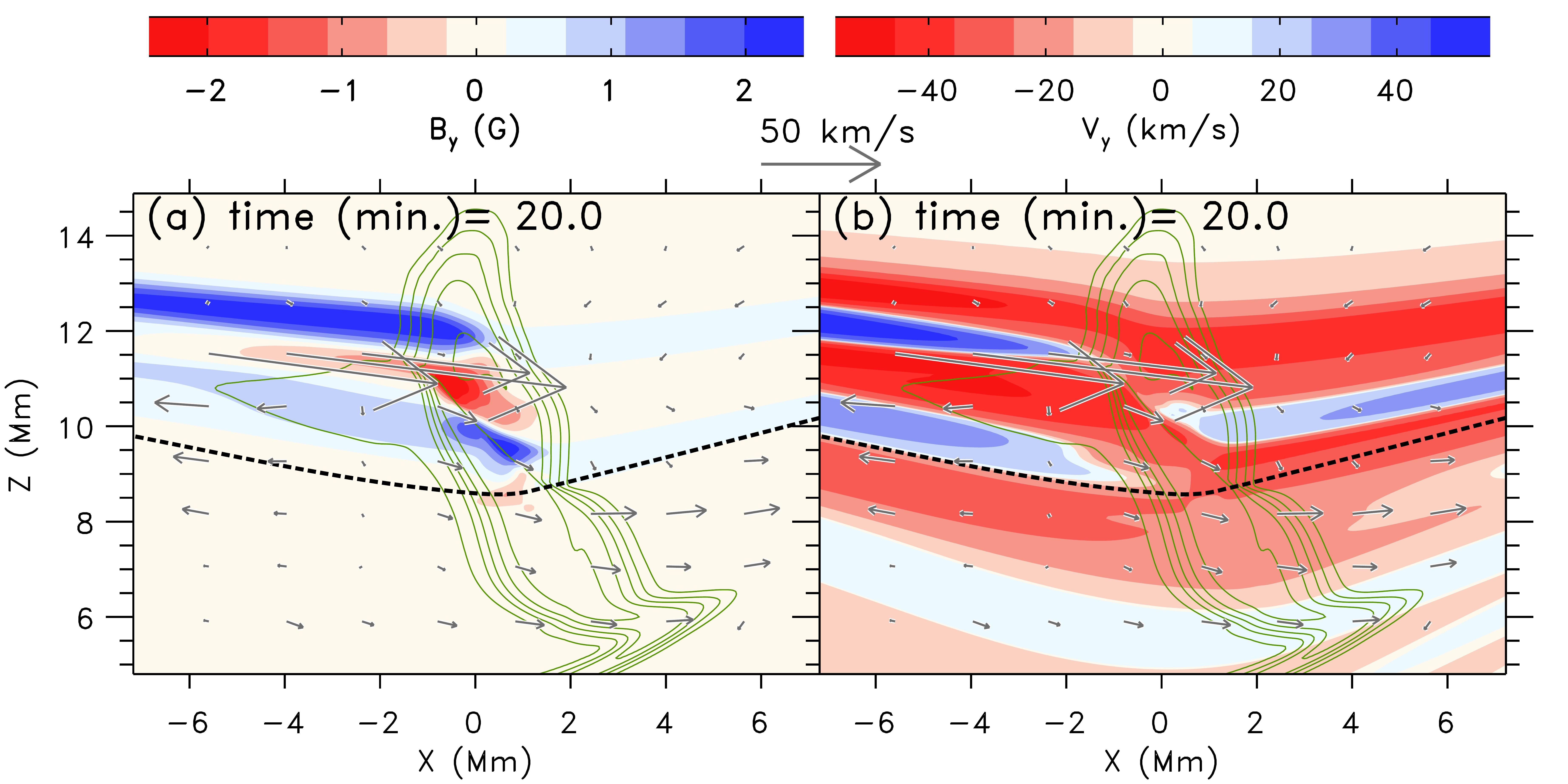}
\caption{Color map of $B_y$ (a) and $v_y$ (b) for case~1, illustrating  the spatial distribution of the Alfv\'enic perturbation at $t=20$. The green contours are the isolines of density and mark the location of the dense prominence. The arrows represent the projected velocity field. The arrow shown at the top of the panels corresponds to a velocity of 50 $\kms$. The dashed line is the field line used in Figure \ref{fig:alfvenic-propagation}.
\label{fig:jetinteraction-by-vy-case1}}
\end{figure*}

In Figure \ref{fig:jetinteraction-by-vy-case1} we show a snapshot of the Alfv\'enic perturbation in the vertical plane taken at $t=20\min$, thus complementing the time-distance diagram of Figure~\ref{fig:alfvenic-propagation}. The various fronts bouncing back and forth along a single field line in Figure~\ref{fig:alfvenic-propagation} are now seen to give rise to a layered structure for the perturbations at different heights (to facilitate the comparison, a dashed line has been drawn coinciding with the field line used in Figure \ref{fig:alfvenic-propagation}). The Alfv\'enic disturbance first reaches the bottom part of the prominence, at an earlier time than shown in Figure~\ref{fig:jetinteraction-by-vy-case1}.
As the BCS is moving to higher positions, the associated Alfv\'enic perturbations also move to higher field lines. At $t=20 \min$, for instance, the time shown in the figure, the Alfv\'enic disturbance has already reached a height of $z=12\Mm$. The maximum perturbation of the $B_y$ field is about $2$ G, to be compared with the original value of about $10$ G for the initial FC field. The $v_y$ perturbation has values above $50 \kms$ including inside the prominence.

\subsection{The acoustic front}\label{sec:acoustic_front}

Additionally to the Alfv\'en perturbation the reconnection causes a sudden pressure increase that leads to an outflow, a jet, propagating along the field lines, with a shock front at its head. The plasma $\beta$ is small enough for this shock to be basically of the acoustic type (it is a slow shock that would be exactly acoustic in the limit $\beta \ll1$); also, the velocity of this shock is well below that of the Alfv\'en perturbation, so the former lags behind the latter. The postshock region behind the acoustic shock is what can be identified as the quiescent jet. 

\begin{figure}
\centering\includegraphics[width=0.48\textwidth]{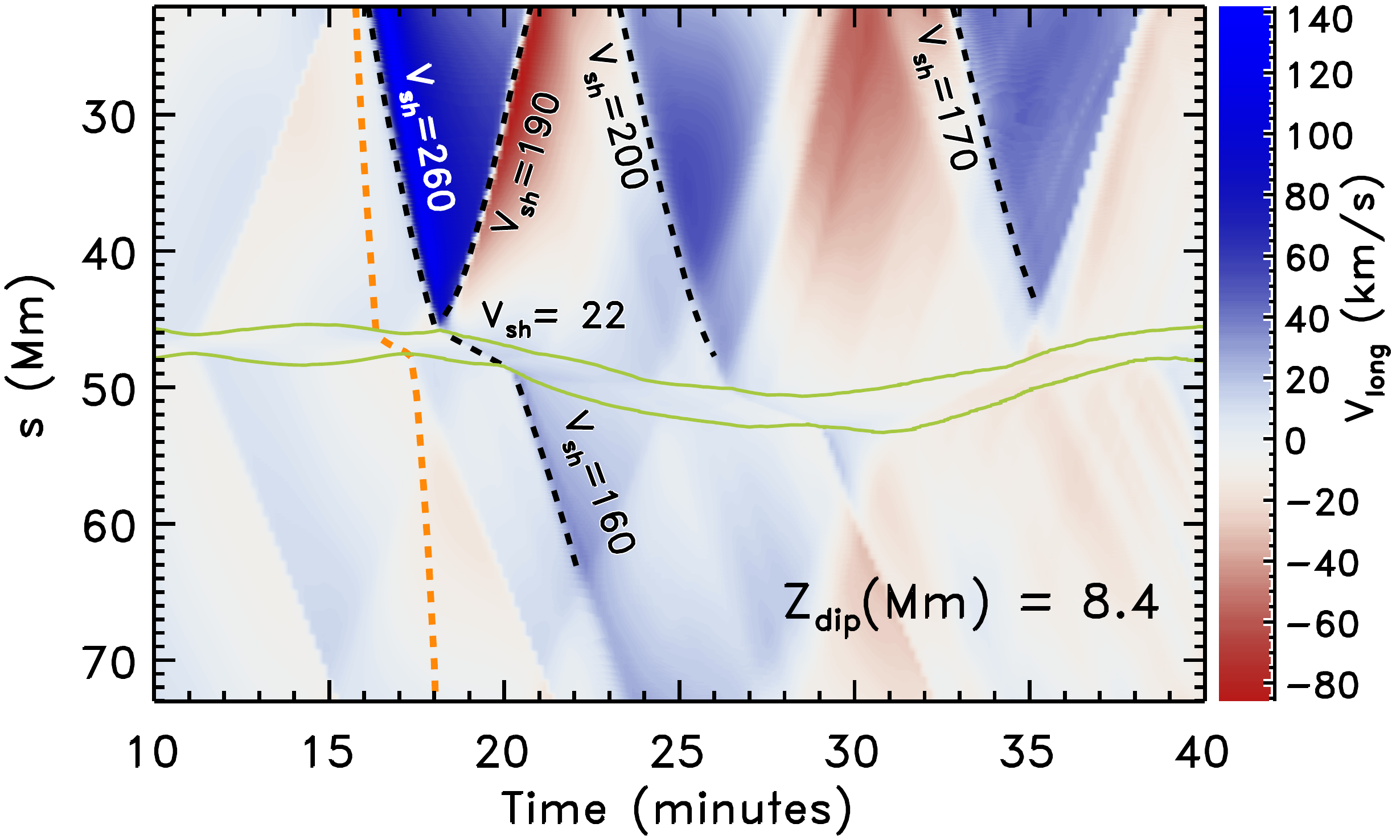}
\caption{Time-distance diagram along the same field line as in Fig.~\ref{fig:alfvenic-propagation} but here for $\vlong$. The thin green contours delimit the cool prominence mass. The black dashed lines mark the position of the fronts; the maximum speed resulting for each segment after fitting the positions with a third-order polynomial is indicated (in $\kms$). The orange dashed line is the position of the Alfv\'enic front shown in Fig.~\ref{fig:alfvenic-propagation}.
\label{fig:time-distance-vlong}}
\end{figure}

Figure \ref{fig:time-distance-vlong} shows a time-distance diagram of the longitudinal velocity, $\vlong$, defined as: 
\begin{equation}\label{eq:vlong}
 \vlong=\vec{v} \cdot \vec{b} \,, \quad\hbox{with} \quad \vec{b}
 = \frac{\vec{B}}{|\vec{B}|} \;, 
\end{equation}
along the same field line selected for Figure~\ref{fig:alfvenic-propagation}.  
The acoustic shock appears in the figure around $t=16.1\mins$ and propagates along the field line toward the prominence; it reaches the prominence at $t=18 \mins$ and $s=49\Mm$. To measure the shock speed, we fit a third-order polynomial to the position of the fronts in the time-distance diagram: the inclination of the front trajectories directly gives the shock speed, $\vshock$. At $t=17\mins$, for instance, $\vshock \approx 240 \kms$, which is clearly above the sound speed of the unshocked plasma, $\csound=147\kms$, and yields an incoming acoustic Mach number of about $1.7$. At that time, the shocked plasma, the jet, has a velocity of about $140\kms$ right after the shock, below the local sound speed there. For comparison, the location of the Alfv\'enic front discussed in the previous subsection has been overplotted as an orange dashed line. 
The fact that the Alfv\'enic front reaches the prominence well before the acoustic one implies that the jet propagates along field lines that have already been modified by the Alfv\'enic perturbation. This is also clear in Figure \ref{fig:3dviews-singleline} where we have plotted $\vlong$ using the color scale over a single field line: the field line is seen to be modified before the arrival of the jet.

To label and identify individual field lines and the motions on them, we use the $z$-coordinate of the dip at the center of the structure at $t=200\secs$, i.e., at the end of the mass loading phase (see Sec.~\ref{sec:model}), which we call $\zdip$. Of course, due to the motions of the plasma the position of the dip of the individual lines changes with time. However, in spite of the large amplitude of the plasma motions apparent in the figures, the change in the underlying magnetic structure is relatively small and the positions of the dips are not significantly changed.

The interaction of this first acoustic front with the prominence at $t \approx18\mins$ produces a reflected shock moving in the opposite direction and a transmitted one inside the prominence, as apparent in Figure~\ref{fig:time-distance-vlong}. The reverse shock tries to stop the advance of the jet. As is often the case with reflected shocks advancing into an expanding shocked medium, the velocity behind the new shock becomes negative [red band in the time range $(20,22)\mins$, between about $s=55$ and the top of the frame], so the plasma ends up moving toward the chromospheric base of the field line, with velocities as negative as $-80\kms$. The transmitted front, in turn, propagates inside the dense prominence plasma with a velocity of $22\kms$, yielding a shock Mach number of about $1.5$, nearly as strong as the incident shock outside of the prominence. After crossing the prominence, the front is transmitted to the other side
in the form of a weak acoustic perturbation. A second incident shock front, moving with a maximum $\vshock = 190\kms$, reaches the prominence at $t=26\mins$. This one is launched through the reflection at chromospheric heights of the reverse acoustic front discussed above.
A third incident shock is apparent on the right side of Figure~\ref{fig:time-distance-vlong}, which reaches the prominence at $t=35\mins$; this is the shock at the head of the eruptive jet. The shock speed reaches $\vshock=170\kms$ in this case; the incoming Mach number is around 1.1. The postshock velocity of the plasma, i.e., the speed of the eruptive jet is around $40\kms$. This value is much smaller than the velocity quoted above for the quiescent jet.
The reason for this low value is that the strong flows associated with the eruptive jet are located mainly in the top part of the prominence in lines with $\zdip>10\Mm$, whereas the field line chosen for Figures~\ref{fig:time-distance-vlong}~and~ \ref{fig:alfvenic-propagation} has $\zdip=8.4\Mm$. For field lines below those, the eruptive jet flows are weak. The opposite happens with the quiescent jet flows: they are weak in the top part of the prominence. The field line selected for Figures~\ref{fig:time-distance-vlong}~and~ \ref{fig:alfvenic-propagation} was chosen to allow both jets to appear in the time-distance diagram.

\begin{figure*}[!ht]
\centering\includegraphics[width=1\textwidth]{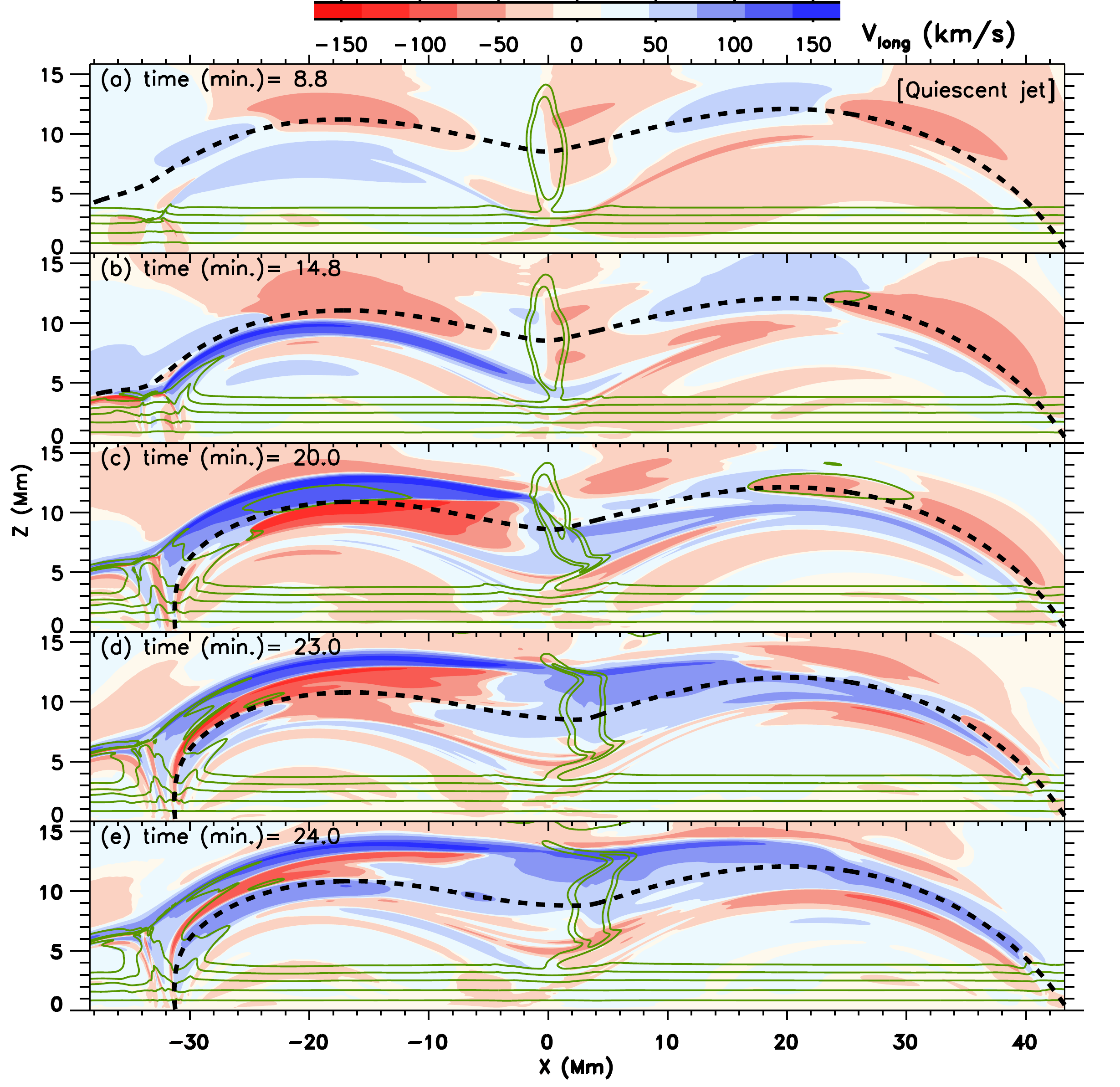}
\caption{Plot of the time evolution of the longitudinal velocity $\vlong$ (Equation~\ref{eq:vlong}) showing the interaction of the ejected plasma with the prominence in case~1. The isolines of the density are shown as green contours. The thick dashed line is the field line used in Figures \ref{fig:alfvenic-propagation} and \ref{fig:time-distance-vlong}. \label{fig:jetflow-vlong-case1-quiescent}} 
\end{figure*}
\begin{figure*}[!ht]
\centering\includegraphics[width=1\textwidth]{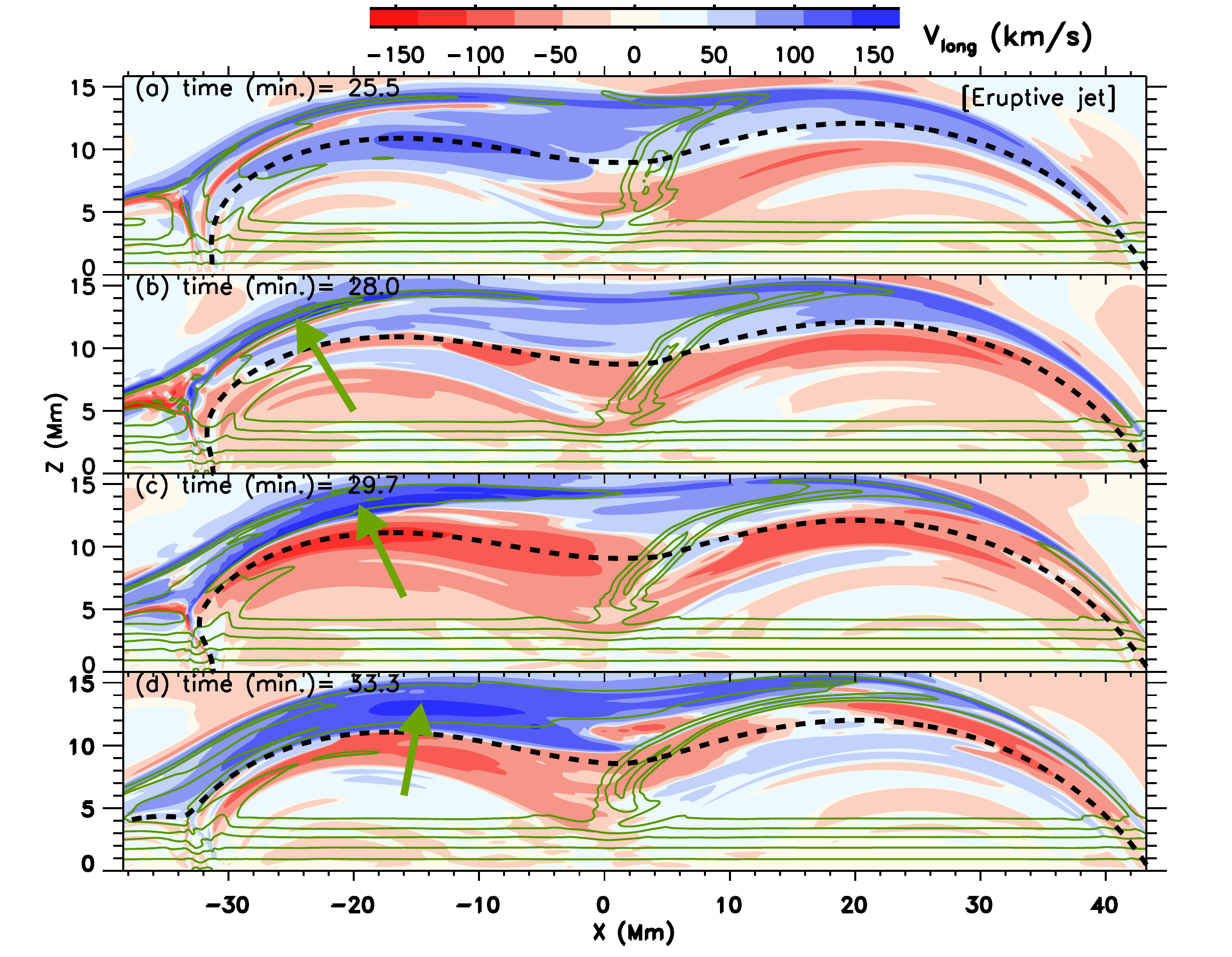}
\caption{As Figure~\ref{fig:jetflow-vlong-case1-quiescent} but for the eruptive jet phase after $t=24$ minutes. The large green arrows points to the region with large flows associated with the eruptive jet.\label{fig:jetflow-vlong-case1-eruptive}} 
\end{figure*}

\begin{figure*}[!ht]
\centering\includegraphics[width=1\textwidth]{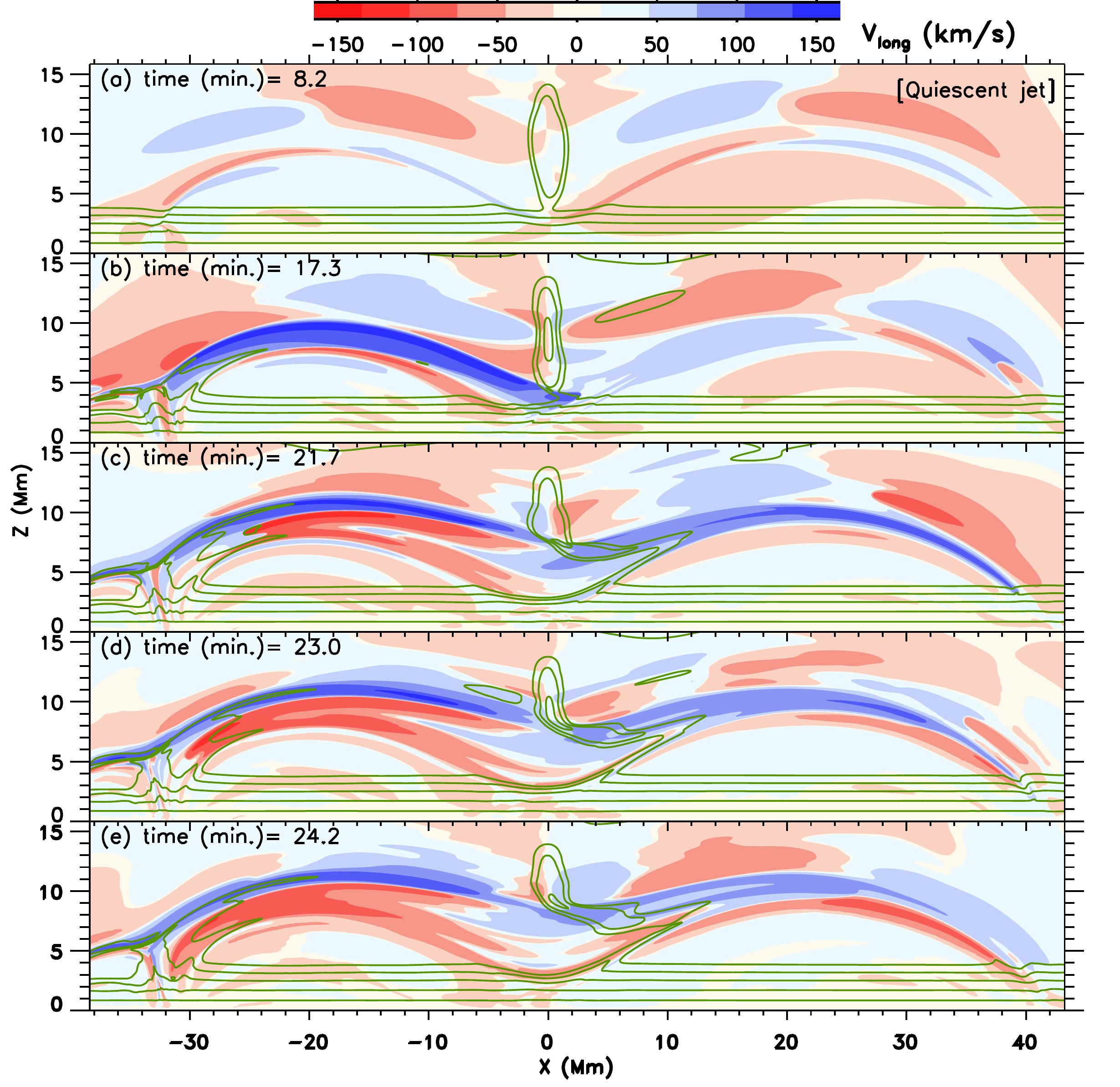}
\caption{Like Figure~\ref{fig:jetflow-vlong-case1-quiescent},
 but for case~2. \label{fig:jetflow-vlong-case2-quiescent}} 
\end{figure*}
\begin{figure*}[!ht]
\centering\includegraphics[width=1\textwidth]{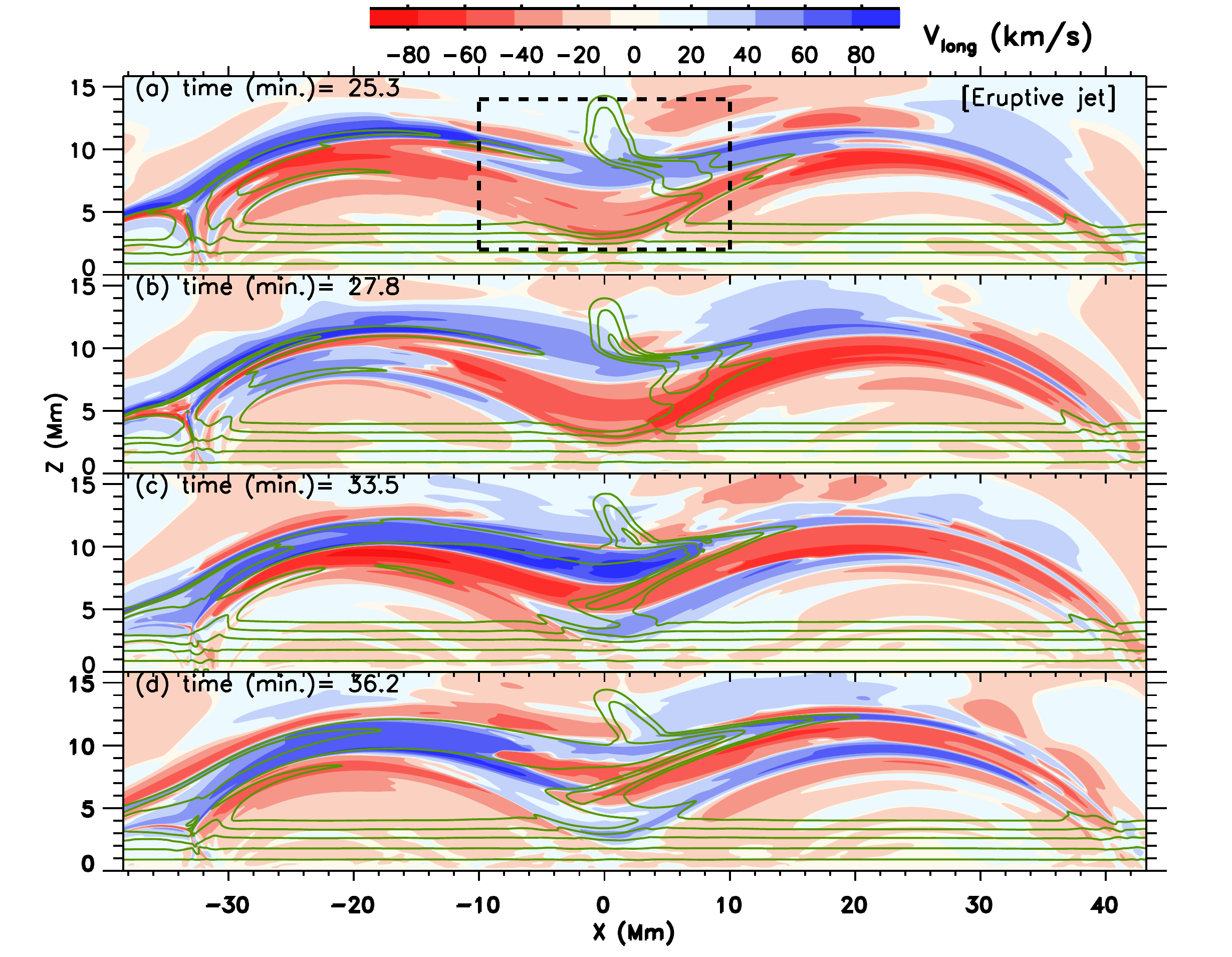}
\caption{Like Figure~\ref{fig:jetflow-vlong-case1-eruptive},
 but for case~2. The dashed box in panel (a) encloses the region used to compute the forces shown in Fig. \ref{fig:forces}.\label{fig:jetflow-vlong-case2-eruptive}} 
\end{figure*}

\subsection{The propagation of the jets}\label{sec:jets}

In this section we focus on how the jets move inside the global FC and interact with the prominence. Figures \ref{fig:jetflow-vlong-case1-quiescent} and \ref{fig:jetflow-vlong-case1-eruptive} show the plasma velocity along the magnetic field, $\vlong$, for case~1 and Figures \ref{fig:jetflow-vlong-case2-quiescent} and \ref{fig:jetflow-vlong-case2-eruptive} for case~2. In Figures~\ref{fig:jetflow-vlong-case1-quiescent} and \ref{fig:jetflow-vlong-case1-eruptive}, the field line used for the description of the evolution of the fronts in the previous subsections (in particular, in Figures~\ref{fig:alfvenic-propagation} and \ref{fig:time-distance-vlong}) has been overplotted as a thick dashed line. The density is also shown using green contours; the prominence is apparent as a vertical structure around $x=0 \Mm$. 

Starting with case~1, Figure~\ref{fig:jetflow-vlong-case1-quiescent} shows the propagation of the quiescent jet. In Figure \ref{fig:jetflow-vlong-case1-quiescent}(a) a slight depression in the density isocontours around $(x,y) =(-33,3)\Mm$ marks the location of the BCS and indicates that breakout reconnection has already started there. The acoustic front traveling at about $240\kms$ described in the previous subsection crosses the distance to the prominence in only a few minutes. By $t=14.8$ [Fig. \ref{fig:jetflow-vlong-case1-quiescent}(b)], the quiescent jet behind the shock is fully developed, has a speed around $150\kms$, and has reached the base of the prominence. As the reconnection in the BCS continues, more field lines lying increasingly higher in the FC are brought to the current sheet and suffer the same process (reconnection, launching of the fronts, development of the jet). In Figure \ref{fig:jetflow-vlong-case1-quiescent}(c), the field line marked as a dashed line has already been reconnected: the quiescent jet is propagating along field lines above it, whereas the phenomenon of reverse jet described in the previous subsection (Fig.~\ref{fig:time-distance-vlong}, $t=20\mins$) is affecting it. As a consequence [Figs. \ref{fig:jetflow-vlong-case1-quiescent}(c)-(e)], the quiescent jet is launched on progressively higher field lines, reaching increasingly higher locations of the prominence. The impact of the forward jets visibly displaces the heavy prominence plasma toward the right.
In the same Figures \ref{fig:jetflow-vlong-case1-quiescent}(c)-(e), we can also see how the front propagates inside the prominence. When it reaches the
other side, it continues propagating in the channel, but, as already said, as a weak acoustic perturbation. When the fronts reach chromospheric heights at either end of the field lines, they bounce back. As this process occurs a few times for each field line, the resulting pattern in the filament channel becomes quite complex.

After $t=24\mins$ the breakout reconnection phase ends and eruptive reconnection starts to take place (Fig.~\ref{fig:jetflow-vlong-case1-eruptive}) in the nearly vertical current sheet
within what remains of the sheared arcade.
This results in stronger flows leading to an eruptive jet [Figs.~\ref{fig:jetflow-vlong-case1-eruptive}(a)-(d)]. In Figure \ref{fig:jetflow-vlong-case1-eruptive}(a) 
the current sheet can be located by identifying a thin, nearly vertical stripe of high velocity (blue color) located at $x \approx 33$ Mm.
In Figure \ref{fig:jetflow-vlong-case1-eruptive}(b), at the top of the reconnection site an extended thin blue region with large velocity (the eruptive jet) extends as an arch that crosses, e.g., $(x,z)=(-25,13)\Mm$ (see the large green arrow), and reaches the top part of the prominence. In this reconnection phase, the field lines are being fed to the current sheet from the sides; successive field lines reaching the vertical current sheet are located in progressively lower levels of the FC, the exact opposite to the BCS reconnection. As a consequence, the eruptive jet is seen to descend in the FC as time advances; this can be seen by comparing the deep blue region (marked with large green arrows) in Figures \ref{fig:jetflow-vlong-case1-eruptive}(b)-(d). In Figure \ref{fig:jetflow-vlong-case1-eruptive}(d), the jet has almost reached the top of the prominence, which is now displaced to the right, at $(x,z)=(20,14)\Mm$, as a result of the previous evolution. The flows pointing to the left apparent in Figures \ref{fig:jetflow-vlong-case1-eruptive}(c) and \ref{fig:jetflow-vlong-case1-eruptive}(d) (red area below the eruptive jet, e.g., covering the point $(x,z)=(-20,9)\Mm$) are return flows from the earlier impact of the quiescent jet apparent in Figure~\ref{fig:jetflow-vlong-case1-quiescent}(b). The return flows of the eruptive jet occur later than those plotted in Figure \ref{fig:jetflow-vlong-case1-eruptive}.

The time evolution of case~2 (Figs.~\ref{fig:jetflow-vlong-case2-quiescent} and \ref{fig:jetflow-vlong-case2-eruptive}) shows many similarities to case~1 but has some significant differences as well. The initial null point is located lower in the corona than for case~1, and so are the BCS and the initial energy release through reconnection. In Figures \ref{fig:jetflow-vlong-case2-quiescent}(b)-(e), one can easily distinguish how the quiescent jet (dark blue region) first reaches the prominence [$t=17.3\mins$, Fig. \ref{fig:jetflow-vlong-case2-quiescent}(b)] and imparts positive horizontal momentum to the latter with the result that it is greatly displaced to the right; the shock front at the head of the jet traverses the prominence and gets to the other side: the postshock flow, the jet, is seen to reach the chromosphere at the opposite end of the FC ($x \approx 45\Mm$) at around $t=21.7\mins$. In those panels the reflected jet is seen as well in, e.g. $(x,z)=(-25,8)\Mm$. In contrast to case~1, the quiescent jet hits the prominence at a maximum height of $z=10\Mm$ (middle height of the prominence), not reaching its top. The eruptive phase starts at $t \sim 25\mins$ [Fig.~\ref{fig:jetflow-vlong-case2-eruptive}(a)]. Like for case~1, the eruptive jet is launched at successively lower heights reaching first the middle height of the prominence, then, as time advances, lower levels, and finally the prominence base. The dark blue regions covering, e.g., the point $(x,z)=(-25,10)\Mm$ in Figures \ref{fig:jetflow-vlong-case2-eruptive}(a)-(d) are part of the eruptive jet complex. As is apparent in Figures~\ref{fig:jetflow-vlong-case2-quiescent} and \ref{fig:jetflow-vlong-case2-eruptive}, the top of the prominence is spared by all jets in this case.

The impact of the quiescent jet on the prominence causes significant displacement of the dense plasma [see the green contours in e.g., Figs. \ref{fig:jetflow-vlong-case1-quiescent}(e) and \ref{fig:jetflow-vlong-case2-quiescent}(e)], which leads to large-amplitude oscillations of the prominence (Section~\ref{sec:after-jet}). The eruptive jet is so energetic that it moves part of the plasma out of the dipped region of the FC [see the green contours around $(x,z)=(25,13)\Mm$ in Figures \ref{fig:jetflow-vlong-case1-eruptive} and \ref{fig:jetflow-vlong-case2-eruptive}.] A fraction of this plasma is drained away from the prominence falling to the chromosphere at the right footpoint.

\begin{figure}[h]
\centering\includegraphics[width=0.47\textwidth]{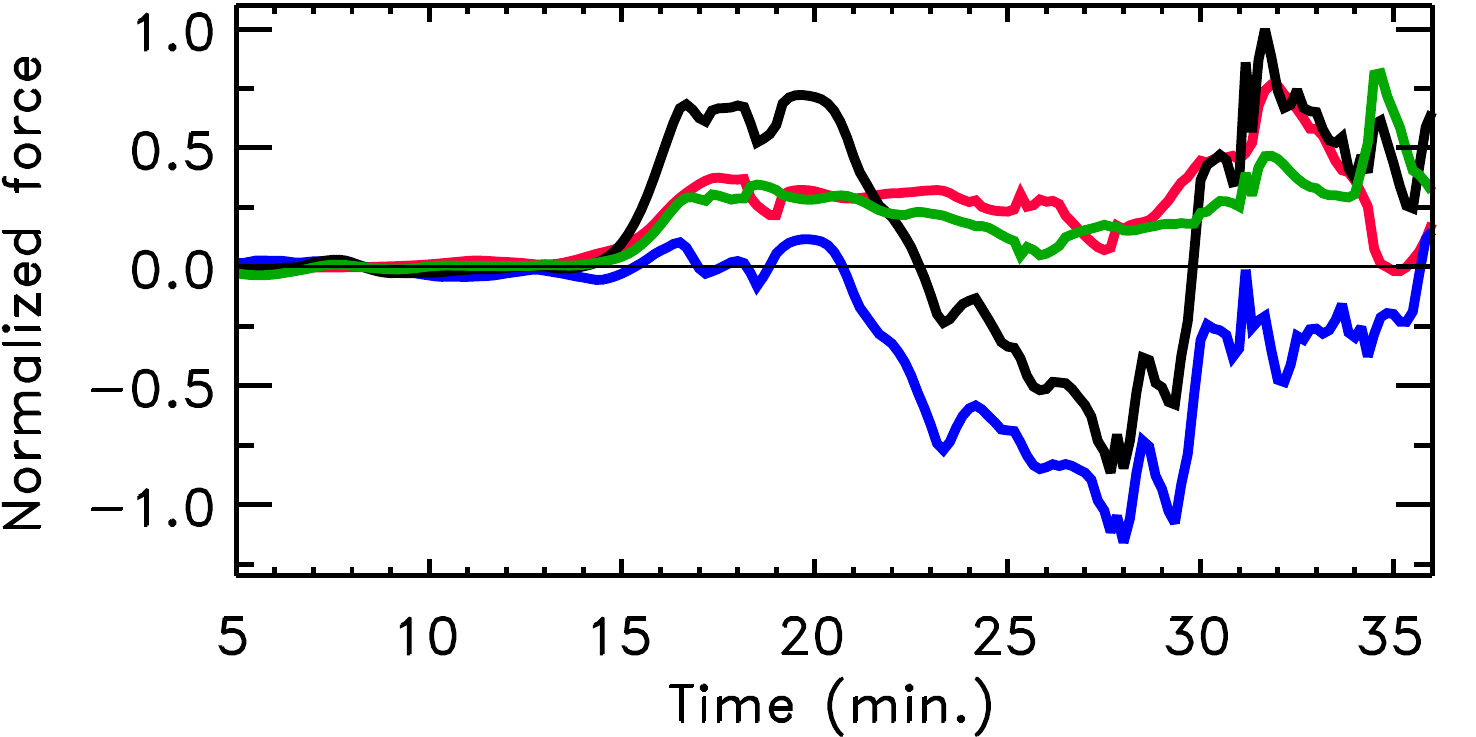}
\caption{
Illustration of the horizontal component of the forces integrated into the domain marked as a dashed rectangle in Fig.~\ref{fig:jetflow-vlong-case2-eruptive}(a) for case~2.
Black: time derivative of the integrated x-component of the
momentum. Red: momentum gain per unit time through the gas
pressure on the sides of the box. Green: x-component of the momentum gain per unit time
through the Reynolds tensor acting on the sides of the box. Blue: volume
integral of the x-component of the Lorentz force. \label{fig:forces} 
} 
\end{figure}

To understand which forces are causing the motion of the prominence during the phases explained in the previous figures, we consider the momentum conservation equation in a fixed volume of the domain around it. Concentrating on case~2, we have selected the box shown as a dashed rectangle in Figure~\ref{fig:jetflow-vlong-case2-eruptive}(a), located at $(-10,10)\Mm \times(2,14) \Mm$. Focusing on the $x$-component, we consider the time derivative of the integral of $\rho\, v_x$ in that domain (black curve in Figure~\ref{fig:forces}). 
That time derivative is exactly given by the difference of the integrated gas pressure force on the vertical boundaries (red curve), the integral of the x-component of the action of Reynolds' stresses on all sides (green curve) and, finally, the integral of the Lorentz force in the volume (blue), which could easily be expressed in terms of the integrated magnetic pressure and tension forces on the boundaries of the box. Significant changes start with the arrival of the jet at $t\approx 15\mins$. The momentum change (black curve) basically reflects the evolution of the prominence, given its large mass compared with the surroundings. The jet first hurls the prominence through the gas pressure and Reynolds stresses (with the latter, in this case, basically given by the ram pressure acting on the vertical side boundaries, $\rho v_x^2$) in similar amounts, which fits the fact that the jet is preceded by a shock which is not highly supersonic. From $t\approx 21\mins$ to $t\approx 30\mins$ the prominence has been shifted so far to the right, up along the field lines of the dip, that the Lorentz stresses act as a recovering force: in that period the blue curve is the dominant one and leads to the negative stretch of the total momentum change apparent in the black curve. From around $t\approx 30 \mins$ to the right end of the frame one can see the effects of the arrival of the eruptive jet: the gas pressure and the Reynolds stresses dominate again and lead to a positive change of the x-component of the momentum.

\section{Evolution following the impact of the jet:
 counter-streaming flows and oscillations}\label{sec:after-jet}

\subsection{Prominence motion: general
 description}\label{sec:counterstreaming_and_flows} \label{subsec:generaldescription-counterstreaming}

\begin{figure*}[!ht]
\hspace{-1.cm}\includegraphics[width=1.15\textwidth]{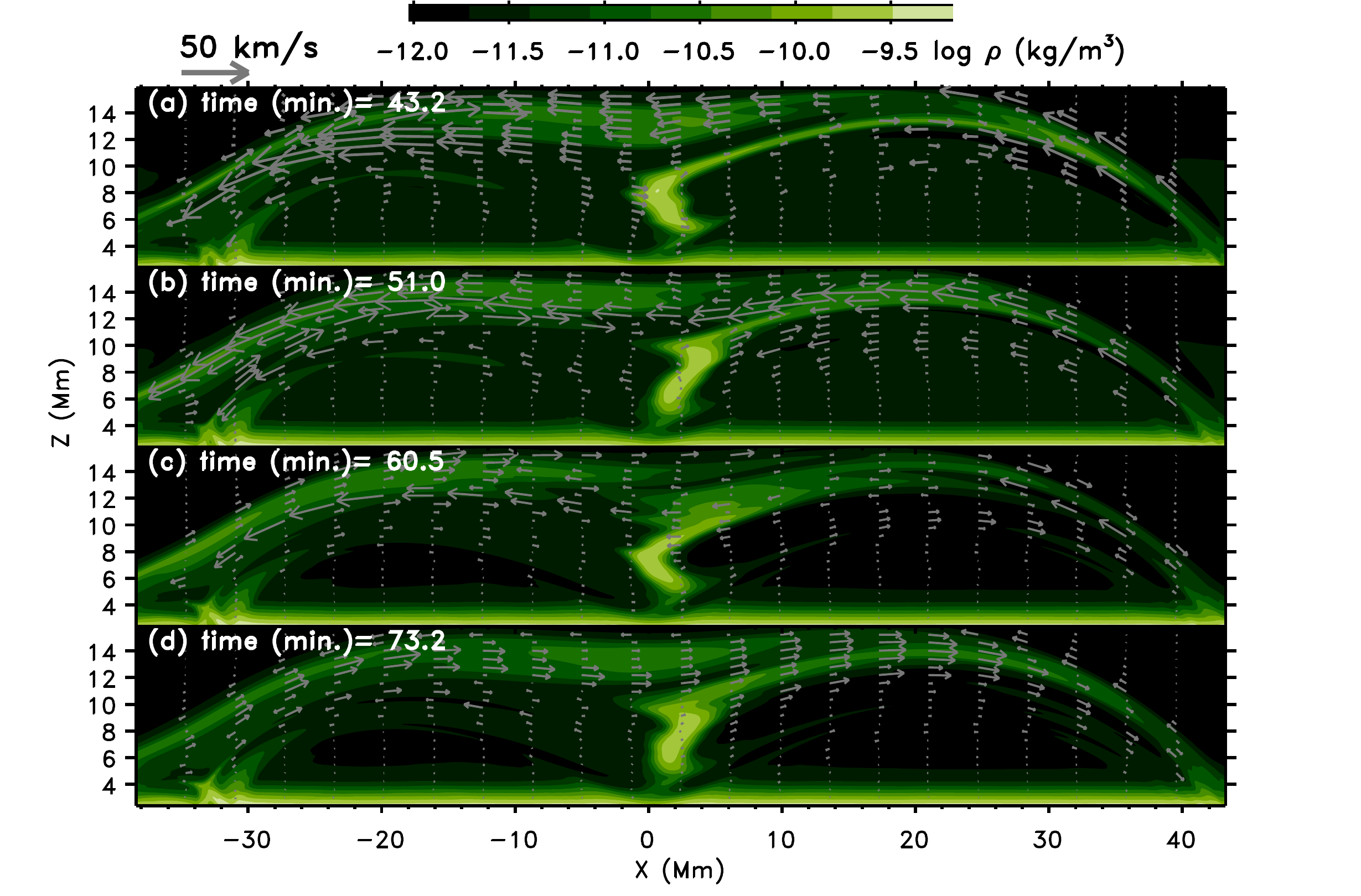}
\caption{Spatial distribution of the density for case~1 at four times
 after the impact of the jets. The arrows show the projected velocity field. The arrow plotted in the top left corner corresponds to a velocity of 50 $\kms$.
\label{fig:density-evolution-after-jet-case1}}
\end{figure*}
\begin{figure*}[!ht]
\hspace{-1.cm}\includegraphics[width=1.15\textwidth]{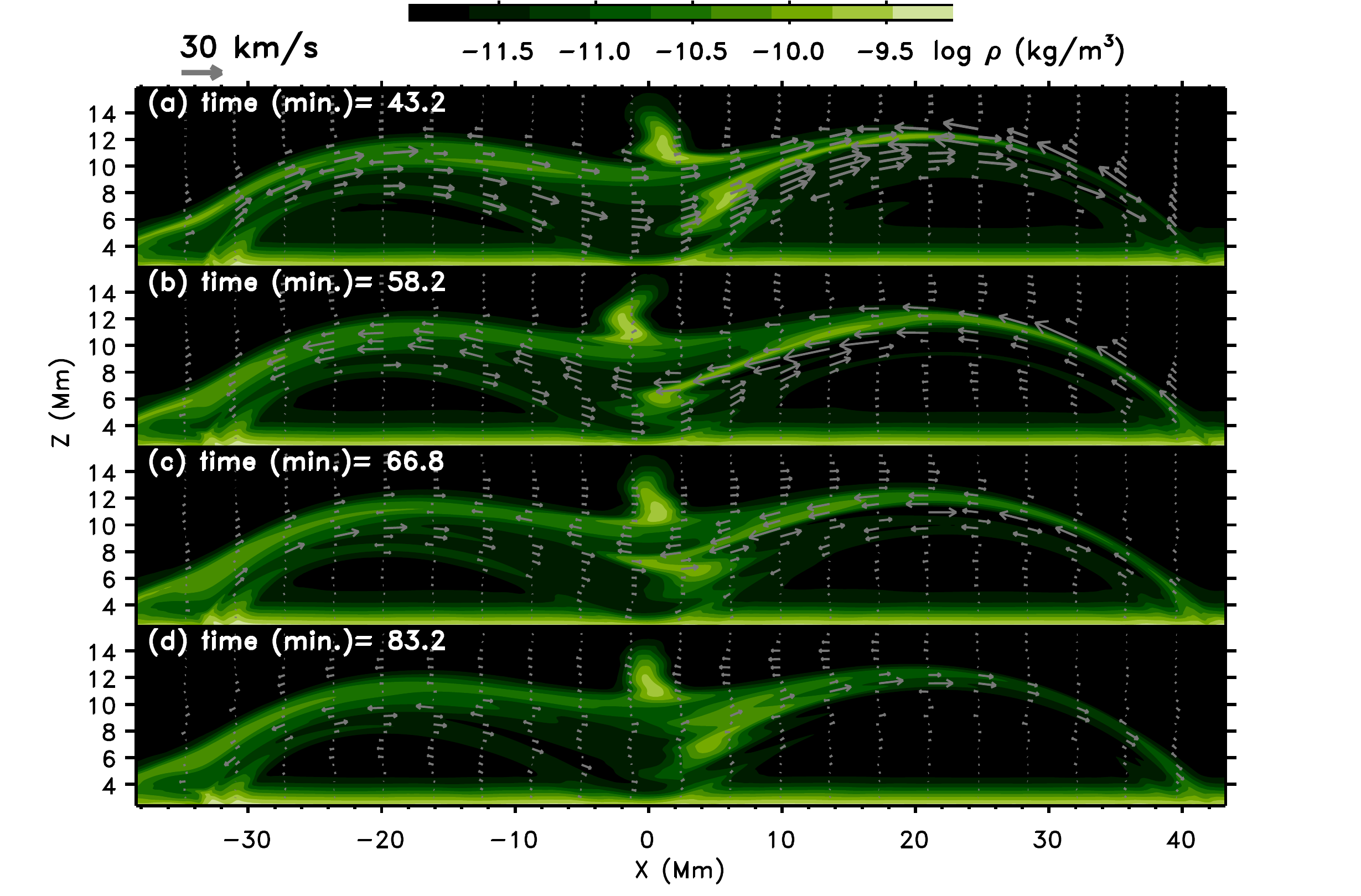}
\caption{Spatial distribution of the density like in
  Figure~\ref{fig:density-evolution-after-jet-case1} but here for case~2.
 \label{fig:density-evolution-after-jet-case2}}
\end{figure*}

In this section we study the dynamics of the plasma in the prominence and filament channel following the impact of the jets; this includes counter-streaming flows and oscillations in different directions. Figures~\ref{fig:density-evolution-after-jet-case1} and \ref{fig:density-evolution-after-jet-case2} show the density of the system for case~1 and 2, respectively, in an advanced phase of the evolution, later than the time range shown in Figures~\ref{fig:jetflow-vlong-case1-eruptive}~and~\ref{fig:jetflow-vlong-case2-eruptive}. In both cases the prominence is greatly displaced from its initial position by the jet flows (see Fig.~\ref{fig:initialconfiguration} for comparison). The maximum displacement is near the top ($\zdip\approx 12\Mm$) for case~1 and near the central height of the prominence ($\zdip\approx 10\Mm$) for case~2. In case~1, Figures \ref{fig:density-evolution-after-jet-case1}(a) and \ref{fig:density-evolution-after-jet-case1}(b) show that the top part of the prominence has been pushed to the right out of the dipped region and beyond the right apex of the field line [located at $(x,z)=(20,14)$] with partial drainage of the mass to the right footpoints. This huge displacement is due to the eruptive phase of the jet. As already seen in Figure~\ref{fig:jetflow-vlong-case1-eruptive}, the eruptive jet has strong flows above $\zdip\approx9\Mm$. Additionally, looking in detail at Figure~\ref{fig:density-evolution-after-jet-case1}, one can clearly see that above the displaced prominence there is a band of enhanced density in lines between $\zdip=10$ to $\zdip=15\Mm$ that extends to the left footpoint and which does not seem to originate in the prominence itself: the density in it is approximately one order of magnitude less than in the prominence. The mass in the band is mainly the result of the eruptive jet: as part of the process, chromospheric-density plasma passes through the reconnection site, enters the filament channel, and is hurled to high levels causing the density enhancement; there is an excess mass on those field lines for the remaining duration of the simulation, falling on either side of the field line apex through the action of gravity. 
In case~2 (Fig. \ref{fig:density-evolution-after-jet-case2}), the situation is similar but, as seen in Chapter~\ref{sec:the-jet}, the jets do not reach above the central height of the prominence. The maximum displacement and the partial drainage are, correspondingly, at the center of the structure, between $\zdip\approx8$ and $\zdip\approx 10\Mm$. Similarly to case~1, in lines above the displaced prominence one sees a band of enhanced density extending to the left footpoint. Below $\zdip\approx10 \Mm$ for case~1 and $\zdip\approx8\Mm$ for case~2, the prominence shows a zig-zag structure associated with the plasma oscillations along the magnetic field which will be analyzed in detail in Section~\ref{subsec:oscillations}.

\begin{figure*}[!ht]
	\hspace{-0.5cm}\includegraphics[width=1.05\textwidth]{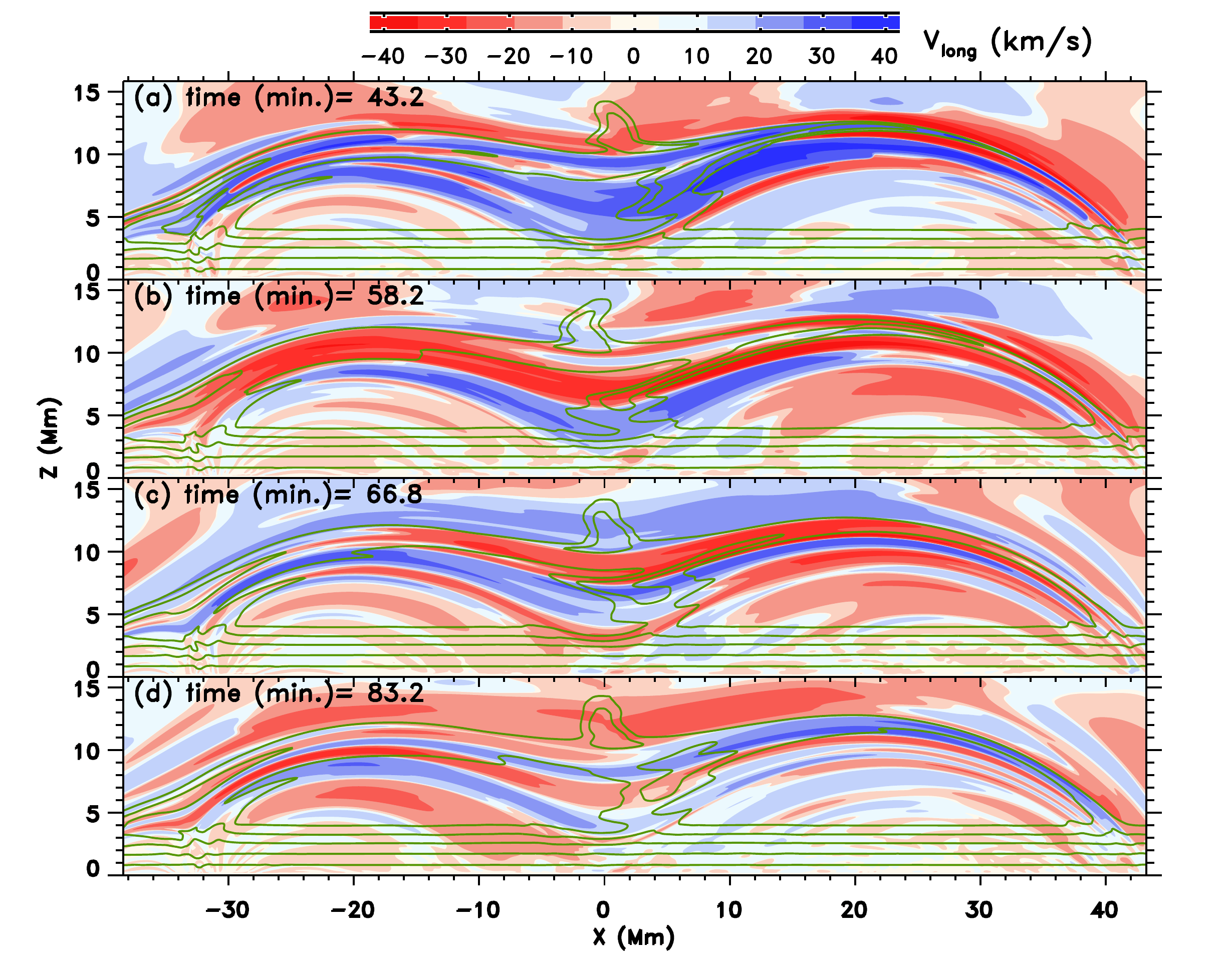}
 \caption{Similar to Fig.~\ref{fig:jetflow-vlong-case1-quiescent}: plot of the time evolution of $\vlong$ for case~2
   and the same times as in
   Fig.~\ref{fig:density-evolution-after-jet-case2}.
\label{fig:vlong-evolution-after-jet-case2}}
\end{figure*}

In the rest of this subsection we focus on case~2 (the discussion here is anyway qualitatively similar for both cases). Figure \ref{fig:vlong-evolution-after-jet-case2} shows the longitudinal velocities inside the filament channel for case~2 for the same times as in Figure~\ref{fig:density-evolution-after-jet-case2}.
In the advanced phase shown in Figure~\ref{fig:vlong-evolution-after-jet-case2} we see that the repeated front reflections that follow the impact of the jets and the longitudinal oscillation of the prominence plasma lead to a complex counter-streaming flow pattern. There is a strong velocity shear between adjacent field lines with alternating positive and negative values of the velocity and the pattern becomes increasingly complicated as time advances. The velocities decrease with time because of numerical dissipation but a physical damping mechanism, like wave leakage \citep{zhang_damping_2019} or compression of the chromosphere later dissipated by viscosity, cannot be discarded. Counter-streaming flows have been observed in solar filaments with typical velocities of approximately $20 \kms$ in opposite directions between neighboring threads\citep[e.g.][]{zirker_counter-streaming_1998,wang_extending_2018}. The counter-streaming flows are also observed in extreme ultraviolet (EUV) spectral lines associated with relatively hot plasma with velocities of up to $100\kms$ \citep[see, e.g.,][]{alexander_anti-parallel_2013}. From the results of our experiments, we conclude that the impact of the jets on the prominence and consequent reverberation of the fronts can significantly contribute to the creation of a pattern of counter-streaming flows. 

\begin{figure*}[!ht]
\hspace{-0.6cm}\includegraphics[width=1.05\textwidth]{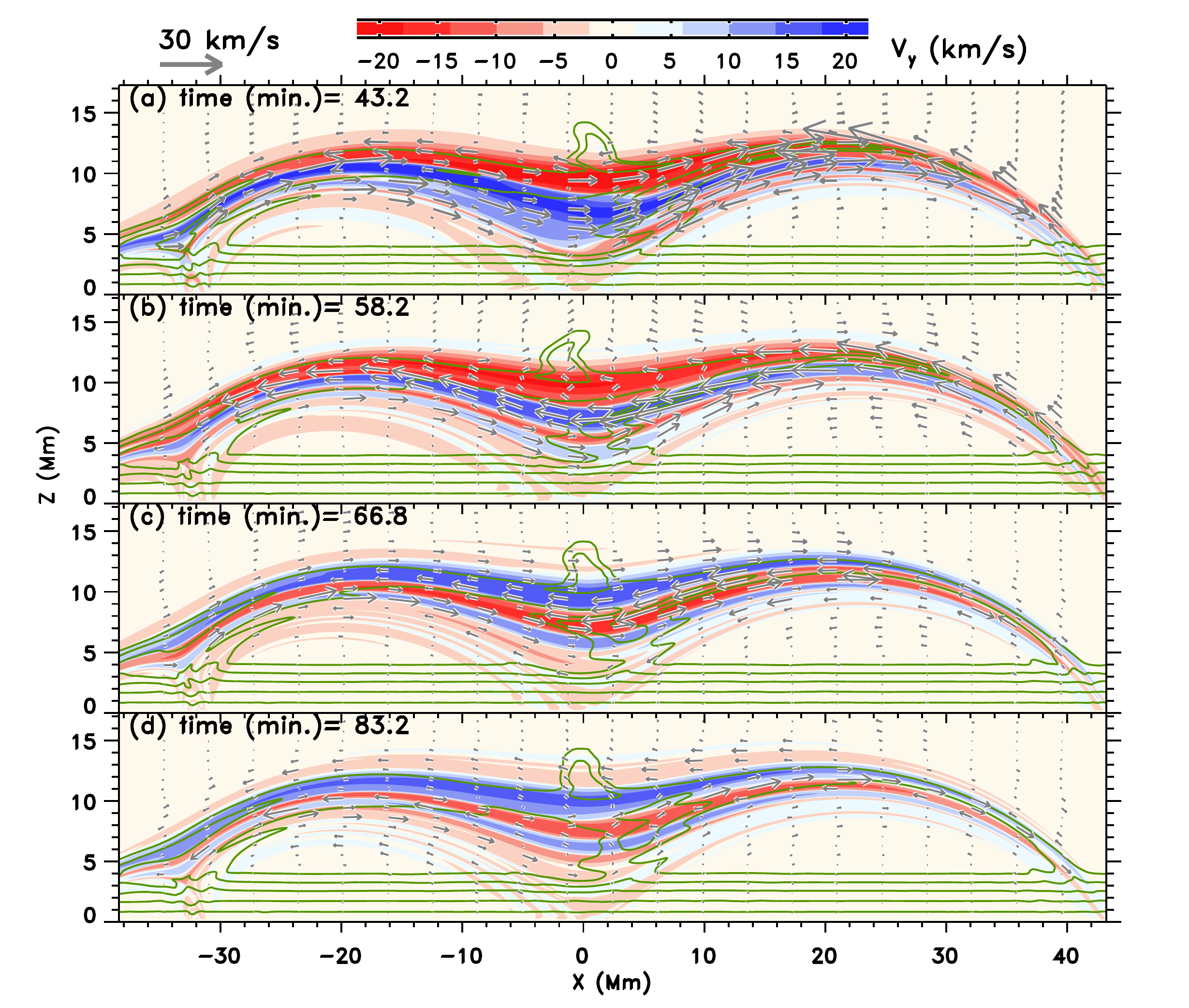}
 \caption{Plot of the time evolution of $\vtransy$ for case~2
   and the same times as in
   Figure~\ref{fig:density-evolution-after-jet-case2}. The arrows show the projected velocity field. The arrow plotted in the top left corner corresponds to a velocity of 30 $\kms$.
\label{fig:vy-evolution-after-jet-case2}}
\end{figure*}

The corresponding map for the transverse velocity (Figure~\ref{fig:vy-evolution-after-jet-case2}) is comparatively simple: as seen in Section~\ref{sec:alfvenic front} (e.g., Figure~\ref{fig:alfvenic-propagation}), the Alfv\'enic perturbation travels with Alfv\'en speed, hence much faster than the jets, crossing the full FC in about $1.5$ minutes. After successive reflections at the prominence and at the chromospheric endpoints a pattern starts to appear for $\vtransy$ in each field line: the motions along $y$ lead to a standing oscillation with a period of 15 minutes. This is apparent in Figure \ref{fig:vy-evolution-after-jet-case2}, especially after all transients have been damped: in panels \ref{fig:vy-evolution-after-jet-case2}(b)-(d) $\vtransy$ has a constant phase along each field line, reaches its maximum value at the center of the line, and is zero at the endpoints. These modes are the Alfv\'enic string or hybrid modes \citep{roberts_waves_1991,joarder_modes_1992,oliver_oscillations_1993} that are the fundamental modes for every field line. These oscillations belong to the so-called Alfv\'en continuum in which each field line oscillates with a period related to its local properties \citep{goossens_existence_1985}. In this sense, they are not global modes of the magnetic structure with a single period and phase for all of the FC. 

\subsection{Oscillation analysis}\label{subsec:oscillations}

\subsubsection{Amplitude and phase}\label{subsec:amplitude-phase}

The plasma and magnetic field oscillations can be analyzed in terms of their polarization properties. Following \citet{luna2016}, we analyze the bulk motion of the prominence considering the three most important polarization directions, namely, the motion parallel to the magnetic field (longitudinal direction), (b) the motion perpendicular to the magnetic field projected onto the $y$-direction; and (c) the motion perpendicular to the field projected onto the $z$-direction (vertical polarization). Since the prominence moves independently along each field line, we study the oscillation by computing the velocities of the center of mass, $\vcm$, of a set of selected field lines (here labeled with the subindex $i$) through
\begin{eqnarray}\label{eq:long-pol}
\vcmpar_i
(t) &=& \frac{\int \rho(s_{i}, t) \, \vlong(s_{i}, t) \, d s_{i}}{\int \rho(s_{i}, t) \, ds_{i}} \, ,\\ \label{eq:y-pol}
\vcmperpy_i
(t) &=& \frac{\int \rho(s_{i}, t) \, \vtransy(s_{i}, t) \, d s_{i}}{\int \rho(s_{i}, t) \, ds_{i}} \, ,\\ \label{eq:z-pol}
\vcmperpz_i
(t) &=& \frac{\int \rho(s_{i}, t) \, \vtransz(s_{i}, t) \, d s_{i}}{\int \rho(s_{i}, t) \, ds_{i}} \, ,
\end{eqnarray}
where $s_{i}$ is the arc-length variable along the $i$-line. We have selected a set of field lines from $\zdip = 2 $ to $\zdip = 16\Mm $, i.e., from the bottom part of the prominence up to a position above it. In order to avoid the contribution of the dense plasma of the chromosphere, Equations~(\ref{eq:long-pol})-(\ref{eq:z-pol}) are integrated above $z=3\Mm$. 

\begin{figure*}[!ht]
\centering\includegraphics[width=\mysize\textwidth]{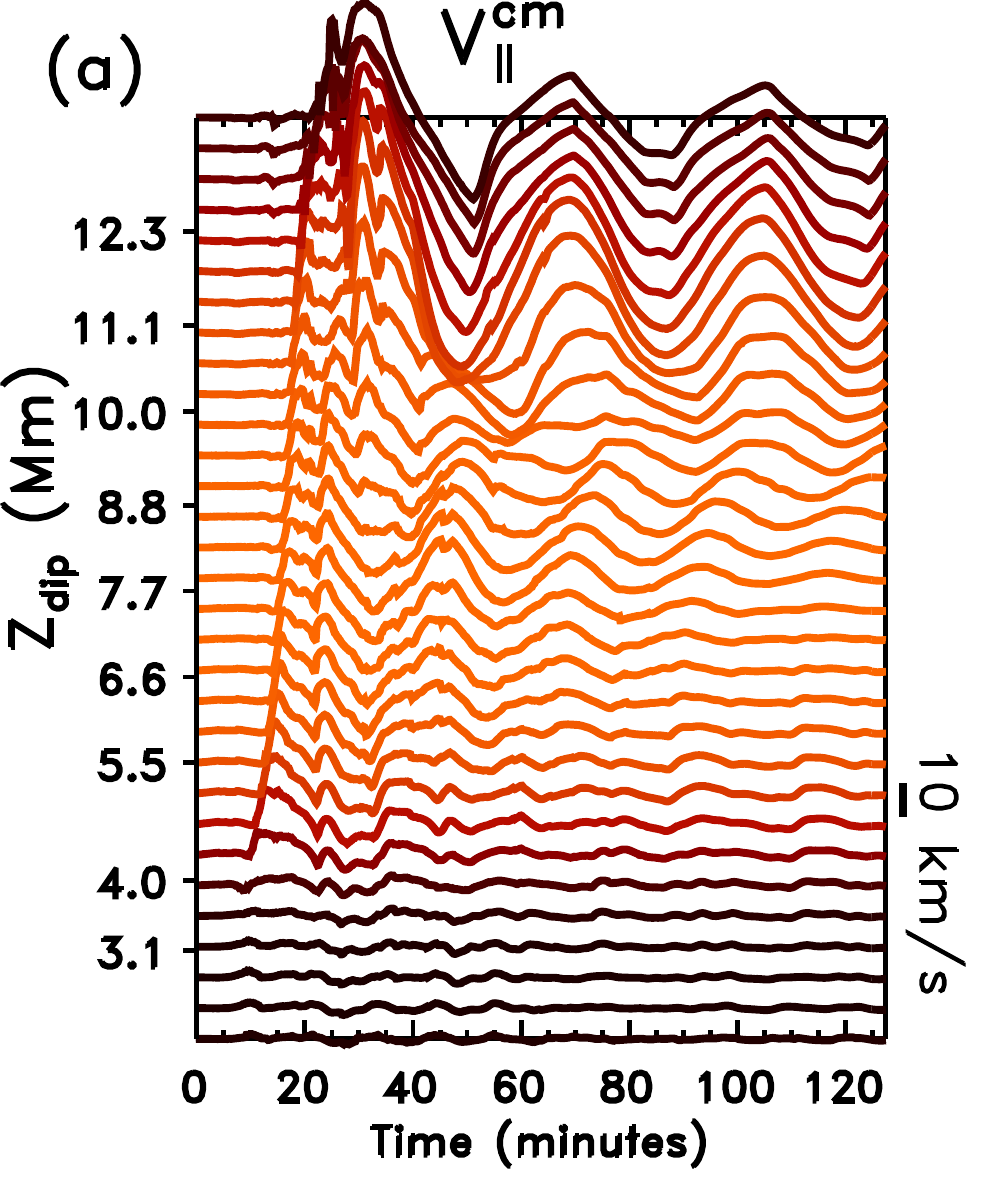}\includegraphics[width=\mysize\textwidth]{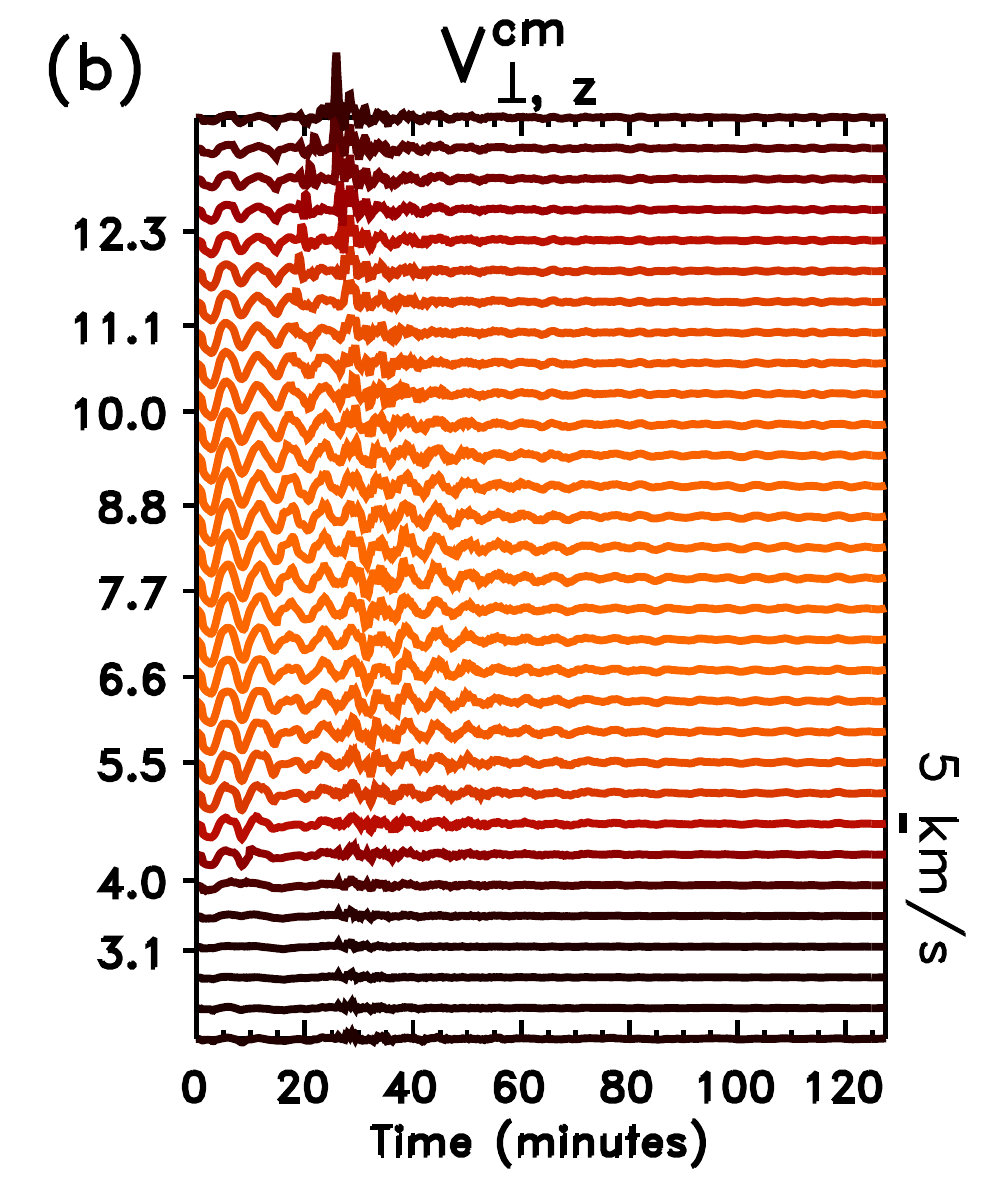}\includegraphics[width=\mysize\textwidth]{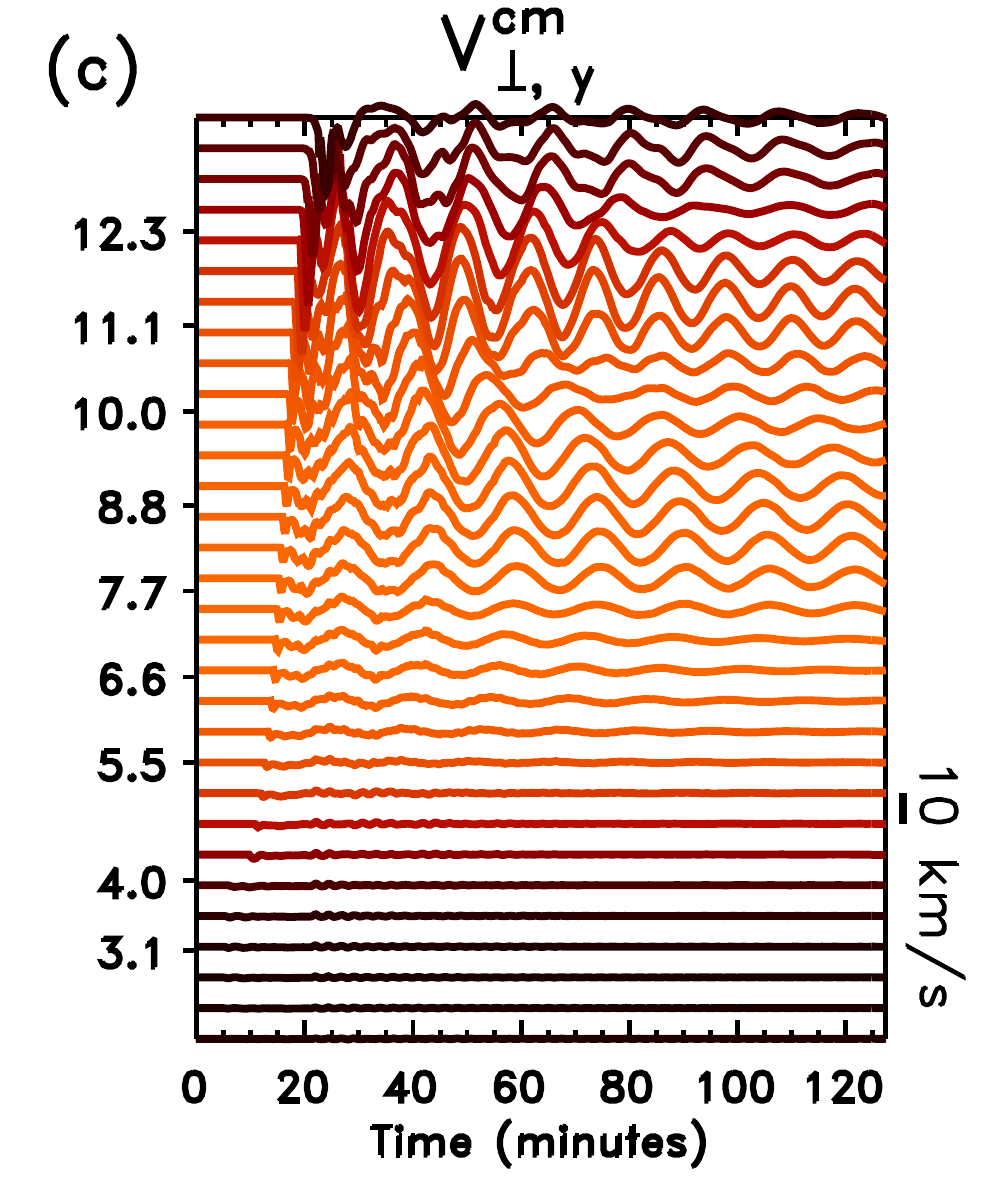}
\centering\includegraphics[width=\mysize\textwidth]{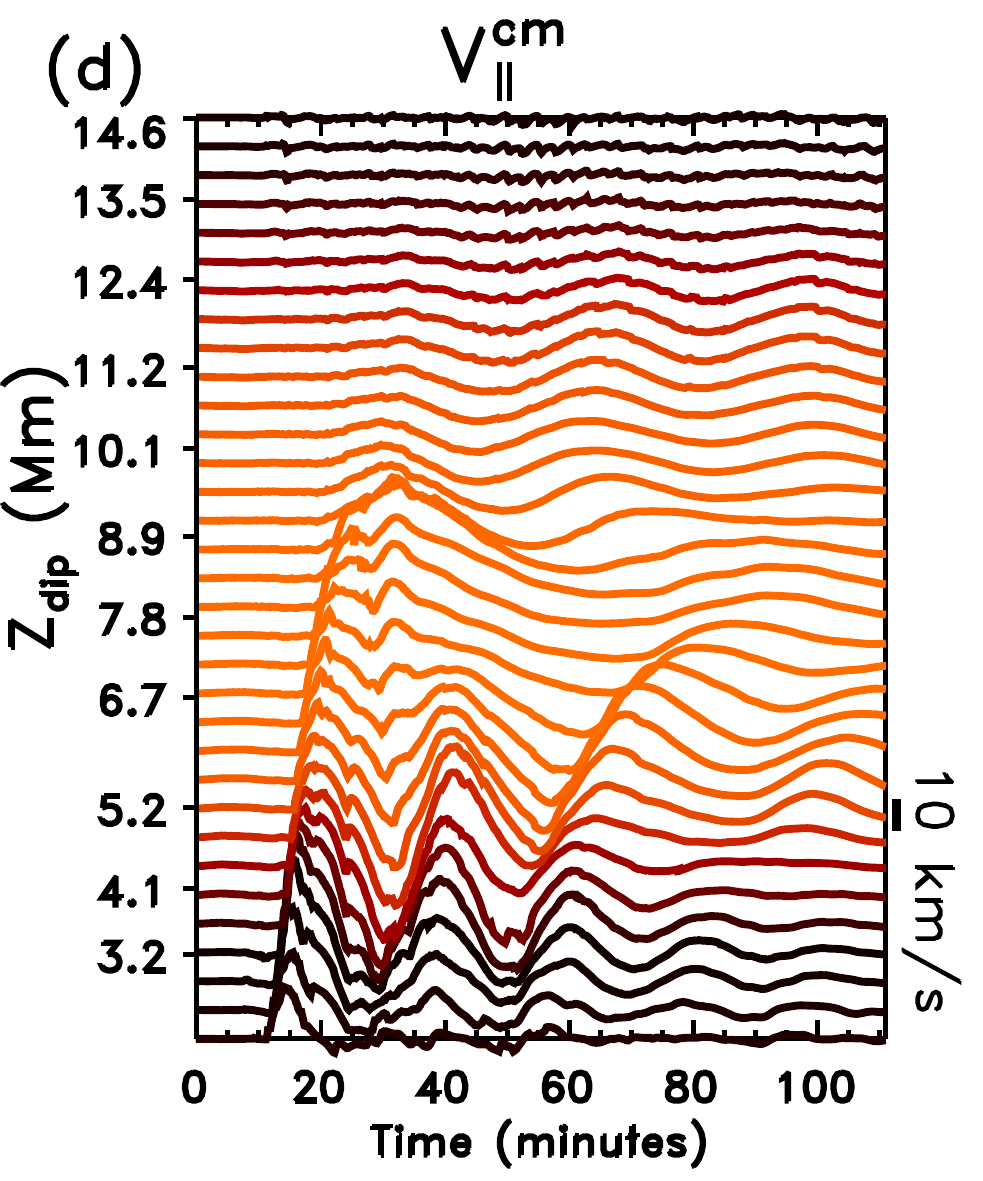}\includegraphics[width=\mysize\textwidth]{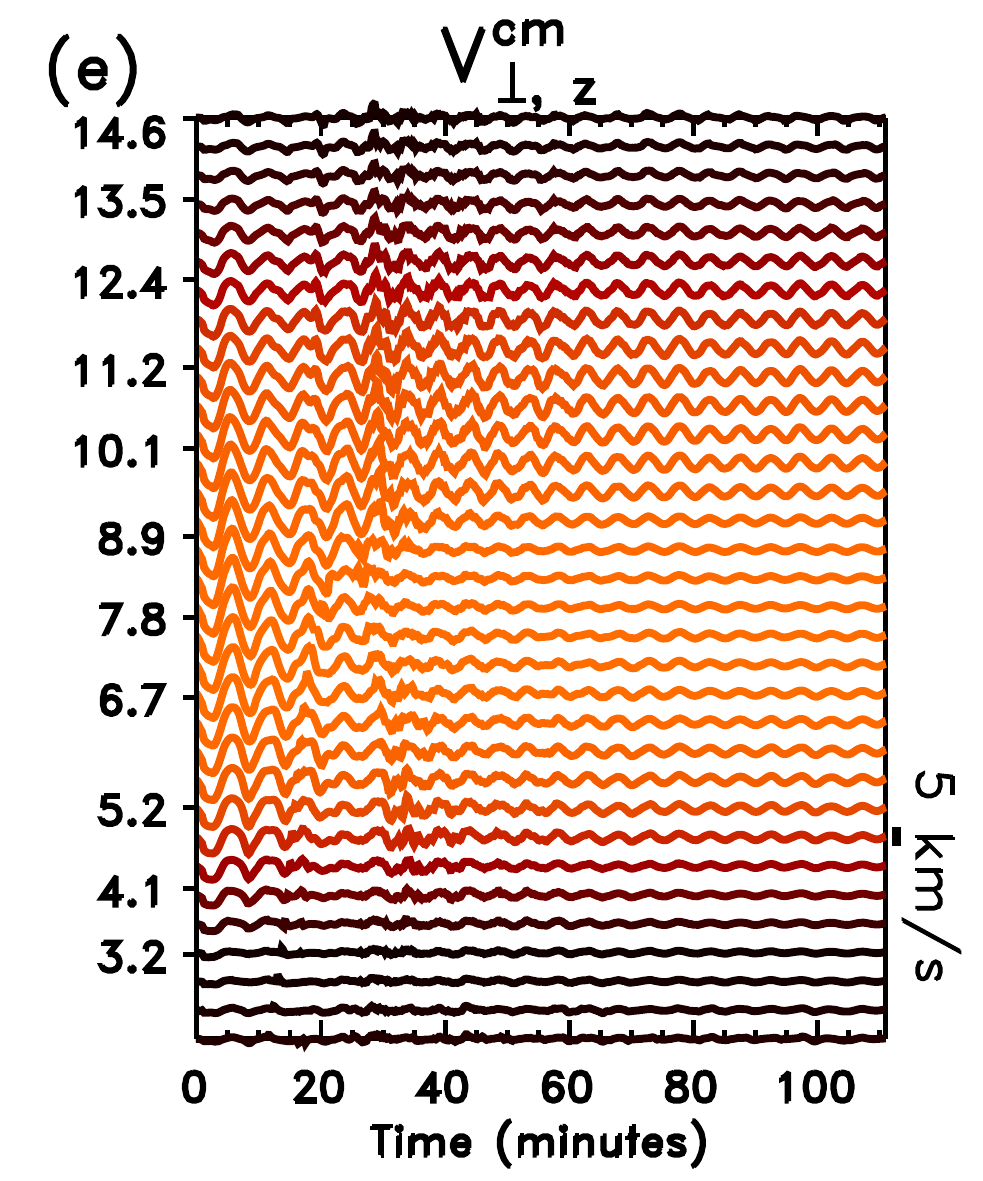}\includegraphics[width=\mysize\textwidth]{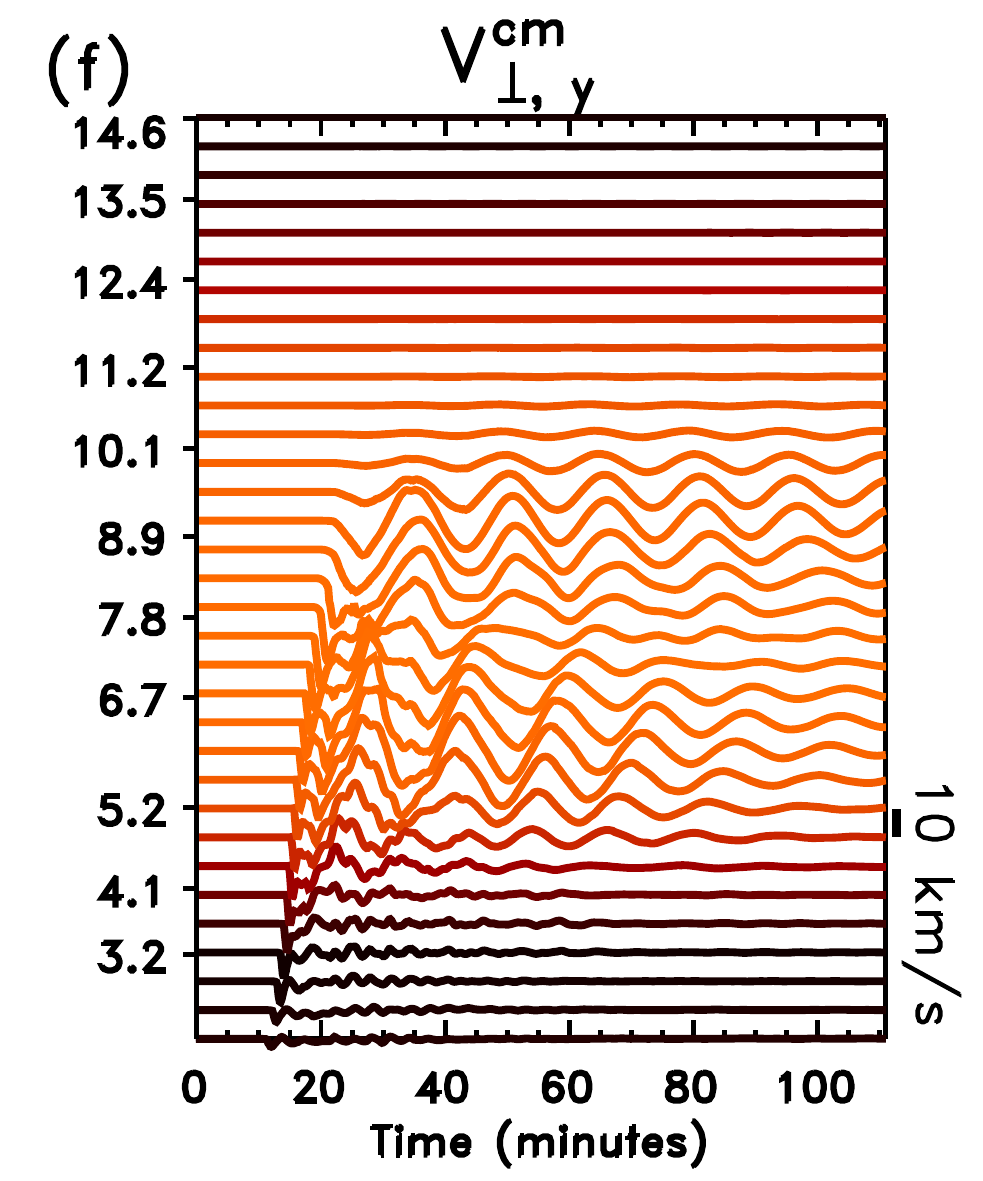}
\caption{Plot of $\vcmpar$ (leftmost panels), $\vcmperpz$
  (central panels) and $\vcmperpy$ (rightmost panels) for case~1 (upper
  row) and case~2 (lower row). The curves show the time evolution of those quantities for different field lines: the lines are labeled with the value of $\zdip$ indicated in the ordinate axis of each panel. The displacement of the curve at any given time with respect to its value at $t=0$ corresponds to the velocity at that time: the velocity unit for the displacements is indicated by a short segment on the right of each panel. The colors correspond to the density contrast $\rho_{max}/\rho_{corona}$ along each field line, from light orange (maximum contrast) to black (no contrast),
  as explained in the text. \label{fig:oscillations}}
\end{figure*}

\begin{figure}
	\centering\includegraphics[width=0.47\textwidth]{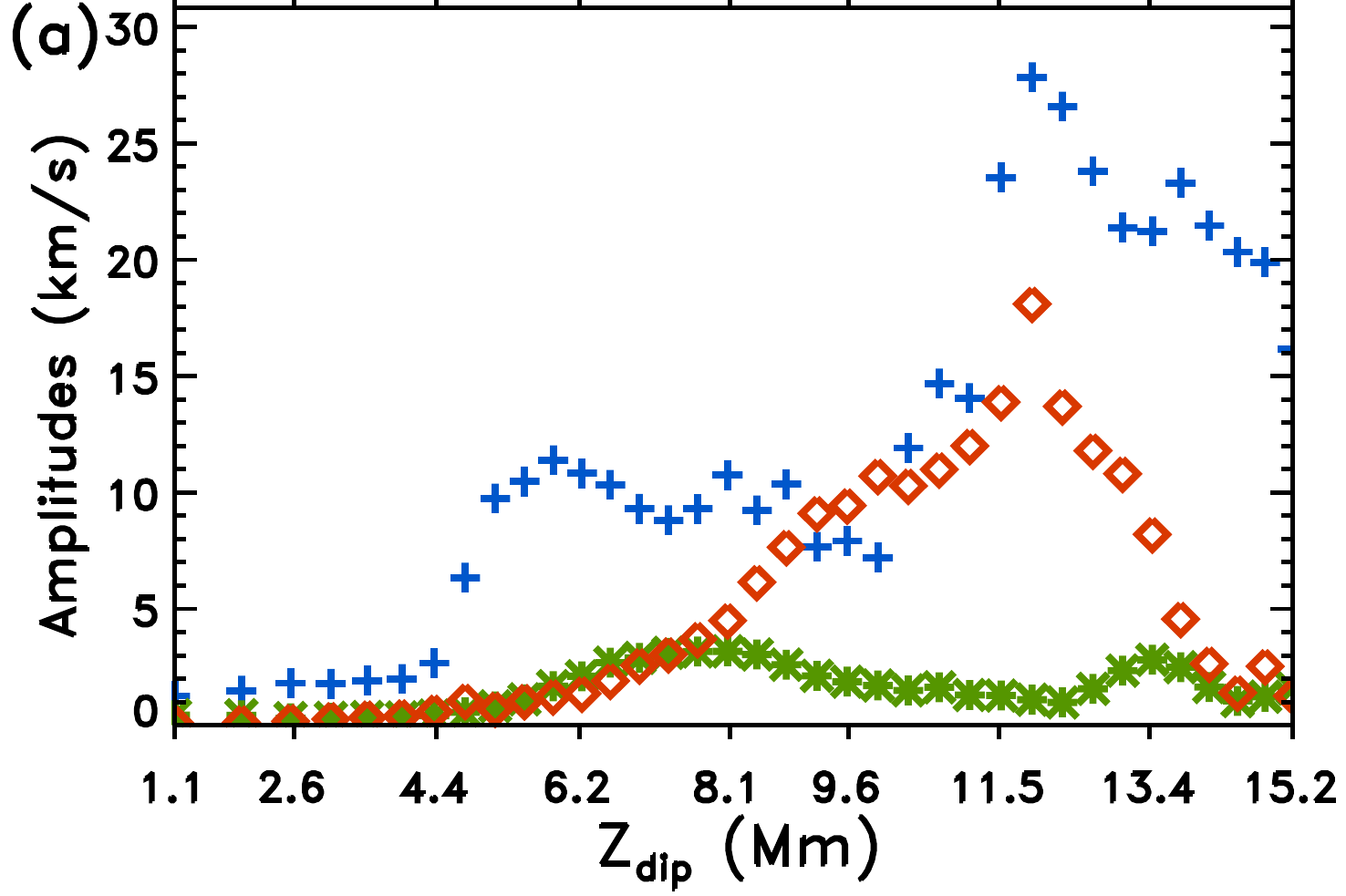}\\
	\centering\includegraphics[width=0.47\textwidth]{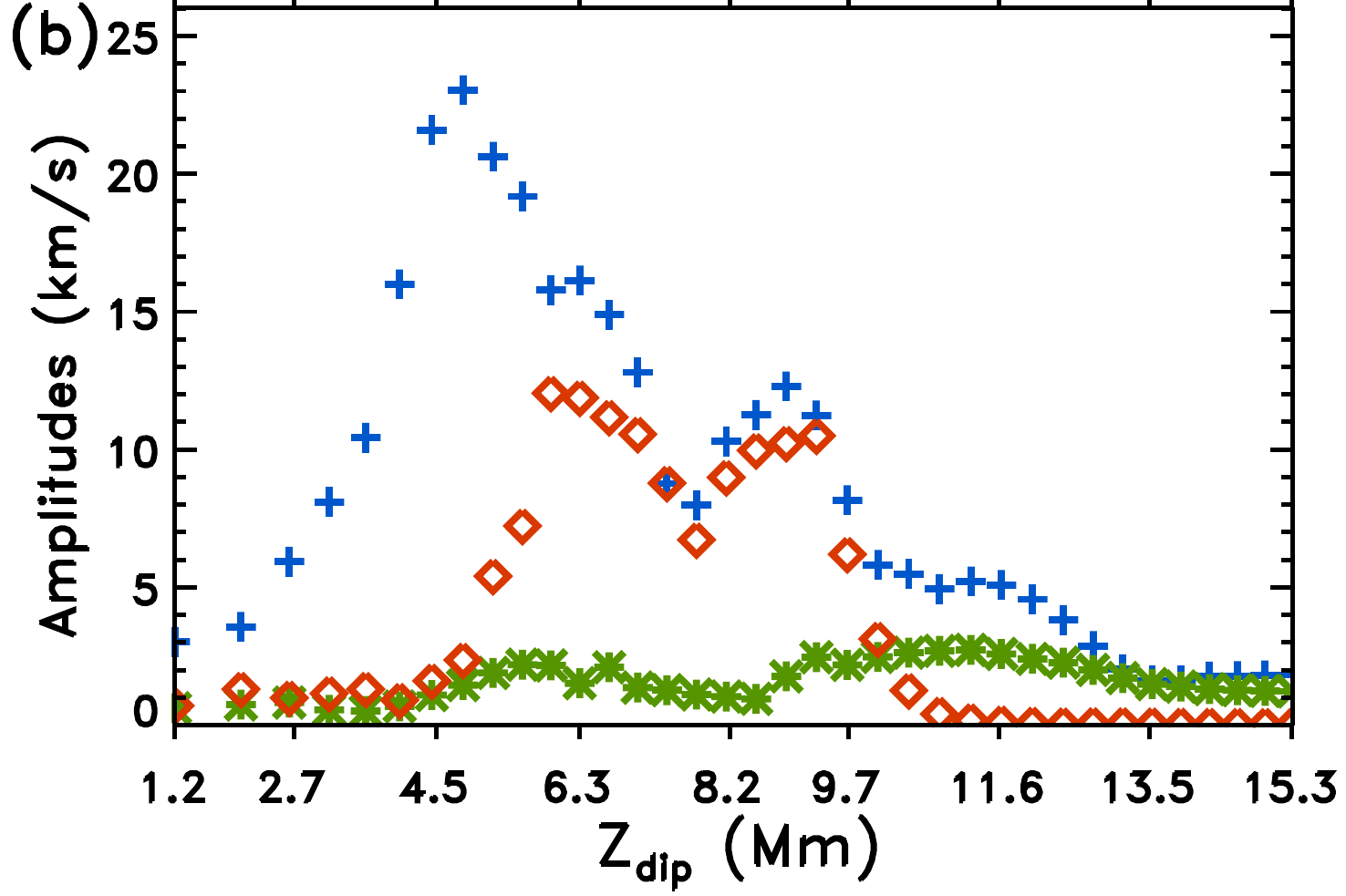}
	\caption{Maximum absolute value of the amplitude of the oscillations shown in
Figure~\ref{fig:oscillations} for (a) case~1 and (b) case~2. Each marker shows the maximum of the absolute value of the velocity of the curves in the previous figure during the oscillation as a proxy for the oscillation amplitude. Blue crosses: $v_{\parallel, i} (t)$; orange diamonds:
$v_{\perp y, i} (t)$; green asterisks: $v_{\perp z, i} (t)$.
\label{fig:amplitudes}}
\end{figure}

Figure \ref{fig:oscillations} shows the center-of-mass velocities from Equations (\ref{eq:long-pol})-(\ref{eq:z-pol}) as a function of time for the selected field lines for cases~1 (upper row of panels) and~2 (lower row). The orange color code measures the density contrast $\rho_{max}/\rho_{corona}$ for each field line, with $\rho_{max}$ the maximum of the density along that field line and $\rho_{corona}$ the initial stratified coronal density outside of the prominence at the height of the field line (see Sec.~\ref{sec:model}). The light-orange hue corresponds to a density contrast of $200$; black indicates a density contrast of $1$. The curves for each field line start at a position along the ordinate axis that corresponds to their $\zdip$ as indicated on the right of each frame. Within each curve, the vertical elongation corresponds to the value of the velocity component, with the unit given as a short segment on the left of each frame.
The most salient feature in all panels is the fact that the velocity components oscillate with various amplitudes and phases. To complement the information provided by this figure, the measured amplitudes are given in Figure~\ref{fig:amplitudes}, again for case~1 at the top and case~2 at the bottom. These amplitudes are the maximum absolute value of the velocity during the oscillation. We shall discuss those two figures jointly. 

Figure \ref{fig:oscillations}(a) shows the longitudinal velocity (Eq. (\ref{eq:long-pol})) for case~1. The velocity is zero until the quiescent jet reaches the cool structure at $t\approx 10\mins$. It then reaches successively higher levels along the following 15 minutes, approximately, until hitting the top of the prominence. The impact of the jet sets the prominence mass in motion, leading to the oscillation apparent in the figure. The quiescent jet is followed by the eruptive one, which, in contrast to the former, first hits the top of the prominence and then successive lower levels, as seen in Section~\ref{sec:the-jet}. The two phases can be clearly distinguished in that panel: between the bottom of the prominence around $\zdip=4.5\Mm$ up to approximately $\zdip=10\Mm$, the oscillations are largely triggered by the quiescent jet. In Figure~\ref{fig:amplitudes}(a), blue crosses, one can see that, for those field lines, the velocity amplitude is between $7$ to $12$ $\kms$. In Figure 20(a), one sees that the longitudinal velocities have phase shifts that depend on the chosen field line; this is associated with the different starting times and periods of the oscillation on each field line. These phase shifts lead to the zig-zag appearance of the prominence visible in Fig. 16.
This effect is typical of the longitudinal oscillations in prominences \citep[e.g.][]{luna2016,Liakh2020aa,adrover-gonzalez_3d_2020}.
In field lines with $\zdip\ge10\Mm$, the oscillations are triggered by the eruptive jet; in Figure~\ref{fig:oscillations}(a), and only for field lines in that range, a secondary peak is apparent at around $t=30\mins$ associated with the arrival of the eruptive jet, followed by the oscillations. The latter is characterized by a very regular phase and longer periods than in the low levels. From Figure \ref{fig:amplitudes}(a), blue crosses, we see that the velocity amplitude in those heights ranges from $13$ to almost $30~\kms$. The largest velocity is measured at approximately $\zdip=12\Mm$. As we have seen in Figure \ref{fig:density-evolution-after-jet-case1} the eruptive jet injects plasma producing an increase of the density on field lines with $\zdip\ge 10 \Mm$. Due to this injection, the physical conditions of the plasma change, thus the nature of the oscillations is different in the regions above and below $\zdip=10\Mm$, as discussed in Section~\ref{subsec:psd}. In the field line with $\zdip=10\Mm$ the prominence has no clear oscillation for the first 80 minutes after the triggering. The reason is that this line is in the transition between the two regimes of the oscillations with a strong velocity shear. The oscillations just above and just below have opposite directions during the first cycles of the oscillation having a small velocity in this field line.

Figure \ref{fig:oscillations}(d) shows the longitudinal center-of-mass velocity for case~2. At $t=10\mins$ the quiescent jet hits the bottom of the prominence, subsequently reaching higher levels for about $15$ minutes up to the central part; the eruptive jet, in turn, starts hitting at the central part thereafter impacting successively lower prominence levels. In this experiment, we also distinguish the two regimes of the oscillation. Below $\zdip=8\Mm$ the oscillations are associated with the quiescent jet; they have phase shifts that depend on the chosen field line. In the region between $\zdip=8$ to $\zdip=10\Mm$ the oscillations are triggered mainly by the eruptive jet. We can also distinguish a secondary peak at $t\approx30\mins$ when the eruptive jet hits the prominence. Like in case~1, the reconnection in the vertical current sheet injects mass into the FC increasing the density on the reconnected field lines (see Fig.~\ref{fig:density-evolution-after-jet-case2}). The oscillation amplitudes for case~2 can be seen in Figure \ref{fig:amplitudes}(b). In the lower region ($\zdip<8\Mm$) the velocities range from $3$ to $23\kms$. Above $\zdip=8\Mm$, interestingly, the amplitudes are smaller, from a few $\kms$ to $12\kms$: the eruptive jet causes a smaller perturbation than the quiescent one in this case, showing the relevant role of the initial position of the null point.

Figures \ref{fig:oscillations}(b) and \ref{fig:oscillations}(e) show the oscillations in the $\vcmperpz$ component, as calculated with Equation~(\ref{eq:z-pol}). The prominence oscillates in the vertical direction from the beginning of the experiment due to the mass loading process. However, when the jet reaches the prominence there is a change of phase and amplitude of the oscillations; the change is particularly clear for some field lines, like for the field lines around $\zdip = 6.6\Mm$ in Figure\ref{fig:oscillations}(b). After the eruptive jet hits the prominence ($t>30\mins$), the oscillations are very regular with almost identical phases and periods in all heights in both cases 1 and 2. The corresponding oscillation amplitudes are shown with green asterisks in Figure \ref{fig:amplitudes}. We see that the amplitude is much smaller than for the other two polarizations we are studying, which follows from the preferred directions of the major perturbations (Alfv\'en, acoustic fronts followed by the jets) caused by the reconnection: the maximum amplitude is some $4\kms$ (at $\zdip=8\Mm$) for case~1 and similar values for case~2.

Figures~\ref{fig:oscillations}(c) and (f), show the oscillations of the $y$-component of the center-of-mass velocity in the transverse direction, $\vcmperpy$, as computed with Equation (\ref{eq:y-pol}). These oscillations are Alfv\'enic in nature and are ultimately caused by the injection of $B_{y}$-component {into the filament channel} during the reconnection process in either one of the reconnection phases, as discussed in Section \ref{sec:the-jet}. The oscillations in this direction share some of the main features of the longitudinal case (Figs. \ref{fig:oscillations}(a) and (d)). On one hand, below a dividing value of $\zdip$ ($\zdip \approx 10.5\Mm$ for case~1 and $\approx 7.1\Mm$ for case~2) the oscillations are mainly caused by the quiescent jet, whereas they are caused by the eruptive jet above it. Correspondingly, the amplitude of the transverse oscillation (orange diamonds in Figure~\ref{fig:amplitudes}) follows a pattern reminiscent of the longitudinal case (blue crosses): the amplitude increases with height until about $\zdip =12\Mm$ in case~1, reaching values of some $18\kms$ there; in case~2 the red crosses have a rough `M' shape located between $\zdip = 4.5\Mm$ and $11\Mm$, with the two maximum amplitudes being $12 \kms$ at $\zdip=6\Mm$ and $11 \kms$ at $\zdip=9.5\Mm$. We remark that, in contrast to the $z$-component discussed above, the maximum amplitudes of the longitudinal and transverse oscillations in the $y$ direction are only about a factor of $2$ apart: the prominence oscillation is not at all confined to the longitudinal direction.

\subsubsection{PSD analysis}\label{subsec:psd}

\begin{figure}[!ht]
	\centering\includegraphics[width=0.47\textwidth]{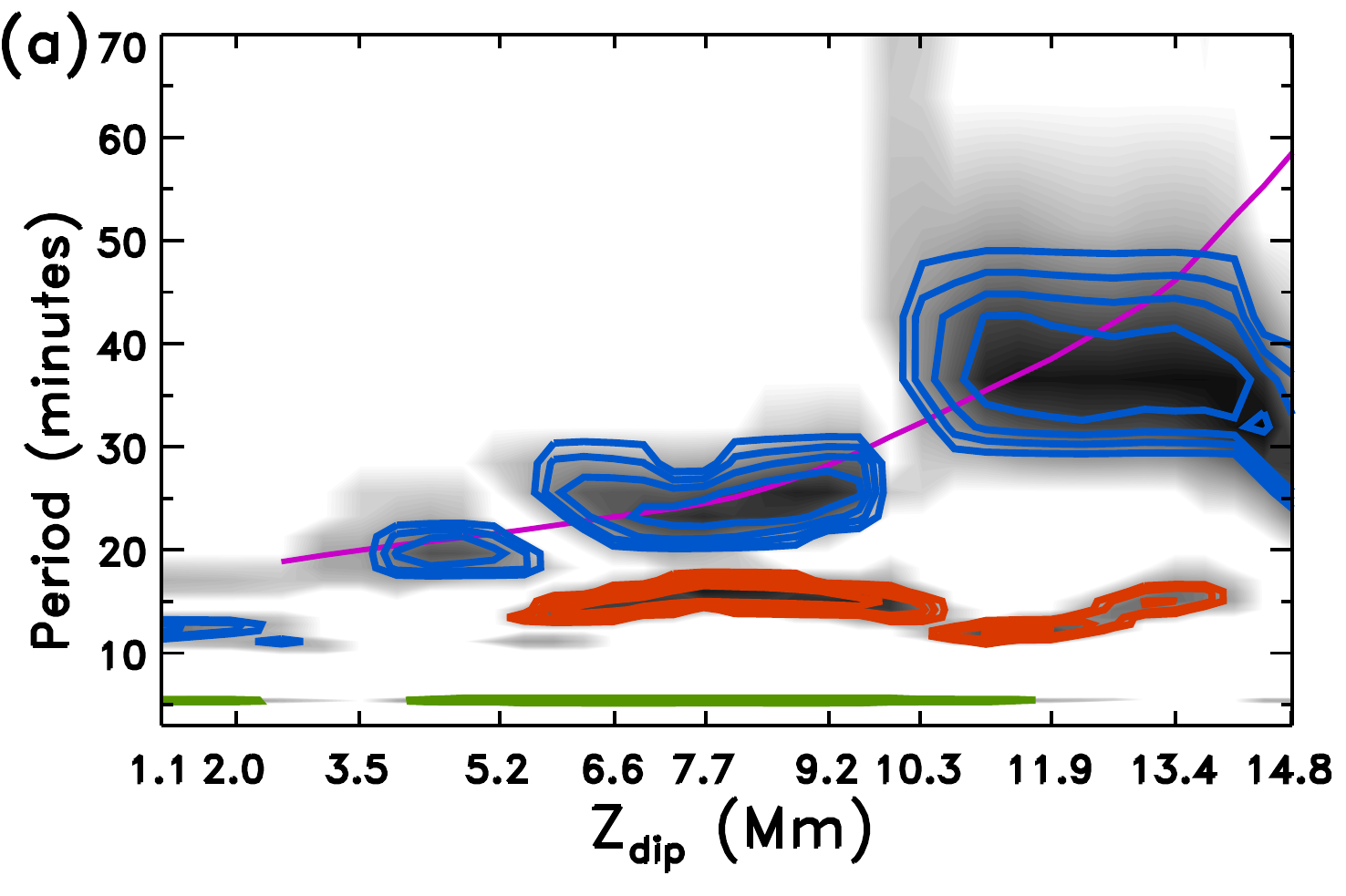}\\
	\centering\includegraphics[width=0.47\textwidth]{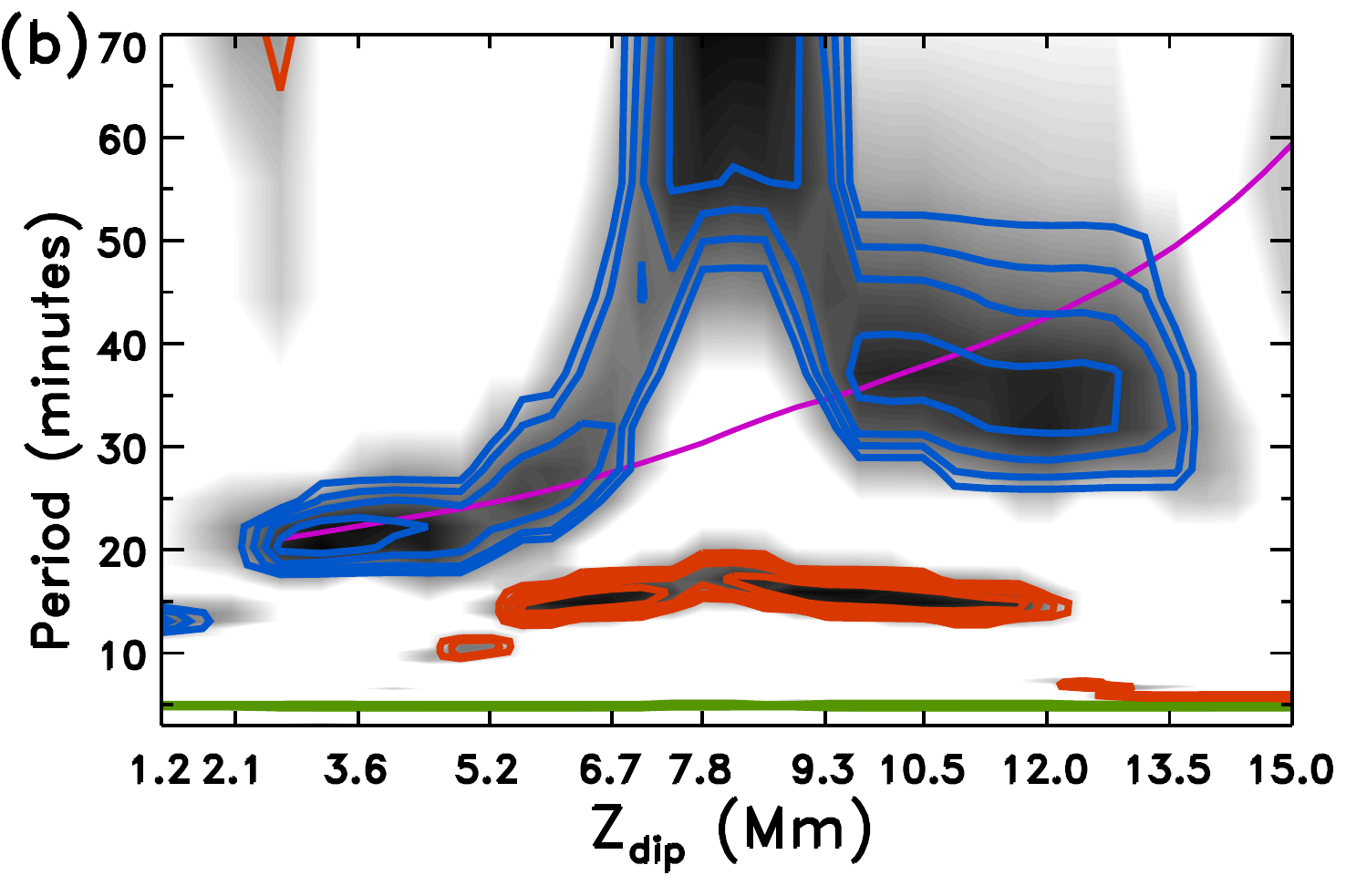}
\caption{PSD diagrams of the different velocity components for
     (a) case~1 and (b) case~2. Shown are isolines for the
      distribution of spectral power in a diagram period (in ordinates)
      versus location of the field line (as measured by $\zdip$, in
      abscissas). 
The blue, red, and green contours are for
$\vcmpar$, $\vcmperpy$, and $\vcmperpz$, respectively. The filled
 background is a grey-scale map of the total power, to enhance the
 identification of the oscillating regions in the diagram. The magenta line is the theoretical period from the pendulum model (see text).\label{fig:periods}}
\end{figure}

Using the $\vcmpar_i (t)$, $\vcmperpy_i (t)$ and $\vcmperpz_i(t)$ signals (Eqs. \ref{eq:long-pol}-\ref{eq:z-pol}) one can compute the power spectral distribution (PSD) using the periodogram method by \citet{lomb1976} and \citet{Scargle1982} with the algorithm created by \citet{carbonell1991}. Note that since the data are regularly spaced, the periodogram and the Fast Fourier Transform power spectrum are equivalent. Figure \ref{fig:periods} shows the PSD as a function of the period (vertical axis) for each of the field lines considered in the previous subsection, with their respective values of $\zdip$ used as abscissas. In the figure, isocontours for the periodograms for the $\vcmpar$ (blue), $\vcmperpy$ (red) and $\vcmperpz$ (green) are shown. Also, to improve the visibility, a grey-level map of the PSD jointly for all the components has been added.

It is possible to see some patterns in the PSD diagrams. For case~1 (Fig. \ref{fig:periods}(a)) between $\zdip=3.5$ and $\zdip=10.3\Mm$ the dark-blue isocontours associated with the longitudinal oscillations form a relatively narrow band of periods. The peak values of the PSD increase smoothly with the height from 20 to 25 minutes. Above $\zdip=10.3\Mm$ the band is wider and it seems that the central period is centered at $38$ minutes and does not change much with $\zdip$. This agrees with Figure \ref{fig:oscillations} where there is a phase shift in the longitudinal oscillation due to the increase of the period with height up to $\zdip=10\Mm$. Above this level, the oscillation seems very regular with a uniform phase. In the longitudinal oscillations, the restoring force is a combination of the gravity projected along the field lines in the dips and the gas pressure gradients \citep{luna2012,luna_effects_2012,zhang_observations_2012}. The gravity dominates over the gas pressure gradient in the so-called pendulum model, when the density contrast of the prominence with respect to the ambient corona is large and when the curvature of the dipped part is relatively large.
In order to compare the results of the simulations with the pendulum model, we compute the radius of curvature of each selected field line. With these radii of curvature and using the Equation (4) by \citet{luna2012} we compute the theoretical period of the longitudinal oscillations in each field line \citep[e.g.][]{luna2016,Liakh2020aa,Fan2020aa}. The theoretical period is shown as a magenta line in Figure \ref{fig:periods}. The theoretical line runs more or less over the blue dark isocontours below $\zdip=10\Mm$. Then we conclude that there is a relatively good agreement between the simulated results and the pendulum model in this region. Above $\zdip=10\Mm$ the magenta line crosses the dark region but it is not parallel to the dark region and diverges; the agreement is not good. A detailed analysis of the periods and the restoring forces involved in the oscillations are out of the scope of this work. However, above $\zdip=10\Mm$ the gas pressure gradient force is not negligible as a restoring force. The eruptive jet injects density and pressure at lines above $\zdip=10\Mm$ changing the physical conditions of the plasma and the pendulum approximation is not valid. In this region the oscillations are dominated by the slow mode. In contrast to the longitudinal oscillations, the periods for the vertical oscillations $\vcmperpz$ are very regular. For almost all $\zdip$ positions the period is the same and around 5 minutes. We found similar behavior in \citet{luna2016}. This oscillation is associated with a fast normal mode of the whole filament channel structure. For the Alfv\'en oscillations, $\vcmperpy$, the periodic motions are in the range $\zdip=5.2-14\Mm$ with a value that varies around 15 minutes. This value agrees with the recent observations of simultaneous longitudinal and transverse oscillations in filament threads \citep{Mazumder2020aa}.

In case~2, the situation is similar to case~1. Between $z=2.1-7\Mm$ the period increases with height. Above this position, the period increases rapidly having a peak between $z=7.8$ to $9.3\Mm$ approximately. The period is much larger than that given by the pendulum model. As in case~1, this discrepancy can be associated with the density injection by the eruptive jet that changes the physical conditions of the plasma and the pendulum approximation is not valid. Similarly to case~1, the vertical oscillations $\vcmperp$ are also very regular with a constant period in all positions of around 5 minutes. Also, the Alfv\'en oscillations are in the region $z=5-12\Mm$ with periods around 15 minutes.

\section{Discussion}\label{sec:conclusions}

In this work, a theoretical model of the interaction of a coronal jet with a prominence is presented for the first time. A parasitic bipolar region is included in the system near one edge of the FC that hosts the prominence. We energize the system in the initial phase of the simulation by shearing one of the magnetic arcades associated with the parasitic polarity, which leads to reconnection and the ejection of both a quiescent, collimated jet and a more violent eruptive one. These two phases are in agreement with the results of previous authors, in particular with those of \citet{Moreno-Insertis_2013}, \citet{Archontis_2013}, and  \citet{Wyper2018aa}, in which the arcade and its shear were obtained through different methods (see introduction).

We have found that an initial Alfv\'enic front propagates along the FC in advance of the jets. The Alfv\'enic disturbance is a natural consequence of the transfer across the reconnection site of the magnetic field component parallel to the arcade axis, a phenomenon already studied in the past \citep[e.g.][]{okubo_observations_1996,yokoyama_magnetic_1998,karpen_dynamic_1998,karpen_reconnection-driven_2017,Pariat09,pariat_model_2016,kigure_generation_2010,archontis_numerical_2013,uritsky_reconnection-driven_2017, Wyper2018aa}. In our case, at the head of the perturbation there is an abrupt change of direction of the field approximately traveling with Alfv\'en speed toward the prominence, with the perturbation causing only minor changes in the density and longitudinal velocity, like in standard nonlinear Alfv\'en waves. The front steepens as it advances, and becomes a switch-on shock when entering the prominence. The shock at the head of the jets, instead, travels at supersonic (but very sub-Alfv\'enic) speed, so the Alfv\'enic and acoustic fronts are well separated from each other: the jet propagates along already perturbed magnetic field. As a precedent for this separation of advancing fronts one can mention the results of \citet{pariat_model_2016} and \citet{Wyper2018aa}. \citet{pariat_model_2016} carried out jet launching experiments for different values of the plasma $\beta$ in the general framework developed by \citet{Pariat09}. Although similar at first sight, there are interesting differences between their results and ours: in their low-$\beta$ cases (see their Figure~6) these authors found an Alfv\'enic front advancing at Alfv\'en speed followed by, and clearly separated in space from, a high-density and -temperature perturbation that moves with the speed of the plasma (not faster, as it should if the front were an acoustic shock, even a strong one). Concerning the first front, both the front itself and the plasma in its trail move approximately with Alfv\'en speed (e.g.: Alfven Mach number close to unity for the plasma). These facts are at variance with our findings, which follow (a) the expected pattern for an Alfv\'enic perturbation that is steepening to later become a switch-on shock, and, (b) concerning the trailing compressive perturbation, the pattern of a mildly supersonic (but strongly sub-Alfv\'enic) slow-mode shock propagating along the field lines. Maybe the discrepancy stems from the necessarily complicated geometry and magnetic topology of the 3D experiment. 
On the other hand, an Alfv\'enic perturbation followed by a compressive one was found by \citealt{uritsky_reconnection-driven_2017}; this structure was located in the turbulent wake following the leading edge of a jet in an experiment with the temperature kept fixed at the (uniform) initial value, which can enhance the density jump at the acoustic front. \citet{Wyper2018aa}, finally, comment on the separation between the Alfv\'enic front and a trailing density enhancement for the eruptive jet phase in their experiment, and mention that it is likely that their scenario in this respect is similar to that of \citet{pariat_model_2016}%

Looking now at the Alfv\'enic front from the point of view of prominence observations, there are no reports that a jet impinging on a prominence is preceded in time by an Alfv\'enic disturbance.
However, checking in detail the published observational material of \citet{zhang_simultaneous_2017} it is possible to see that the prominence threads started to move conspicuously approximately $15$ minutes before the arrival of the bright jet. We see that the flux rope structure is executing what looks like a rolling motion that could tentatively be associated with the arrival of an Alfv\'enic perturbation reaching the prominence before the hot plasma.
In fact, in the observations by \citeauthor{zhang_simultaneous_2017} the jet is seen to reach the prominence traveling with a projected speed of approximately $224 \kms$ from its source, located some $225\Mm$ away. Assuming the Alfv\'en speed to be about $1000\kms$, one can estimate a value of some $13\min$ for the delay between the arrival of the two perturbations, which agrees very well with the observation. 
We have already observed this behavior in other events which, so far, have gone unreported.
Conversely, if a precise determination of the temporal delay between the arrival of the Alfv\'enic disturbance and the jet could be carried out, then one could obtain an estimate for the average Alfv\'en velocity, possibly also for the field strength, in the FC.
This method could provide a new seismological tool to infer the hard-to-measure FC magnetic field. This method will be valid only in jets produced in the FC that host the prominence and not with distant triggers that are magnetically disconnected from the prominence.

The successive partial reflections and transmissions of the jet fronts in the prominence and the reflections at the chromospheric ends of the field lines lead to a complex pattern of counter-streaming flows both in the hot (coronal) part of the FC and, as part of the longitudinal oscillations, in the prominence. In the hot region adjacent to the prominence the flow speed decreases with time: it is $130\kms$ during the first arrival of the jet and around $30\kms$ at the end of the simulation. The cool prominence plasma moves with velocities ranging from a few $\kms$ up to a maximum of $30\kms$ in some regions of the prominence. Counter-streaming flows with speeds of 5-20 $\kms$ have been routinely observed in cool prominence threads \citep{zirker_counter-streaming_1998,wang_extending_2018} but also, with
velocities up to $100\kms$, in EUV lines in the surroundings of a
filament \citep[e.g.][]{alexander_anti-parallel_2013}. Our results agree with the values observed in the threads indicating that the longitudinal oscillations can contribute to the observed counter-streaming flows in the
cool prominence plasma as suggested by \citet{chen_imaging_2014}
and \citet{zhou_simulations_2020}.
Following the results of the present paper, we suggest that the flows established in the FC in the aftermath of the impact of the jets could contribute to the observed counter-streaming flows in the hot surroundings of the filament threads. 
However, other mechanisms, like the process of evaporation and condensation in prominences \citep[e.g.,][]{Antiochos_etal_1999}, could also lead to alternating flows in the hot parts of the FC as \citet{zhou_simulations_2020} proposed.
%
The switch-on Alfv\'enic disturbance and both phases of the jet trigger large-amplitude oscillations of the prominence in different directions.
This is the first time that the generation of LAOs is explained in a self-consistent model, in contrast to previous works, which imposed artificial perturbations \citep{terradas_magnetohydrodynamic_2013,luna2016,zhou_three-dimensional_2018,Liakh2020aa}.
The Alfv\'enic disturbance excites the fundamental mode called string or hybrid mode with a period of around 15 minutes in agreement with recent observations \citep{Mazumder2020aa}. 
Concerning the longitudinal oscillations, the two phases of the jet, quiescent or eruptive, trigger oscillations with different characteristics in the prominence.
The initial position of the NP determines the maximum height of impact of the jets in the prominence. It also leads to different relative intensities of the impact of the quiescent and eruptive jets. The study of the amplitudes of the longitudinal LAOs may also shed light on the jets that initiated the periodic motion. The periods of the longitudinal oscillations range from 20 to 70 
minutes, which is in agreement with the observed values \citep[e.g.,][]{luna2018}.
The literature on LAOs \citep[see review by][]{Arregui2018aa} distinguishes two groups of agents that trigger LAOs in prominences. The first group is associated with MHD shock waves from distant flares that trigger longitudinal, transverse, or mixed polarity LAOs in prominences. The second group is associated with subflares, microflares, or jets that push the mass in the direction of the field, thus producing longitudinal LAOs. In the second group, there is a magnetic connection between the prominence and the source of the perturbation. Our model is only applicable to this second group of events, and the perturbations propagate from the reconnection site to the prominence along the FC structure.
Our work theoretically demonstrates that a nearby jet can trigger longitudinal oscillations in the prominence as in \citet{luna_observations_2014}. A more recent observation by \citet{zhang_simultaneous_2017} shows a much larger structure, with the jet source located very far from the prominence, 225 Mm. Yet, both are magnetically connected and this event can therefore be classified as a disturbance in the second group. The jet reaches the prominence and triggers both longitudinal and transverse modes and not only movements along the field.
Our theoretical findings also support this observational evidence: the jet reaches the prominence and triggers not just transverse, but also longitudinal modes in it.
This indicates that the observed LAOs triggered by jets and identified as longitudinal oscillations can be actually the superposition of two perpendicular modes. The combined direction of the motion is not aligned with the local magnetic field. This misalignment of the motion with the magnetic field could lead to an error in the determination of the field direction using seismological tools.
However, when the periods of both modes are well separated, the average direction of the motion in several oscillation cycles gives the averaged direction of the magnetic field. 

The physics of coronal jets and prominences have been studied separately in the past. However, observations in the past several years have shown that coronal jets can impact prominences and set them in motion. In this work, we explore the physics behind this interaction: the launching of the jets along the filament channel, the impact onto the prominence with consequent oscillations of the dense plasma, and the resulting flows on the other side of the prominence. We conclude that the jets can significantly perturb the prominence, which can end up executing large-amplitude oscillations.
Additional theoretical work is necessary to understand the influence of the different parameters of the FC, prominence, and jet in the resulting oscillations. In addition, three-dimensional simulations must be carried out to have a more realistic experiment. All these points will be the subject of future research.


\section*{Acknowledgements}
This research has been partially supported by the Spanish Ministry of Economy, Industry and Competitiveness (MINECO) through projects AYA2014-55078-P, PGC2018-095832-B-I00 and through the 2015 Severo Ochoa Program SEV-2015-0548. The authors are also grateful to the European Research Council for support through the Synergy Grant number 810218 (ERC-2018-SyG). We thankfully acknowledge the technical expertise and assistance provided by the Spanish Supercomputing Network (Red Espa\~nola de Supercomputaci\'on), as well as the use of the LaPalma Supercomputer, located at the Instituto de Astrof\'isica de Canarias. Resources partially supporting this work were also provided by the NASA High-End Computing (HEC) Program through the NASA Center for Climate Simulation (NCCS) at Goddard Space Flight Center. M.~Luna acknowledges support from the MINECO through the Ramón y Cajal fellowship RYC2018-026129-I and from the International Space Sciences Institute (ISSI) through the team 413 on ``Large-Amplitude Oscillations as a Probe of Quiescent and Erupting Solar Prominences''. The authors are grateful to the anonymous referee for the thorough revision of the manuscript and many interesting and useful suggestions. 

%
%

\end{document}